\documentclass[aps,pre,twocolumn,superscriptaddress,floatfix]{revtex4-1}

\usepackage{bm}
\usepackage{subfigure}
\usepackage{epsfig}
\usepackage{subfigure}
\usepackage{float}
\usepackage{amsmath,amsfonts,amssymb,graphicx}
\usepackage{graphicx}
\usepackage{dcolumn}
\usepackage{xcolor}

\begin{document}

\title{Anomalous in-gap edge states in two-dimensional pseudospin-1 Dirac insulators}

\author{Hong-Ya Xu}
\affiliation{School of Electrical, Computer, and Energy Engineering, Arizona State University, Tempe, Arizona 85287-5706, USA}

\author{Ying-Cheng Lai} \email{Ying-Cheng.Lai@asu.edu}
\affiliation{School of Electrical, Computer, and Energy Engineering, Arizona State University, Tempe, Arizona 85287-5706, USA}
\affiliation{Department of Physics, Arizona State University, Tempe, Arizona 85287-5706, USA}

\date{\today}
\begin{abstract}

Quantum materials that host a flat band, such as pseudospin-1 lattices and magic-angle twisted bilayer graphene, can exhibit drastically new physical phenomena including unconventional superconductivity, orbital ferromagnetism, and Chern insulating behaviors. We report a surprising class of electronic in-gap edge states in pseudospin-1 materials without the conventional need of band-inversion topological phase transitions or introducing magnetism via an external magnetic type of interactions. In particular, we find that, in two-dimensional gapped (insulating) Dirac systems of massive spin-1 quasiparticles, in-gap edge modes can emerge through only an {\em electrostatic potential} applied to a finite domain. Associated with these unconventional edge modes are spontaneous formation of pronounced domain-wall spin textures, which exhibit the feature of out-of-plane spin-angular momentum locking on both sides of the domain boundary and are quite robust against boundary deformations and impurities despite a lack of an explicit topological origin. 
The in-gap modes are formally three-component evanescent wave solutions, akin to the Jackiw-Rebbi type of bound states. Such modes belong to a distinct class due to the following physical reasons: three-component spinor wave function, unusual boundary conditions, and a shifted flat band induced by the external scalar potential.
Not only is the finding of fundamental importance, but it also paves the way for generating highly controllable in-gap edge states with emergent spin textures using the traditional semiconductor gate technology. Results are validated using analytic calculations of a continuum Dirac-Weyl model and tight-binding simulations of realistic materials through characterizations of local density of state spectra and resonant tunneling conductance.

\end{abstract}
\maketitle

\section{Introduction} \label{sec:intro}

The physics of quantum materials hosting a flat band, such as the
magic-angle twisted bilayer graphene, has become a forefront area of research.
These materials can generate surprising physical phenomena such as
unconventional superconductivity~\cite{Caoetal:2018,YCPZWTGYD:2019}, orbital
ferromagnetism~\cite{Sharpeetal:2019,Luetal:2019}, and the Chern insulating
behavior with topological edge states. The purpose of this paper is to report
the surprising emergence of a class of in-gap edge states in two-dimensional 
Dirac/Weyl pseudospin-1 materials, which cannot be fit into any of the known 
scenarios for producing such states. The uncovered states, at their birth, 
exhibit topologically nontrivial domain-wall like pseudospin textures.

In modern physics, the emergence of low-dissipation or dissipationless
topological surface or edge states in condensed matter systems is a
fascinating phenomenon~\cite{KDP:1980,TKNN:1982,Haldane:1988}, as
exemplified by topological insulators (TIs)~\cite{BHZ:2006,FK:2007,ZLQDFZ:2009,KWBRBMQZ:2007,HQWXHCH:2008,XQHWPLBGHC:2009,Moore:2010,HK:2010,QZ:2011}. A TI
has a bulk band gap so its interior is insulating but there are gapless
surface states within the bulk band gap, which are protected by the
time-reversal symmetry that renders the states robust against backscattering
from nonmagnetic impurities. These topologically protected surface or edge
states possess a perfect spin-momentum locking characterized by the invariance
of spin orientation with respect to the direction of the momentum. Quite
recently, high-order TIs hosting, e.g., robust in-gap excitations of
zero-dimensional corner modes have been uncovered~\cite{Benalcazar2017,Song2017,Schindlereaat2018}. Topological states of matter, in addition to their
importance to fundamental physics, have potential applications in electronics
and spintronics~\cite{PM:2012}. For electronic systems, current understanding
of the physical mechanisms behind the topological edge states requires
a discontinuous change in the associated bulk topological invariants across the
interface/edge rendered by, e.g., a strong external magnetic field in a
two-dimensional electron gas~\cite{TKNN:1982}, band inversion driven by
spin-orbit coupling~\cite{Kane2005a,Kane2005b,BHZ:2006}, introduction of
ferromagnetism in topological insulators~\cite{Chang2013}, presetting
domain walls in gapped Dirac materials~\cite{Semenoff2008,Martin2008,Qiao2011,Yasuda2017}, stacking order in layered two-dimensional
materials~\cite{Zhang2011}, and particular spatial crystalline
symmetries~\cite{Fu2011b}.

Pseudospin-1 type of low-energy excitations beyond the Dirac-Weyl-Majorana
paradigm have recently been realized in electronic lattice
systems~\cite{MaJL2012,RomhMannyi2015,Zhong2017,Slot2017,Bradlynaaf2016,Takane2019}. In a broader context, two-dimensional massive spin-1 bulk
excitations can arise in classical {\em nonlinear} physical systems such as
rotating shallow water in a horizontally unbounded plane~\cite{Delplace2017}
and the wave system of magnetoplasmons, where the corresponding Hamiltonian
representations~\cite{Jin2016} can be effectively reduced to the Dirac-like
equation for spin-1 particles. Applying the sign-changing Dirac mass scenario
to the systems leads to an extension of the Jackiw-Rebbi mechanism that serves
to ascertain the topological origin of, e.g., the equatorial
waves~\cite{Delplace2017}, as well as rich topological phenomena in
bosonic and classical systems~\cite{Gomes2012,Tarruell2012,Klembt2018,Yang2019,Kane2013}.

The subject of our study is pseudospin-1 relativistic quantum systems described
by the generalized Dirac-Weyl equation which are fundamentally linear.
Specifically, low-energy excitations in condensed matter systems such as
graphene~\cite{Novoselovetal:2005} and topological insulators~\cite{BHZ:2006,FK:2007,ZLQDFZ:2009,KWBRBMQZ:2007,HQWXHCH:2008,XQHWPLBGHC:2009,Moore:2010,HK:2010,QZ:2011} and in analogous physical systems of molecules, cold atoms,
cavity polaritons, light and even mechanical waves in judiciously designed
lattices~\cite{Gomes2012,Tarruell2012,Klembt2018,Yang2019,Kane2013} are
described by the Dirac-Weyl equation. In those circumstances, if the
corresponding quasiparticles are massless, the energy band structure
contains a pair of Dirac cones characteristic of the relativistic
energy-momentum dispersion relation. A finite mass leads to a band gap,
giving rise to unconventional topological phases in Dirac material
systems~\cite{Wehling:2014} with unusual physical properties associated with
tunneling, confinement and transport, which have no analogies in quantum
systems described by the Schr\"{o}dinger equation. Among those, the physics of
edge states and robust in-gap excitations are of fundamental interest.
Jackiw and Rebbi~\cite{Jackiw1976} predicted a surprising zero-energy bound
state solution of the Dirac equation in the presence of a kink-shaped mass
profile that generates a domain wall separating regions with sign-changing
Dirac mass. The realization in polyacetylene~\cite{SSH1979,Heeger1988} and
the theoretical studies of narrow-gap semiconductors~\cite{Volkov1985,Pankratov1987} led to the discovery of the phenomenon of band-gap inversion
enabling topologically protected conducting interface states and localized
sub-gap excitations in TIs~\cite{BHZ:2006,FK:2007,ZLQDFZ:2009,KWBRBMQZ:2007,HQWXHCH:2008,XQHWPLBGHC:2009,Moore:2010,HK:2010,QZ:2011,Benalcazar2017,Song2017,Schindlereaat2018}. In the description based on the massive Dirac
equation, band-gap inversion is equivalent to a sign change in the mass.
The topological edge states give rise to appealing physical properties and
phenomena such as robust low-power-dissipation wave transport~\cite{Qiao2011},
electrically tunable magnetism~\cite{Wang2015b}, and quasiparticles analogous
to elementary fermionic particles in high-energy physics~\cite{Shen2013}.

Our main finding is that, in pseudospin-1 systems with an energy gap, a
surprising class of in-gap edge bound states can arise {\em without}
band or mass inversion based domain-walls that separate the regions with
different kinds of bulk band topology and any external magnetic interaction, 
but these states are remarkably robust against geometric deformations and 
impurities. In fact, they are generated through {\em only} a local
electrostatic potential barrier of the repulsive type in the underlying 
insulating spin-1 systems. We uncover a number of remarkable, quite unusual 
spectral properties of these modes. Unlike the topological edge states 
previously discovered and studied, the states reported here require no 
established topological restrictions such as interfacing domains/systems of 
different bulk topological invariants and any particular type of discrete 
symmetries. 
In fact, through self-inducing topological spin textures, the uncovered states 
possess the degree of robustness enjoyed by conventional topological states 
but they belong to a distinct class due to the following physical reasons: 
three-component spinor wave function, unusual boundary conditions, and a 
shifted flat band induced by the external electrical potential.
Experimentally, these states can be generated readily 
through routine electrostatic gating within the same material (or within a 
single device), rendering them promising in applications, e.g., a 
gate-controlled spin-1 Dirac electron transistor of high on/off ratio.

\section{Results from continuum Dirac-Weyl Hamiltonian} \label{sec:Dirac_Weyl}

\begin{figure}[ht!]
\centering
\includegraphics[width=\linewidth]{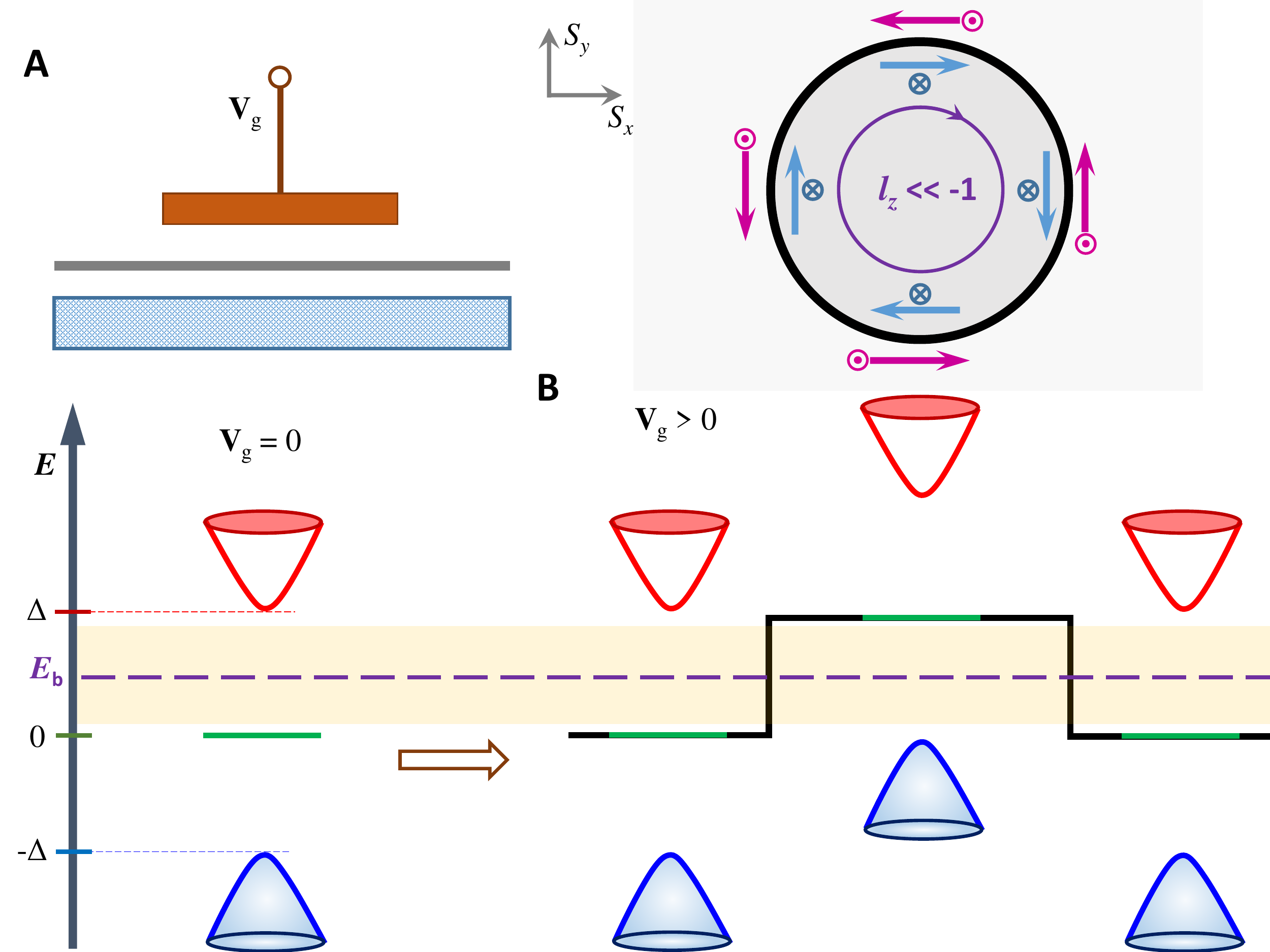}
\caption{ Schematic illustration of the system setting and main finding.
(A) A side view of the setting leading to in-gap edge modes without
magnetism and the conventional band inversion topological phase transition, 
where a gapped two-dimensional system (thick gray line) hosting
Dirac-like low-energy excitations of massive spin-1 is subject to a locally
applied electrostatic potential. The energy band diagram in the absence of
the potential is shown on left side of the bottom panel [below (A)]. (B) In-gap
edge bound modes with an emergent domain-wall like spin ordering/texture (top
panel) arise in the presence of a repulsive type of potential, which defines 
an antidot profile as shown in the bottom panel.
The criterion for the stable emergence of the in-gap 
states is $|V_g-\Delta|\lesssim\Delta/2$.	
}
\label{fig:fig1}
\end{figure}

\subsection{Illustration of finding}

Figure~\ref{fig:fig1}A presents a schematic illustration of the system setting,
whose effective Hamiltonian is
$H_{eff} = v_F\bm{\hat{S}}\cdot\bm{\hat{p}}+\Delta \hat{S}_z + U(\bm{r})$,
where the first term describes the bulk low-energy excitation of a massive
spin-1 particle with quasi-momentum $\bm{\hat{p}}=(p_x,p_y)$, the second term
represents the generalization of the Dirac mass with $\hat{S}_z$ being a
component of the spin-1 matrix vector $\bm{\hat{S}}$, and the last term is
the locally applied electrostatic potential of height $V_g$ which defines a
closed interface at the boundary. As we will establish, this magnetism-free
configuration permits in-gap edge states, and the states with higher angular 
momenta possess highly organized domain-wall like spin textures, as illustrated
in Fig.~\ref{fig:fig1}B. In general, for the in-gap states to emerge 
and be stable, the perturbation in the form of the applied gate potential 
$V_g$ cannot be negligibly small in comparison with the pristine band gap 
$\Delta$. Neither can the perturbation be too large to result in a 
substantially reduced effective band gap size. In fact, the inequality 
$V_g<2\Delta$ is required and the reduced band gap ($2\Delta-V_g$) should 
be comparable to the pristine one. In terms of $\Delta$ and $V_g$ as 
defined in Fig.~\ref{fig:fig1}, the criterion for the stable emergence of 
the in-gap states is $|V_g-\Delta|\lesssim\Delta/2$. The case shown in 
Fig.~\ref{fig:fig1} is for $V_g\simeq\Delta$.

\subsection{Emergence of in-gap edge states}

\begin{figure*}[ht!]
\centering
\includegraphics[width=\linewidth]{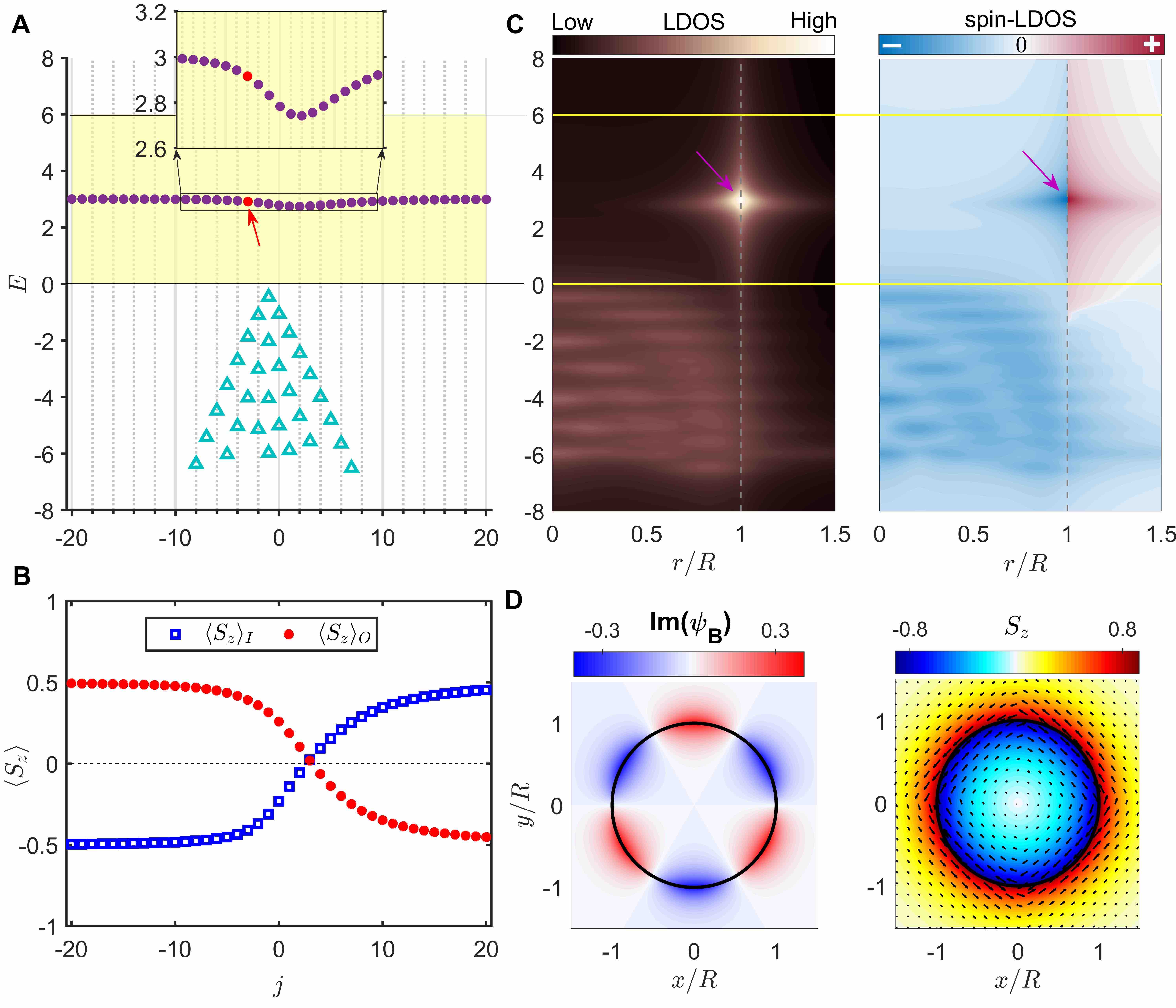}
\caption{Emergence of in-gap edge modes.
(A) Eigenenergy $E$ (in units of $\hbar v_F/R$) as a function of the total
angular momentum $j$ for $V_g=\Delta=6\hbar v_F/R$. The light yellow shaded
area represents the band gap. The inset shows the in-gap modes within the same 
energy range as that of Fig.~\ref{fig:fig3}B. The light blue triangles 
denote the common eigenstates due to the induced quantum dot confinement of 
bulk valence band carriers, where all the corresponding wavefunctions are
localized within the dot, see, e.g., complementary Fig.~\ref{fig:SIfig0} 
in Appendix~\ref{appendix_B}. (B) Expectation values of $S_z$ versus $j$
for the in-gap modes marked by the purple dots in (A). The values are
evaluated on both sides of the boundary, which are denoted by
$\langle S_z\rangle_I$ (blue squares; inside the domain) and
$\langle S_z\rangle_O$ (red dots; outside of the domain), respectively.
(C) LDOS and spin-resolved LDOS maps versus energy $E$ and the radial
spatial position $r/R$ associated with the spectra in (A), where an empirical
parameter value $\Gamma/\epsilon_*=0.2$ is used to characterize the energy
broadening effect as in an experimental situation. (D) Spatial profiles of
wave (left panel) and spin texture (right panel) distributions of the in-gap
mode indicated by the red arrow in (A).}
\label{fig:fig2}
\end{figure*}

\begin{figure*}[ht!]
\centering
\includegraphics[width=\linewidth]{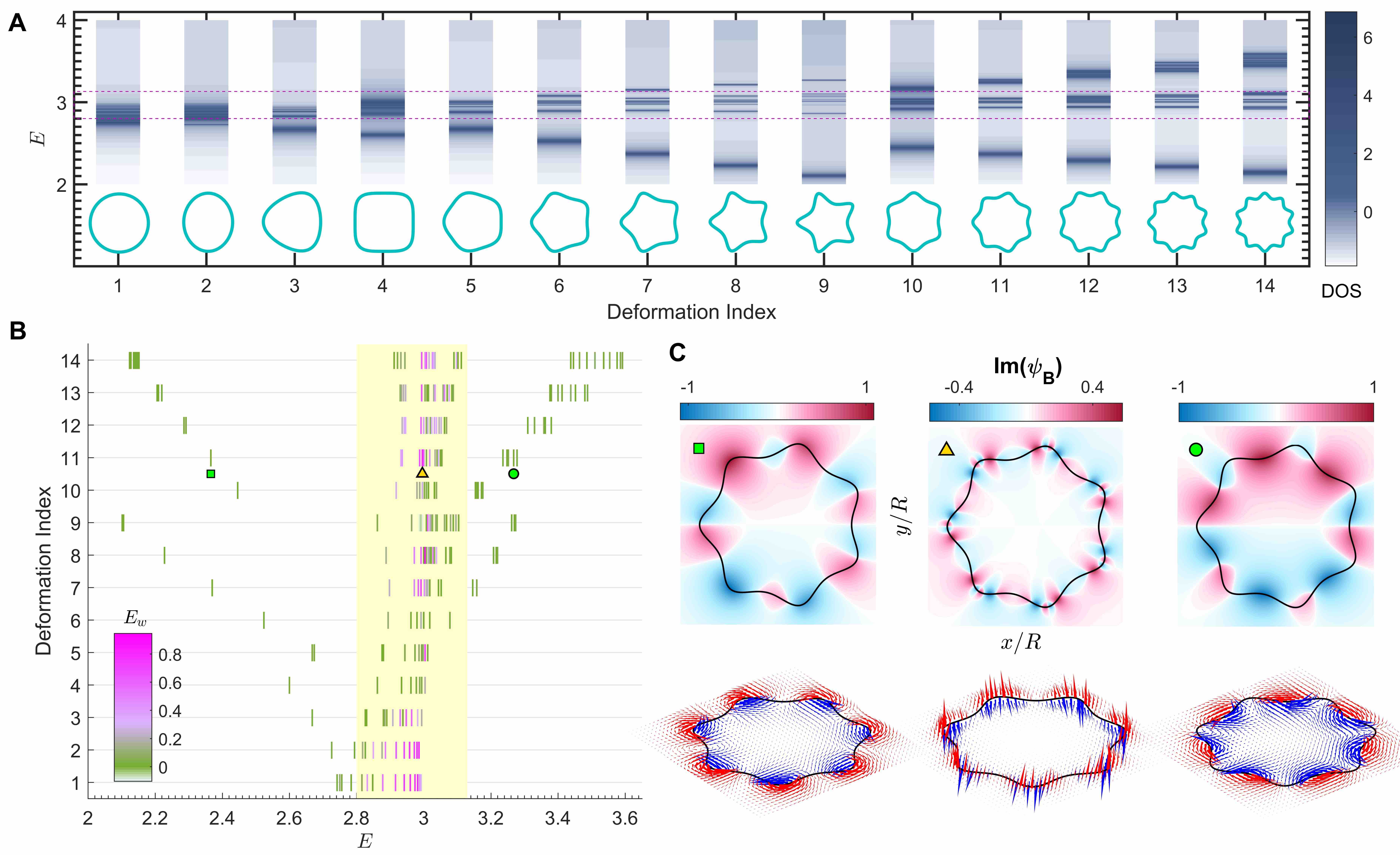}
\caption{Robustness of in-gap modes against geometric deformations
of the domain. (A) DOS based spectral lines for 14 boundary shapes (inset).
(B) Dependence of the energies of the in-gap edge modes on the deformed shape
as revealed by a color-coded map of the effective (exchange) energy penalty
$E_w$ for forming a globally organized domain-wall spin texture, defined as
the dot product of the spin expectation values inside and outside
of the domain for each mode. The penalty attains large values for edge modes
with a strong domain-wall ordering but has small values for ones with a
dominant in-plane vortex spin texture. 
The yellow shaded region is for eye guidance of the 
approximately invariant energy range in which the in-gap modes arise in
the presence of systematically varying geometric deformations. 
(C) Representative real-space wave (top panel) and spin texture (bottom panel) 
profiles of the categorized in-gap edge modes indicated by the corresponding 
color-filled markers in (B) for three distinct energy values.}
\label{fig:fig3}
\end{figure*}

For a circular domain of radius $R$, the electrostatic potential is given by
$U(\bm{r}) = V_g\Theta(R-r)$, where $\Theta$ is the Heaviside function.
The system as governed by $H_{eff}\psi=E\psi$ can be solved analytically in
the polar coordinates $\bm{r}=(r,\theta)$ to yield closed-form solutions
of the form
\begin{equation}
\psi_j^\mu(r,\theta)= \frac{1}{\sqrt{2}}
\left(\begin{array}{c}
\frac{\hbar v_F k_\mu}{E-\Delta}Z_{j-1}^\mu(k_\mu r)e^{-i\theta} \\
i\sqrt{2}Z_j^\mu(k_\mu r) \\
-\frac{\hbar v_F k_\mu}{E+\Delta}Z_{j+1}^\mu(k_\mu r)e^{i\theta}
\end{array}\right)e^{ij\theta},
\end{equation}
where $\mu=I, O$ labels the inner and outer regions as defined by the
interface, $\hbar v_Fk_\mu=\sqrt{E_\mu^2 - \Delta^2}$, and $F_j^I=J_j$ and
$F_j^O=H_j^{(1)}$ are the Bessel and the Hankel functions of the first kind
with $j$ being the integer angular momentum quantum number. As explained in 
Appendices~\ref{appendix_A} and \ref{appendix_B}, 
the in-gap modes uncovered take the form of a three-component 
evanescent edge state. In comparison with known edge states, either 
topological or non-topological, the states uncovered here belong to a 
distinct class due to the following physical reasons: three-component spinor 
wave function, unusual boundary conditions, and a shifted flat band induced by
the external scalar potential.
Particularly, for $|j|\gg1$, we calculate the eigenenergy $E\approx V_g/2$ 
and the resulting spin textures
\begin{eqnarray}
	\bm{S}&=&[S_x,S_y,S_z] \nonumber \\
	&\approx& [\sin\theta\sin\Phi(r),-\cos\theta\sin\Phi(r),\cos\Phi(r)],
\end{eqnarray}
where 
\begin{displaymath}
\cos\Phi(r) = j/\sqrt{j^2+\xi^2}[2\Theta(R-r)-1] 
\end{displaymath}	
with $\xi=(V_g+2\Delta)r/2\hbar v_F$. Concretely, for a representative 
parameter setting, e.g., $V_g=\Delta=6\hbar v_F/R$, we calculate the 
resulting energy spectra as a function of the angular momentum quantum 
number $j$, as shown in Fig.~\ref{fig:fig2}A. We see that additional
bounded eigenstates arise in the gap, i.e., those in the shaded area in
Fig.~\ref{fig:fig2}A. The striking feature is that these states emerge for
$E_\mu < \Delta$, where the system is an insulator. In this case, without any
change in the band topology (e.g., due to band inversion), conventional
understanding of TIs stipulates that such states are impossible.

A feature of the spin textures is worth mentioning. If we calculate the 
topological number defined as 
\begin{displaymath}
\mathcal{N}=\frac{1}{4\pi}\iint\bm{n}\cdot\left(\frac{\partial\bm{n}}{\partial x}\times\frac{\partial\bm{n}}{\partial y}\right)dxdy,
\end{displaymath}
where $\bm{n}=\bm{S}/|\bm{S}|$, we get
\begin{displaymath}
\mathcal{N}=-\textrm{sign}(j)/2,
\end{displaymath}
signifying vortex-like spin textures that can arise from in-gap excitations 
of meron-like skyrmions~\cite{Nagaosa2013}. Similar features have been 
predicted in chiral $p$-wave superconductors~\cite{SV:1987,GB:2012} that have
the same symmetry class as the spin-1 Dirac Hamiltonian studied in this paper 
[Eq.~(\ref{eq:H1})].

To gain further insights, we characterize the energy spectra using two
experimentally relevant quantities: the local density of state (LDOS) and
spin-LDOS defined as
$D(E,r) = \sum_{\nu}\langle{\nu}|{\nu}\rangle\delta(E-E_{\nu})$ and 
$D_s(E,r) = \sum_{\nu}\langle{\nu}|S_z|{\nu}\rangle\delta(E-E_{\nu})$, 
respectively, where $\nu$ is the eigenstate label. As shown in
Fig.~\ref{fig:fig2}C, the in-gap modes are localized at the boundary and
exhibit distinct domain wall spin textures, where the energy broadening
effect (e.g., caused by measurement) has been taken into account by
approximating the delta function as ${\Gamma}/{\pi [(E-E_\nu)^2 + \Gamma^2]}$
with $\Gamma=0.2\epsilon_*$. Figure~\ref{fig:fig2}D shows the spatial
distributions of the corresponding wave density and spin texture for a
representative state (indicated by the red arrow in Fig.~\ref{fig:fig2}A). A
calculation of the associated spin projection $\langle S_z\rangle$ versus $j$
in the inner and outer regions reveals that the domain wall spin ordering is
more pronounced for states with higher angular momenta, as shown in
Fig.~\ref{fig:fig2}B. Associated with the strengthening of the spin ordering,
the energy flow tends to decrease, as revealed by a nearly dispersionless
dependence of the energy level on the angular momentum quantum number, as
shown in Fig.~\ref{fig:fig2}A. Figure~\ref{fig:fig2}B demonstrates the
emergence of spin-angular momentum locking that depends on the side of the
interface in which the state is located, suggesting that these states are
robust.

\subsection{Robustness}

The robustness of the QSH and QAH edge states are known to be protected by
the presetting discontinuous change in the associated bulk topological
invariants across the interface, such as the $Z_2$ index and Chern number,
all requiring some sort of magnetic interaction. However, for the edge modes
demonstrated in Fig.~\ref{fig:fig2}, there is no such {\em a priori}
topological origin/restriction. The question is whether the modes are 
protected or stable against irregular perturbations. To address this question,
we consider a general type of perturbation: geometric deformation of the 
potential domain. A significant challenge is to obtain accurate eigensolutions 
of the massive spin-1 Dirac equation as, with an irregular domain, analytic 
solutions are no longer possible. We have developed an accurate and efficient 
numerical method to find solutions for arbitrarily shaped domain interfaces 
(Appendix~\ref{appendix_C}). As an illustration, we create deformed domains 
via the superformula in botany that can generate a great diversity of natural 
shapes with only a few parameters~\cite{Gielis2003}. Figure~\ref{fig:fig3}A 
shows, for thirteen deformed boundary shapes (insets), the corresponding 
energy spectra resolved by the total density of states (DOS). With respect to 
the eigenstates of the circular geometry, there are considerable shifts (up or
 down) in some eigenenergies of the strongly deformed domains but, importantly,
 there are stable states with virtually no changes in their energies in spite 
of the severe deformations.

To ascertain the nontrivial feature of the in-gap states, we examine the 
associated spin properties. In particular, we introduce an effective 
exchange energy penalty:
\begin{displaymath}
E_w = -\langle \bm{S}\rangle^I\cdot\langle \bm{S}\rangle^O, 
\end{displaymath}	
to identify a domain wall like spin ordering structure between the inner and 
outer regions. It can be seen from Fig.~\ref{fig:fig3}B that the stable modes 
insensitive to deformation attain large energy penalties, a strong
indication of the emergence of domain wall spin ordering, while the modes 
with small values of $E_w$ are sensitive to deformations. 
Figure~\ref{fig:fig3}C shows the real-space wave density and the corresponding 
spin texture patterns of three representative states as indicated in 
Fig.~\ref{fig:fig3}B. The wave density topography associated with the strong 
domain wall spin texture is mainly contributed by the high angular momentum 
states (those with distinctly more angular nodes - c.f., middle panel of 
Fig.~\ref{fig:fig3}C). This agrees with the prediction of the continuum theory
that a nearly perfect out-of-plane spin-angular momentum locking should
emerge for the high orbital angular momentum states, as shown in 
Fig.~\ref{fig:fig2}B, providing the physical reason for the robustness.   
(Intuitively, this behavior can be understood that a faster spinning egg is 
able to stand upright in a more stable manner.) The unambiguous signature 
of spin-angular momentum locking can greatly circumvent mode coupling due to 
backscattering caused by the deformation. For those modes, the 
conventionally anticipated level repulsion/shifting effect due to geometric 
deformation is greatly suppressed, an unequivocal indication that the modes 
with spin-angular momentum locking are robust with self-induced protection.

As in most studies of TIs~\cite{BHZ:2006,BZ:2006}, we have employed a
sharp potential boundary to demonstrate the findings. However, by performing
calculations using a finite difference method for realistic and smoothly
varying potential profiles, we find that the topological states as
exemplified in Figs.~\ref{fig:fig2} and \ref{fig:fig3} persist 
(Appendix~\ref{appendix_C}). We also find that these states can tolerate 
strong disorders.

{
\section{Results from tight-binding calculations of an experimentally 
relevant lattice model}

\begin{figure}[ht!]
\centering
\includegraphics[width=\linewidth]{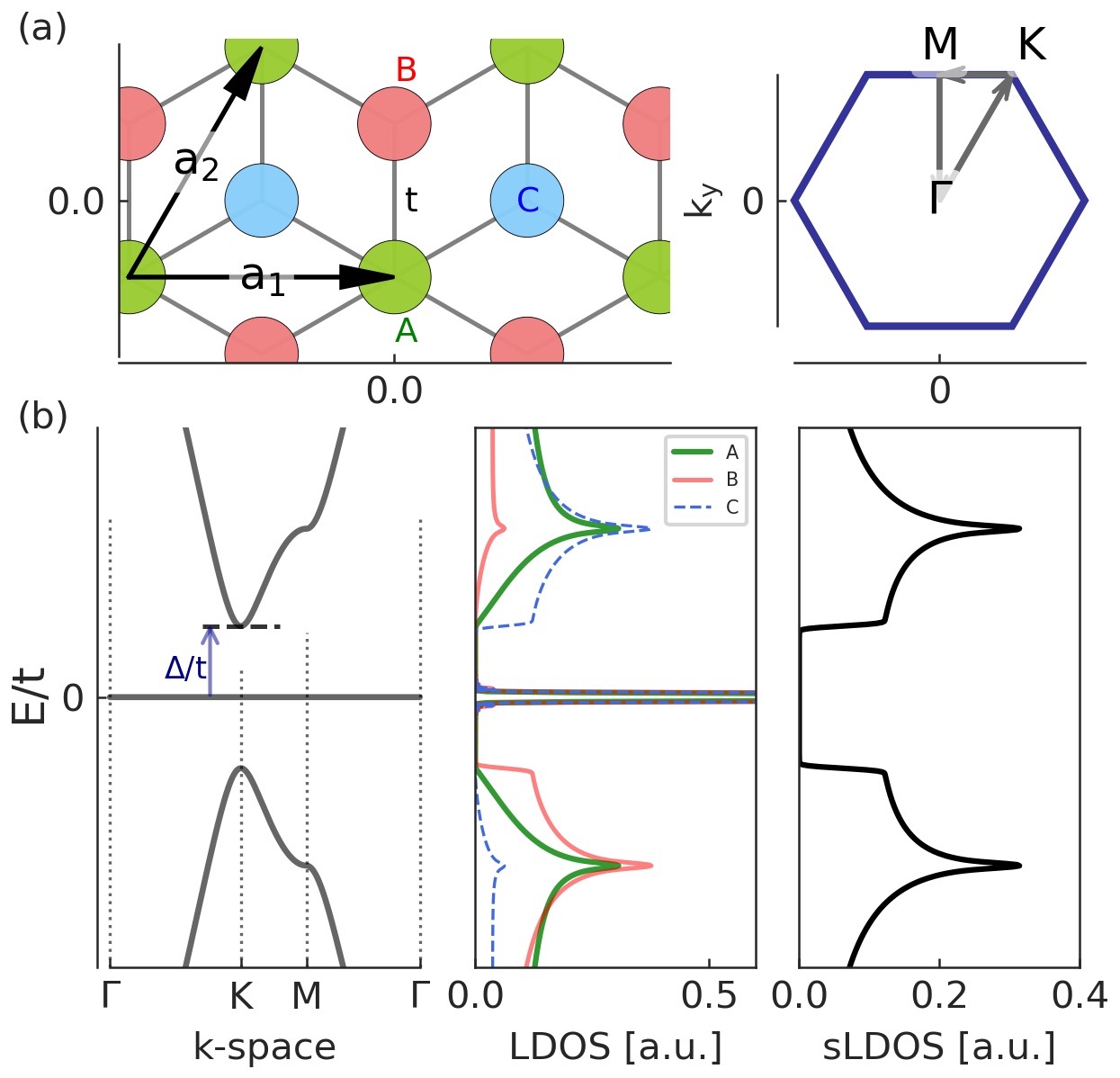}
\caption{ {
Tight-binding Dice lattice model of a 2D spin-1 Dirac insulator.
(a) Left: schematic of a Dice lattice consisting of three sublattices denoted 
by A, B and C with a nearest neighbors hopping $t$ (between them) 
and primitive vectors $\bm{a}_1=(a, 0),$ $\bm{a}_2=(a/2, \sqrt{3}a/2)$, 
given $a$ the primitive lattice constant. Right: the corresponding first 
Brillouin zone. (b) Left: bulk band structure plotted along the lines 
connecting points of high symmetry indicated in right panel of (a). Middle 
and right show the resulting LDOS and pseudospin polarized LDOS (sLDOS) 
spectra, respectively.}}
\label{fig:fig4}
\end{figure}

The in-gap excitations predicted have the striking physical properties of 
dispersionless spectral flow and spontaneous domain wall spin ordering. 
They manifest themselves as distinct real-space topographies of LDOS and 
spin-LDOS, which can be experimentally mapped out using the low-temperature 
scanning tunneling spectroscopy technique~\cite{Hamal2011,Subra2012}. 
With advances in Dirac materials in recent years, realizing the spin-1 
generalization of ordinary Dirac/Weyl fermions in the form of low-energy 
collective states or quasiparticles is experimentally possible in condensed 
matter systems~\cite{Raoux2014,RomhMannyi2015,Bradlynaaf2016,Drost2017,
Slot2017}, photonic crystals~\cite{Vicencio:2015,Drost2017}, and even 
classical systems~\cite{Jin2016}. 

\begin{figure*}[ht!]
\centering
\includegraphics[width=\linewidth]{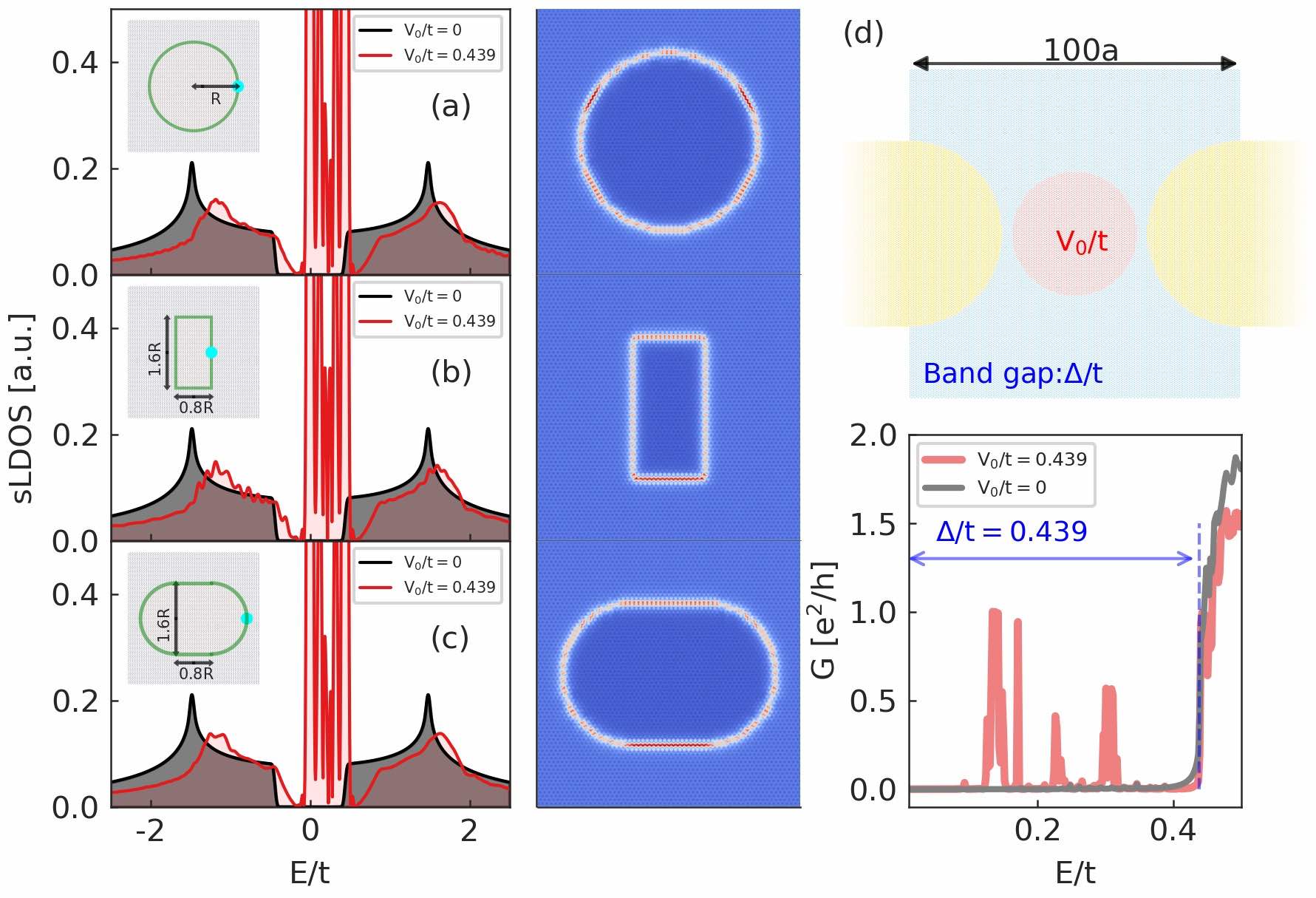}
\caption{{
In-gap edge modes in the Dice lattice based material system. 
Pseudospin polarized LDOS at the position of the domain boundary 
(marked by the cyan dot) as function of energy for a uniformly gated region 
with a shape of (a) disk, (b) rectangle and (c) stadium via an electrostatic 
gate potential $V_0/t$. Middle panels display typical real space patterns 
of associated in-gap states. (d) Top: schematic illustration of a 
gate-controlled spin-1 Dirac electron transistor setup. Bottom: the simulation 
result of of transport conductance versus energy.}}
\label{fig:fig5}
\end{figure*}

Our theoretical prediction is general for gapped systems of massive spin-1 
particles subject to an electrostatic potential applied to a finite domain.
The band-gap associated Dirac-like mass generation can be implemented in 
alternative ways. For example, for a two-dimensional lattice with three 
sublattices~\cite{Raoux2014,Mukherjee:2015,Vicencio:2015,Drost2017,Slot2017},
such as a Lieb or a dice lattice, the generalized mass term can be induced 
via a staggered sublattice potential that breaks the inversion symmetry, 
which is an extension of the standard Dirac mass term in, e.g., graphene. 
As a way of example, we consider the case of a dice lattice model as 
illustrated in Fig.~\ref{fig:fig4}(a), which are relevant to emerging 2D 
Dirac materials such as transition metal dichalcogenide/dihalide 
monolayers~\cite{Li2014}, monolayer Mg$_2$C (MXene)~\cite{Wangss2018}, 
decorated graphene~\cite{Gio2015} etc. Its tight-binding Hamiltonian in real 
space is given by 
\begin{eqnarray}
H_{Dice}=&-t\sum_{\langle i,j\rangle}\left(c_{Bi}^\dag c_{Aj}+c_{Bi}^\dag c_{Cj} + H.c.\right) \nonumber \\
&+\Delta\sum_{i}\left(c_{C i}^\dag c_{C i}-c_{A i}^\dag c_{A i}\right),
\end{eqnarray}
where $c_{\nu i}^\dag$ ($c_{\nu i}$) with $\nu= A,B,C$ are creation 
(annihilation) operators of the localized states $|\nu i\rangle$ at site 
$i$ belonging to the sublattice $\nu$, $\langle i,j\rangle$ denotes pairs of 
nearest-neighbor sites with the tunneling strength (hopping energy) of $t$. 
The last term represents a staggered sublattice potential that is responsible 
for the Dirac-type mass based gap opening. In the absence of any external 
field, we obtain the bulk energy band structure and corresponding LDOS 
spectra, as shown in Fig.~\ref{fig:fig4}(b). We see that, near the $K$ point, 
the system behaves as a band insulator hosting Dirac-like quasiparticles of 
massive spin-1. Notably, the flat band leads to a sharp peak in the LDOS, but 
has a vanishing group velocity as well as a vanishing out-of-plane pseudospin 
polarization/orientation [c.f. right panel of Fig.~\ref{fig:fig4}], i.e. 
$\textrm{sLDOS}\equiv|\mathcal{D}_B-\mathcal{D}_C|=0$ with $\mathcal{D}_\mu$ 
the LDOS occupied at sublattice $\mu$. 

An electrostatic potential of height $V_0/t$ is locally applied to a small 
region of an undoped dice lattice sheet to realize the gate controlled 
quantum dot structures. Concretely, for $\Delta/t=0.439$ and 
$V_0/t=\Delta/t$($<2\Delta/t$), we calculate the sLDOS measured at the 
boundary of the gated region for three different domain shapes with a 
characteristic size parameter $R=5$nm as depicted in insets of 
Figs.~\ref{fig:fig5}(a-c). The results are displayed by red curves, while 
those for the (ungated) case of $V_0/t=0$ (black curves) are also shown 
for comparison. Signified by dramatic changes in the sLDOS spectra with 
large amplitudes, a number of in-gap states emerges. As displayed in 
the middle panel of Fig.~\ref{fig:fig5}, they are highly localized edge 
modes. This result agrees with that obtained from the analytic continuum 
spin-1 Dirac model in Sec.~\ref{sec:Dirac_Weyl}. 

We also consider a lead-contacted dice lattice flake with a circular 
gate-defined quantum dot as schematically illustrated in the top panel of 
Fig.~\ref{fig:fig5}(d) for a possible experimental detection via transport 
measurements. One typical simulation result is given in bottom panel of 
Fig.~\ref{fig:fig5}(d). Remarkably, the emerging in-gap modes acting as 
``doorway'' states can actuate resonant tunneling through the device with 
large conductance. Because of the east of realizing control with an 
electrostatic gate potential, the setup can act as a novel quantum switch 
or transistor of high on/off ratio with spin-1 Dirac electrons.

Alternatively, associated with triple point semimetals of bulk massless 
spin-1 excitations described by a three-band extension of the Weyl 
Hamiltonian~\cite{Bradlynaaf2016}, i.e., $H_3\propto k_xS_x+k_yS_y+k_zS_z$, 
a thin film structure of thickness $L$ in the $z$ direction can host the 
two-dimensional spin-1 quasiparticles with an analogous finite mass 
$\propto \pi/L$ due to the confinement effect. This provides another potential 
experimental platform. In addition, the massive spin-$1$ physics turns out to 
be accessible in a dimerized quantum magnet~\cite{RomhMannyi2015} and is even 
relevant to classical systems of two-dimensional magnetoplasmon~\cite{Jin2016},
where the mass term is induced by an applied magnetic field.  

\begin{figure}[ht!]
\centering
\includegraphics[width=\linewidth]{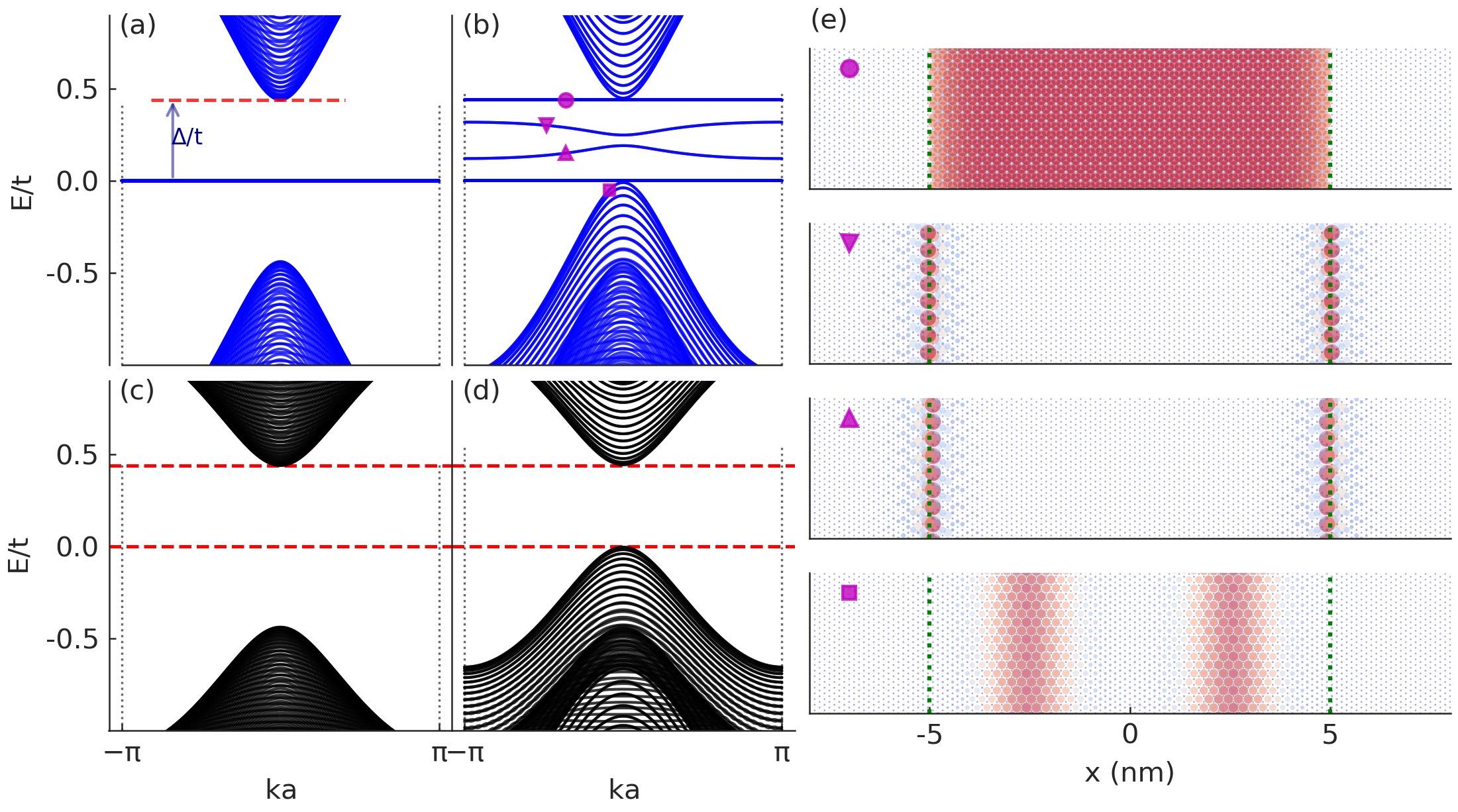}
\caption{{
Energy spectra and electronic states for a semi-infinite geometry
of gapped dice lattice. The spin-1 low energy excitations carry an    
effective mass $\Delta/t$. (a) States in the absence of any applied gate   
potential and (b) in the presence of a potential $V/t$. (c,d) The 
corresponding results from a gapped graphene lattice. (e) Spatial LDOS 
patterns of the respective states as indicated by different markers in (b). 
The dotted vertical green lines mark the boundaries of the locally applied 
gate potential along the $x$-axis. A translational symmetry is imposed on 
the $y$ axis.}}
\label{fig:Rfig}
\end{figure}

We have also solved a gapped Dice lattice in a semi-infinite geometry in the 
presence or absence of a locally applied gate potential, which represents a 
trivial bulk band insulator with low energy, massive, pseudospin-1 excitations.
For comparison, we have also included the known case of gapped graphene. The 
results are shown in Fig.~\ref{fig:Rfig}. It can be seen that, in contrast to 
the well studied graphene case [(c) and (d)], in the Dice lattice with massive
pseudospin-1 quasiparticles, the in-gap states emerge as the result of simply
applying an electrostatic potential to a trivial bulk band insulator. As shown
in (e), they are localized edge states that are distinct from the 
dispersionless flat band states [top panel in (e)] and from the typical
quantum well bound states [bottom panel in (e)] as well. These results agree 
with the prediction from the general continuum model.
}

\section{Conclusion and discussion} \label{sec:discussion}

To summarize, we have predicted a class of in-gap edge excitations with 
spontaneous domain-wall spin textures in insulating Dirac-type systems of 
massive spin-1 particles with only a locally applied electrostatic potential.
Despite the absence of magnetism and any {\em a priori} topological origin,
these states are extremely robust against boundary deformation and disorders.
The remarkable property of these states is the self-induced 
emergence of domain-wall spin ordering that renders distinct spin-angular 
momentum locking on different sides of the domain interface. Consequently, 
the states are stable against impurities and/or geometric deformation. 
The in-gap modes are formally three-component evanescent wave solutions, 
bearing certain resemblance with the Jackiw-Rebbi type of bound states.
The modes belong to a distinct class due to the following physical reasons: 
three-component spinor wave function, unusual boundary conditions, and a 
shifted flat band induced by the external scalar potential.
Our findings provide a fully electrostatic based route to generating protected, 
robust spin ordering edge states without requiring any sort of magnetism, 
extrinsic or intrinsic. The states can be exploited for spintronics and 
quantum information processing applications, e.g., realization of a 
gate-controlled spin-1 Dirac electron transistor or quantum switch. 
With rapid advances in generalized Dirac materials, especially those hosting
the spin-1 generalization of ordinary Dirac/Weyl fermions, and with the
state-of-the-art measurement technologies, experimental confirmation of the
states discovered here is possible.

We note a distinct feature of the system studied: the inherent mid-gap flat
band hosting macroscopically degenerate states. Without the applied
electrostatic potential ($V_g=0$), we obtain the flat band states, i.e.,
$E(\bm{p})=0$, given by (nonnormalized)
\begin{equation}
\Psi_{\bm{k},0}(\bm{r}) \sim \frac{1}{\sqrt{2}}\left[v_F|\bm{p}| e^{-i\zeta}, -\sqrt{2}\Delta, -v_F|\bm{p}| e^{i\zeta}\right]^T
e^{i\bm{k\cdot r}},
\end{equation}
with the wavevector $\bm{k}=(k_x, k_y)\equiv\bm{p}/\hbar$ making an angle
$\zeta=\arctan(k_y/k_x)$ with the $x$ axis. The states result in a vanishing
current and a trivial spin distribution over the space as well as a vanishing
Chern number~\cite{Green2010,Dora2011,Jin2016}. 
Our finding is that a locally
applied potential shifts the flat band relative to the surrounding and
surprisingly leads to a class of exotic edge excitations that inherit the
(quasi)flat dispersionlessness but attain a nontrivial feature
associated with the emerging domain-wall like spin ordering. 
Due to the vanishing Chern number of the flat band, in the configuration in 
Fig.~\ref{fig:fig1}B, the regions with different applied potential $V_g$ 
possess the same Chern number. This indicates that the uncovered in-gap states 
do not have a topological origin. 
It has been known that flat bands can lead to exotic physical phenomena such as zero-refractive
index, unconventional Anderson localization~\cite{Goda2006,Bodyfelt2014},
itinerant ferromagnetism~\cite{Taiee2015}, and unconventional
superconductivity~\cite{Julku2016,Caoetal:2018,Roy2019}.
Moreover, the finite gap opening makes it possible to categorize the
unperturbed bulk system into the phase of class-D with a particle-hole
symmetry and a broken time-reversal symmetry, which also arises in $p+ip$
superconductors~\cite{Jin2016}. In this regard, the two-dimensional gapped
pseudospin-$1$ system represents a paradigm to investigate high-spin
topological phases with exotic edge excitations and flat-band physics. With
enriched pseudospin degrees of freedom, graphene-based heterostructures,
such as graphene-In$_2$Te$_2$ bilayer~\cite{Gio2015} and twisted bilayer
graphene superlattice~\cite{Guo2018}, can also be exploited for possible
experimental realization of the topological edge states uncovered in this
paper.

Taken together, the main contributions of this paper are:
(1) in-gap edge modes can arise in a topologically trivial spin-1 Dirac 
insulators with local electrical gating or nonmagnetic doping,
(2) the in-gap edge modes possess pseudospin polarized textures akin to 
localized domain walls of either the hedgehog or the vortex type without 
requiring any external pseudospin resolved field,
(3) the edge modes are robust against boundary deformations and disordered 
scalar impurities,
(4) the edge modes are nearly dispersionless in energy and intrinsically 
possess the capability of strong charge and spin confinement/localization,
and (5) all these features of the in-gap edge modes can be electrically 
controlled within the same material setting.
We note that, the existing mechanisms for in-gap bound modes
or excitations can be either topological or nontopological. Examples are
the extensively studied topological in-gap edge modes~\cite{HK:2010,QZ:2011}, 
the nontopological Yu-Shiba-Rusinov bound states associated with magnetic 
impurities in superconductors~\cite{Yu:1965,Shiba:1968,Rusinov:1969}, vacancy 
defects or particular lattice terminations induced bound states in crystalline 
lattice systems~\cite{CPLNG:2008,CLV:2010}, and modes induced by nonmagnetic
impurities in topologically nontrivial band insulators~\cite{LSLS:2011}. 
There was also a recent work~\cite{SL:2018} hinting that the multicomponent 
character of the Dirac-Bloch wavefunction and the associated boundary 
conditions would enable nontopological Dirac materials, through proper 
engineering of the graphene lattice boundaries, to potentially host robust 
surface states. 
The system of pseudospin-1 Dirac insulators that we have studied does not
require any special lattice engineering, does not involve any magnetic-type
of perturbations or defects either, nor does it have a nontrivial band
topology. Yet, robust in-gap edge modes can arise. Our system thus does not 
fall into any known category of systems in which in-gap bound modes can arise, 
and the edge modes uncovered belong to a distinct class due to the 
three-component spinor wave function and the unusual boundary conditions 
as well as an electrically induced shift of the flat band.

\begin{acknowledgments}

This work was supported by the Pentagon Vannevar Bush Faculty Fellowship
program sponsored by the Basic Research Office of the Assistant Secretary of
Defense for Research and Engineering and funded by the Office of Naval
Research through Grant No.~N00014-16-1-2828.

\end{acknowledgments}

\appendix
 
\section{Basics} \label{appendix_A}

In the position representation $\bm{r}=(x,y)$, the Hamiltonian for a massive 
spin-$1$ generalization of Dirac/Weyl fermion reads   
\begin{equation} \label{eq:H1}
\hat{H} = v_F\bm{\hat{S}\cdot \hat{p}} + \Delta\hat{S}_z + U(\bm{r}), 
\end{equation}
where $v_F$ is the Fermi velocity, $\bm{\hat{p}}$ is the momentum operator, 
$\bm{\hat{S}}=(S_x, S_y)$ and $\hat{S}_z$ are spin-$1$ matrices, $\Delta$ 
denotes a Dirac-type mass, and $U(\bm{r})$ is a scalar type of perturbation
(e.g., an electrostatic potential). The energy eigenstates 
$\Psi(\bm{r})=[\psi_1(\bm{r}),\psi_2(\bm{r}),\psi_3(\bm{r})]^T$ can be 
determined by the generalized Dirac-Weyl equation 
\begin{equation}
\hat{H}\Psi(\bm{r}) = E\Psi(\bm{r}).
\end{equation}
For a spatially homogeneous/constant potential, e.g., $U(\bm{r})=V_0$, the 
eigenenergies are $E=V_0$ and $V_0+s\sqrt{\Delta^2+\hbar v_F|\bm{k}|^2}$ 
with $s=\pm$ being the dispersion band index. The corresponding plane wave 
solutions can be written as 
\begin{displaymath} 
\Psi_{\bm{k},0}(\bm{r}) = \frac{1}{\sqrt{2}}\left[k e^{-i\zeta}, -\sqrt{2}\delta, -k e^{i\zeta}\right]^T
e^{i\bm{k\cdot r}}, 
\end{displaymath}
and 
\begin{equation}\label{eq:PW}
\Psi_{\bm{k},s}(\bm{r}) = \frac{1}{2}\begin{pmatrix}
\alpha e^{-i\zeta} \\
\sqrt{2} \\
\beta e^{i\zeta}
\end{pmatrix}e^{i\bm{k\cdot r}},
\end{equation} 
where the wavevector $\bm{k}=(k_x, k_y)$ has length 
$k=\sqrt{\epsilon^2-\delta^2}$ with 
$\epsilon=(E-V_0)/\hbar v_F, \delta=\Delta/\hbar v_F$, which makes an angle 
$\zeta=\arctan(k_y/k_x)$ with the $x$ axis. Other factors are 
$\alpha=k/(\epsilon-\delta)$ and $\beta=k/(\epsilon+\delta)$. The current 
operator is defined based on Eq.~(\ref{eq:H1}) as 
\begin{equation} \label{eq:current_operator}
\bm{\hat{u}}=\bm{\nabla}_{\bm{p}}\hat{H}=v_F\bm{\hat{S}}.
\end{equation} 
The local current associated with state $\Psi(\bm{r})=[\psi_1,\psi_2,\psi_3]^T$
can be calculated from the local expectation value of $\bm{\hat{u}}$ as 
\begin{eqnarray}
\bm{u}(\bm{r}) &= v_F(\psi_1^*, \psi_2^*, \psi_3^*)\bm{\hat{S}}
\begin{pmatrix}
\psi_1 \\
\psi_2 \\
\psi_3
\end{pmatrix} \nonumber \\
&=\sqrt{2}v_F\left(\Re[\psi_2^*(\psi_1+\psi_3)],-\Im[\psi_2^*(\psi_1-\psi_3)]\right). 
\end{eqnarray}
By definition, the local current is the local probability density of spin 
vector $(S_x, S_y)$. Using the plane wave solution (\ref{eq:PW}), we obtain
\begin{displaymath}
\bm{u}=v_F\frac{\epsilon}{\sqrt{\epsilon^2-\delta^2}}\frac{\bm{k}}{k}.
\end{displaymath}

The effects of the applied scalar potential are to shift the Dirac point 
($k=0$) in the energy domain, to tune the kinetic energy 
$\epsilon=(E-V_0)/\hbar v_F$, and to alter the particle attributes from hole- 
to electron-type, and vice versa. 

The time-reversal symmetry operator is 
$$\mathcal{T}=
\left(\begin{array}{ccc}
0 & 0 & -1 \\
0 & 1 & 0 \\
-1 & 0 & 0
\end{array}\right)\mathcal{K}\Big|_{\bm{k}\rightarrow-\bm{k}},$$
where $\mathcal{K}$ is the operator for complex conjugation. Due to the 
Dirac-like mass term, the time-reversal symmetry is broken.

\section{Eigensolutions of type-II quantum dots of massive spin-1 particles}
\label{appendix_B}

\begin{figure*}[ht!]
\centering
\includegraphics[width=\linewidth]{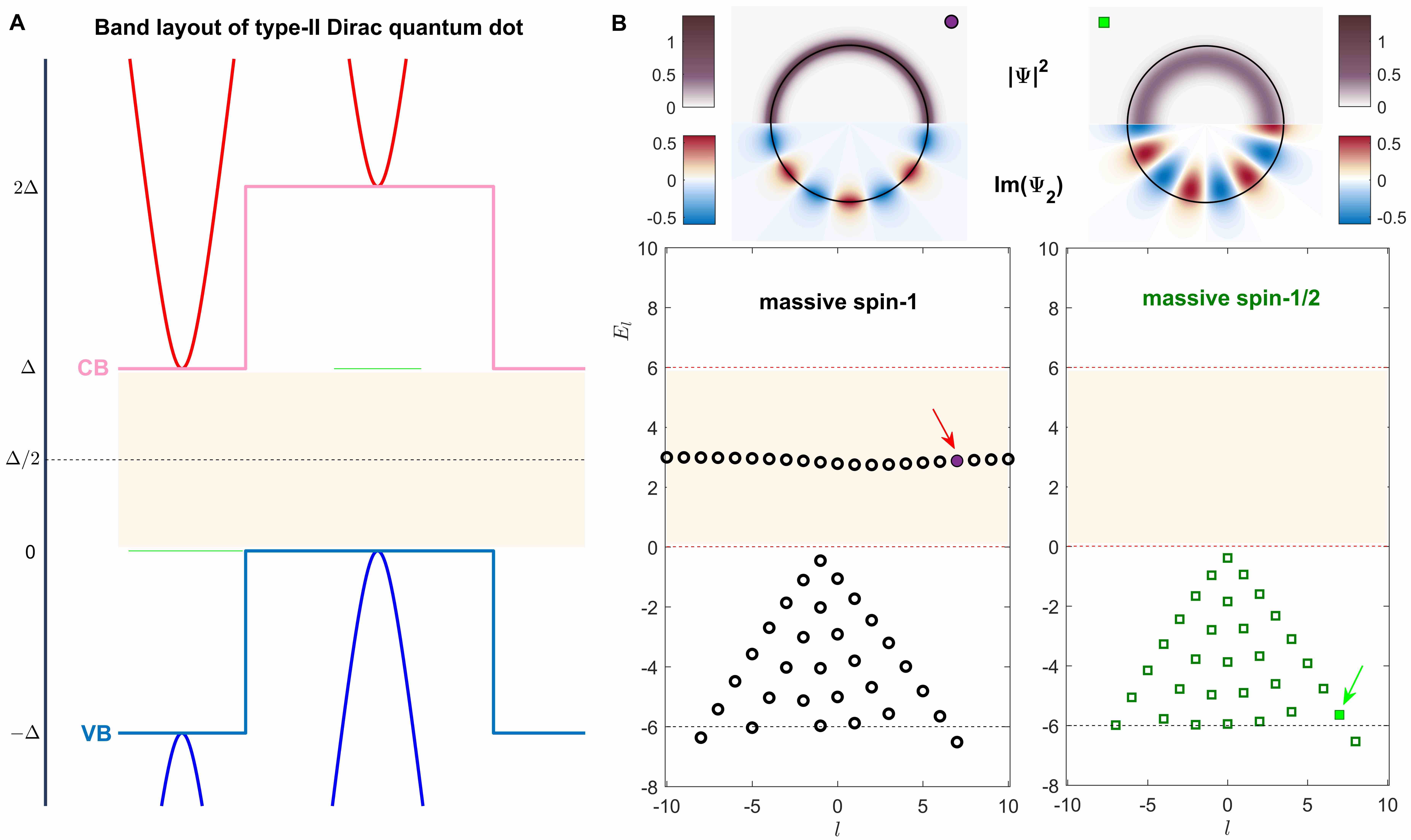}
\caption{ A type-II Dirac material quantum dot for massive spin-1 
generalization of Dirac fermions and the associated eigenstates. 
(A) Energy band diagram of a type-II quantum dot for Dirac-type massive 
spin-1 particles. (B) Top: wave probability patterns for the eigenstates 
indicated by the corresponding colored arrows in the bottom panel for both
massive spin-1 and massive spin-1/2 particles. Bottom: eigenenergies versus 
angular momentum. Parameters are $\Delta=V_0=6\hbar v_F/R$ for both cases.}
\label{fig:SIfig0}
\end{figure*}

We obtain the eigensolutions of the spin-1 massive Dirac system where an 
electrostatic potential is applied to a circular domain: $U(r)=V_0\Theta(r-R)$.
This is effectively a type-II quantum (anti-)dot configuration for Dirac-like 
massive spin-1 particles. Because of the rotational symmetry, it is convenient
to use polar coordinates $\bm{r}=(r,\theta)$, where the eigenequation is
\begin{equation}
\hat{H}\Psi(\bm{r})= \hat{H}
\begin{pmatrix}
\psi_1 \\
\psi_2 \\
\psi_3
\end{pmatrix}
=E\begin{pmatrix}
\psi_1 \\
\psi_2 \\
\psi_3
\end{pmatrix},
\end{equation}
where 
$$\hat{H}=\left[\frac{\hbar v_F}{\sqrt{2}}\left(\begin{array}{ccc}
0 & \hat{\mathcal{L}}_- & 0\\
\hat{\mathcal{L}}_+ & 0 & \hat{\mathcal{L}}_- \\
0 & \hat{\mathcal{L}}_+ & 0
\end{array}\right)+\Delta\hat{S}_z+U(r)
\right],$$
with
$$\hat{\mathcal{L}}_\pm = -ie^{\pm i\theta}\left(\partial_r\pm\frac{i}{r}\partial_\theta\right).$$ 
Because the total angular momentum operator 
$\hat{J}_z = -i\partial_\theta+\hat{S}_z$ commutes with the Hamiltonian 
$\hat{H}$, the common set of eigenstates has the general form
\begin{equation}
\Psi_l(\bm{r}) = [\mathcal{R}_1(r)e^{i(l-1)\theta}, \mathcal{R}_2(r)e^{il\theta}, \mathcal{R}_3(r)e^{i(l+1)\theta}]^T 
\end{equation}	
with $l\in\mathbb{Z}$. For the dispersive bands, we have
\begin{equation}
\Psi_l^\mu(\bm{r})= \frac{C_\mu}{\sqrt{2}}
\left(\begin{array}{c}
\alpha_\mu Z_{l-1}^\mu(k_\mu r)e^{-i\theta} \\
i\sqrt{2}Z_l^\mu(k_\mu r) \\
-\beta_\mu Z_{l+1}^\mu(k_\mu r)^{i\theta}
\end{array}\right)e^{il\theta},
\end{equation}
where the index $\mu=I,O$ labels the inner and outer regions of the circular 
domain boundary, $\alpha_\mu=\hbar v_Fk_\mu/(E_\mu-\Delta)$ and 
$\beta_\mu = \hbar v_Fk_\mu/(E_\mu+\Delta)$ with 
$\hbar v_Fk_\mu = \sqrt{E_\mu^2-\Delta^2}$ and $(E_I,E_O)=(E-V_0,E)$, and 
$Z_l^I(x) = J_l(x)$ and $Z_l^O(x)=H_m^{(1)}(x)$ are the Bessel and the Hankel 
functions of the first kind, respectively. Matching the spinor wavefunctions
$\Psi_l^I$ and $\Psi_l^O$ at the domain boundary (interface) $r=R$ yields the 
following transcendental equation 
\begin{eqnarray} \label{eq:traeq}
&&J_l(k_IR)\left[\alpha_OH_{l-1}^{(1)}(k_OR)-\beta_OH_{l+1}^{(1)}(k_OR)\right] = \nonumber\\
&&H_l^{(1)}(k_OR)\left[\alpha_IJ_{l-1}(k_IR)-\beta_IJ_{l+1}(k_IR)\right],  
\end{eqnarray}
which can be calculated numerically to yield the eigenenergies and eigenstates 
with high accuracy. Figure~\ref{fig:SIfig0} shows some representative 
results. For reference, we have also included the corresponding results for
the standard massive, spin-1/2 Dirac fermion system. We see that, for the 
massive spin-1 system, apart from the conventional quantum dot bound states, 
an additional group of modes emerge in the gap. While edge states can arise
in the band gap as in conventional topological insulators, some kind of 
magnetic perturbations are required~\cite{BHZ:2006,FK:2007,ZLQDFZ:2009,KWBRBMQZ:2007,HQWXHCH:2008,XQHWPLBGHC:2009,Moore:2010,HK:2010,QZ:2011}. As 
there is no magnetic perturbation of any sort in our quantum dot system for 
massive spin-1 Dirac particles, the emergence of the states in the band gap 
is quite counterintuitive and striking.

We show analytically that the modes in the band gap possess a unique spectral 
peculiarity and are in fact edge states with domain-wall like, topologically 
nontrivial spin textures. In particular, in the gap $|E_\mu|<|\Delta|$, the 
radial wavenumbers are purely imaginary, which can be redefined as
\begin{eqnarray}
k_OR &= \frac{\sqrt{E^2-\Delta^2}}{\hbar v_F/R}=\sqrt{\epsilon^2-\delta^2} = ip, \\
k_IR &= \sqrt{(\epsilon-v_0)^2-\delta^2} = iq.
\end{eqnarray}
With the substitutions 
$$K_l(x)=\frac{\pi}{2}i^{l+1}H_l^{(1)}(ix), I_l(x) = i^{-l}J_l(ix),$$ 
we rewrite the eigenvalue equation Eq.~(\ref{eq:traeq}) as 
\begin{eqnarray}
&I_l(q)\left[\frac{p}{\epsilon-\delta}K_{l-1}(p)+\frac{p}{\epsilon+\delta}K_{l+1}(p)\right] = \nonumber\\
&-K_l(p)\left[\frac{q}{\epsilon-v_0-\delta}I_{l-1}(q)+\frac{q}{\epsilon-v_0+\delta}I_{l+1}(q)\right],  
\end{eqnarray}
with the associated eigenstates given by 
\begin{eqnarray}
&\Psi_l(\bm{r})=\langle O|\Psi_l\rangle+\langle I|\Psi_l\rangle, \nonumber\\
&= \frac{\sqrt{2}i^{-l}C_O}{\pi}
\left(\begin{array}{c}
\frac{ip}{\epsilon-\delta} K_{l-1}(p\rho)e^{-i\theta} \\
\sqrt{2}K_l(p\rho) \\
\frac{ip}{\epsilon+\delta} K_{l+1}(p\rho)e^{i\theta}
\end{array}\right)e^{il\theta}\Theta(r-R)  + \nonumber\\
&\frac{i^lC_I}{\sqrt{2}}
\left(\begin{array}{c}
\frac{q}{\epsilon-v_0-\delta} I_{l-1}(q\rho)e^{-i\theta} \\
i\sqrt{2}I_l(q\rho) \\
\frac{q}{\epsilon-v_0+\delta} I_{l+1}(q\rho)e^{i\theta}
\end{array}\right)e^{il\theta}\Theta(R-r), \nonumber\\
&= \frac{\sqrt{2}i^{-l}C_O}{\pi}\left\{\left(\begin{array}{c}
\frac{ip}{\epsilon-\delta} K_{l-1}(p\rho)e^{-i\theta} \\
\sqrt{2}K_l(p\rho) \\
\frac{ip}{\epsilon+\delta} K_{l+1}(p\rho)e^{i\theta}
\end{array}\right)e^{il\theta}\Theta(\rho-1)  + \right. \nonumber\\
&\left.\frac{K_l(p)}{I_l(q)}
\left(\begin{array}{c}
\frac{-iq}{\epsilon-v_0-\delta} I_{l-1}(q\rho)e^{-i\theta} \\
\sqrt{2}I_l(q\rho) \\
\frac{-iq}{\epsilon-v_0+\delta} I_{l+1}(q\rho)e^{i\theta}
\end{array}\right)e^{il\theta}\Theta(1-\rho)\right\},
\end{eqnarray}
where $\rho=r/R$, $I_l(x)$ and $K_l(x)$ are modified Bessel functions. Making 
use of asymptotic expansions of high order Bessel functions~\cite{AbrSte2012}, 
i.e., $l\gg1$:   
$$I_l(x) \sim \frac{1}{\sqrt{2\pi l}}\left(\frac{ex}{2l}\right)^l; K_l(x)\sim \sqrt{\frac{\pi}{2l}}\left(\frac{ex}{2l}\right)^{-l},$$
we obtain, from the eigenvalue equation Eq.~(\ref{eq:traeq}), the following
relation
\begin{eqnarray}
\lim_{l\rightarrow\infty}&\left[\frac{1}{\epsilon+\delta}\sqrt{\frac{l+1}{l}}\left(1+\frac{1}{l}\right)^l+\right. \nonumber\\
&\left.\frac{1}{\epsilon-v_0-\delta}\sqrt{\frac{l}{l-1}}\left(1+\frac{1}{l-1}\right)^{l-1}\right] \rightarrow 0. 
\end{eqnarray} 
Using the identity $\lim_{n\rightarrow\infty}(1+1/n)^n=e$, we arrive at an 
equation that can be solved to yield the asymptotic eigenenergies:
\begin{equation}
\frac{2\epsilon-v_0}{(\epsilon+\delta)(\epsilon-v_0-\delta)} \rightarrow 0 \Longrightarrow \epsilon\rightarrow\frac{v_0}{2}.
\end{equation} 
The eigenenergies are independent of the angular momentum and are thus 
in-gap (energy) dispersionless excitations. The associated eigenstates are
approximately given by 
\begin{eqnarray}
\Psi_l(\bm{r})\approx &C_l\left\{\rho^{-l}\left(\begin{array}{c}
\frac{\rho(\epsilon+\delta)}{4il^2}e^{-i\theta} \\
\frac{\sqrt{2}}{2l} \\
\frac{i}{\rho(\epsilon+\delta)} e^{i\theta}
\end{array}\right)\Theta(\rho-1)  +\right. \nonumber\\
&\left.\rho^l
\left(\begin{array}{c}
\frac{i}{\rho(\epsilon+\delta)}e^{-i\theta} \\
\frac{\sqrt{2}}{2l} \\
\frac{\rho(\epsilon+\delta)}{4il^2}e^{i\theta}
\end{array}\right)\Theta(1-\rho)\right\}e^{il\theta},
\end{eqnarray}
where 
\begin{equation}
C_l = \frac{\sqrt{2}i^{-l}C_O}{\pi}\sqrt{2\pi l}(e\sqrt{\delta^2-v_0^2/4}/2l)^{-l}.
\end{equation}
So, inside the domain $\rho<1$, we have 
\begin{eqnarray}
\langle I|\Psi_l\rangle \approx &C_l\rho^l
\left(\begin{array}{c}
\frac{i}{\rho(\epsilon+\delta)}e^{-i\theta} \\
\frac{\sqrt{2}}{2l} \\
\frac{\rho(\epsilon+\delta)}{4il^2}e^{i\theta}
\end{array}\right)e^{il\theta} \xrightarrow{l\gg1} \nonumber\\
&C_le^{-l(1-\rho)}\left(\begin{array}{c}
\frac{i}{\rho(\epsilon+\delta)}e^{-i\theta} \\
0 \\
0
\end{array}\right)e^{il\theta},
\end{eqnarray}
Outside of the domain $\rho>1$, we have 
\begin{eqnarray}
\langle O|\Psi_l\rangle \approx &C_l\rho^{-l}\left(\begin{array}{c}
\frac{\rho(\epsilon+\delta)}{4il^2}e^{-i\theta} \\
\frac{\sqrt{2}}{2l} \\
\frac{i}{\rho(\epsilon+\delta)} e^{i\theta}
\end{array}\right)e^{il\theta}  \xrightarrow{l\gg1} \nonumber\\
&C_le^{-l(\rho-1)}\begin{pmatrix}
0 \\
0 \\
\frac{i}{\rho(\epsilon+\delta)}e^{i\theta}
\end{pmatrix}e^{il\theta}.
\end{eqnarray}
We thus have that the in-gap excitations are localized edge modes and exhibit 
domain-wall like spin textures for high angular momentum values. 

\begin{figure*}[ht!]
\centering
\includegraphics[width=\linewidth]{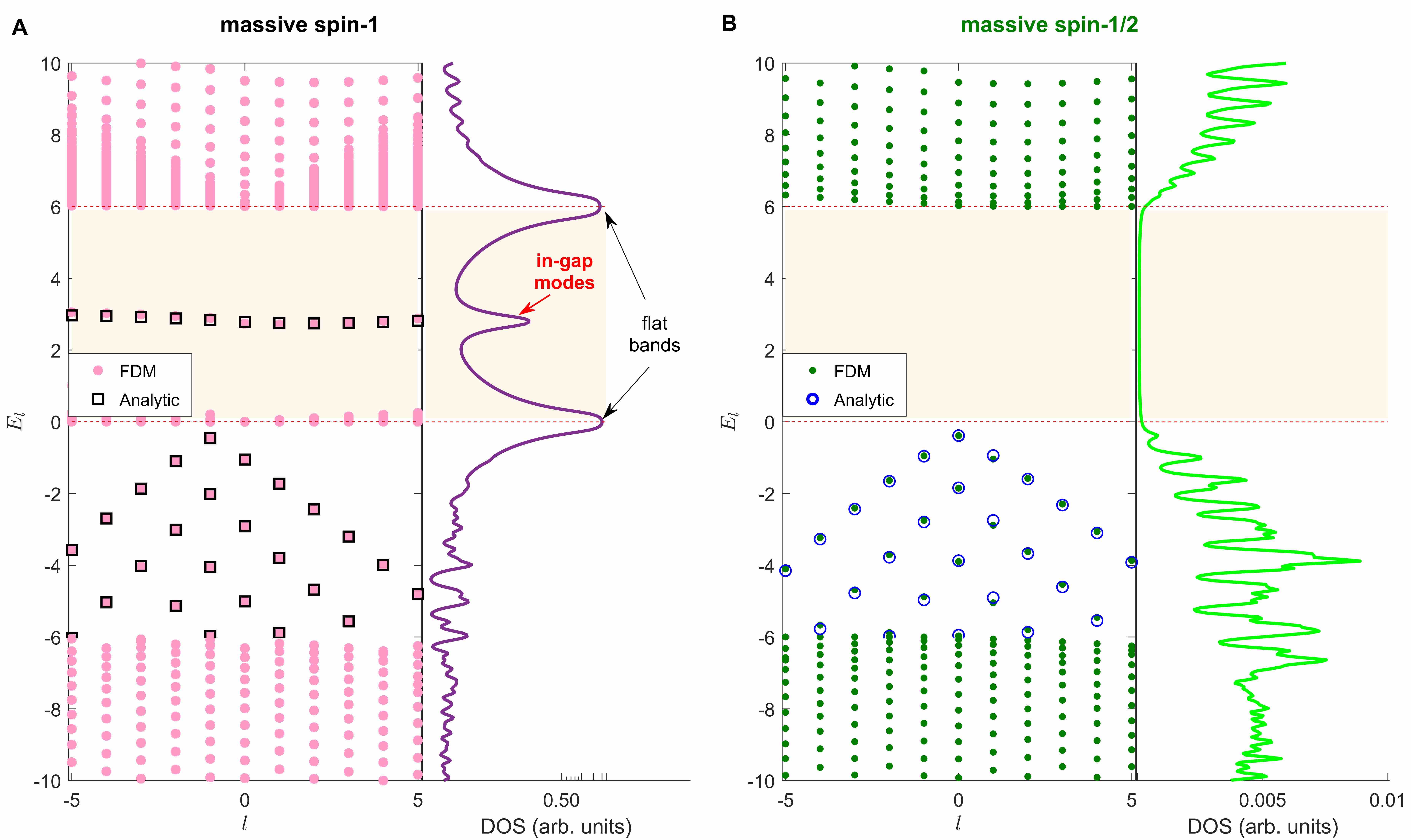}
\caption{ Eigenenergy spectra numerically calculated from the finite 
differential solver for massive spin-1 Dirac systems with a smooth potential 
domain boundary. (A) For massive spin-1 Dirac particles, eigenenergy versus 
angular momentum (left panel) and the resulting local DOS versus energy (right
panel). (B) Results for the corresponding massive spin-1/2 Dirac fermion 
system for comparison.}
\label{fig:SIfig2}
\end{figure*}

\begin{figure*}[ht!]
\centering
\includegraphics[width=\linewidth]{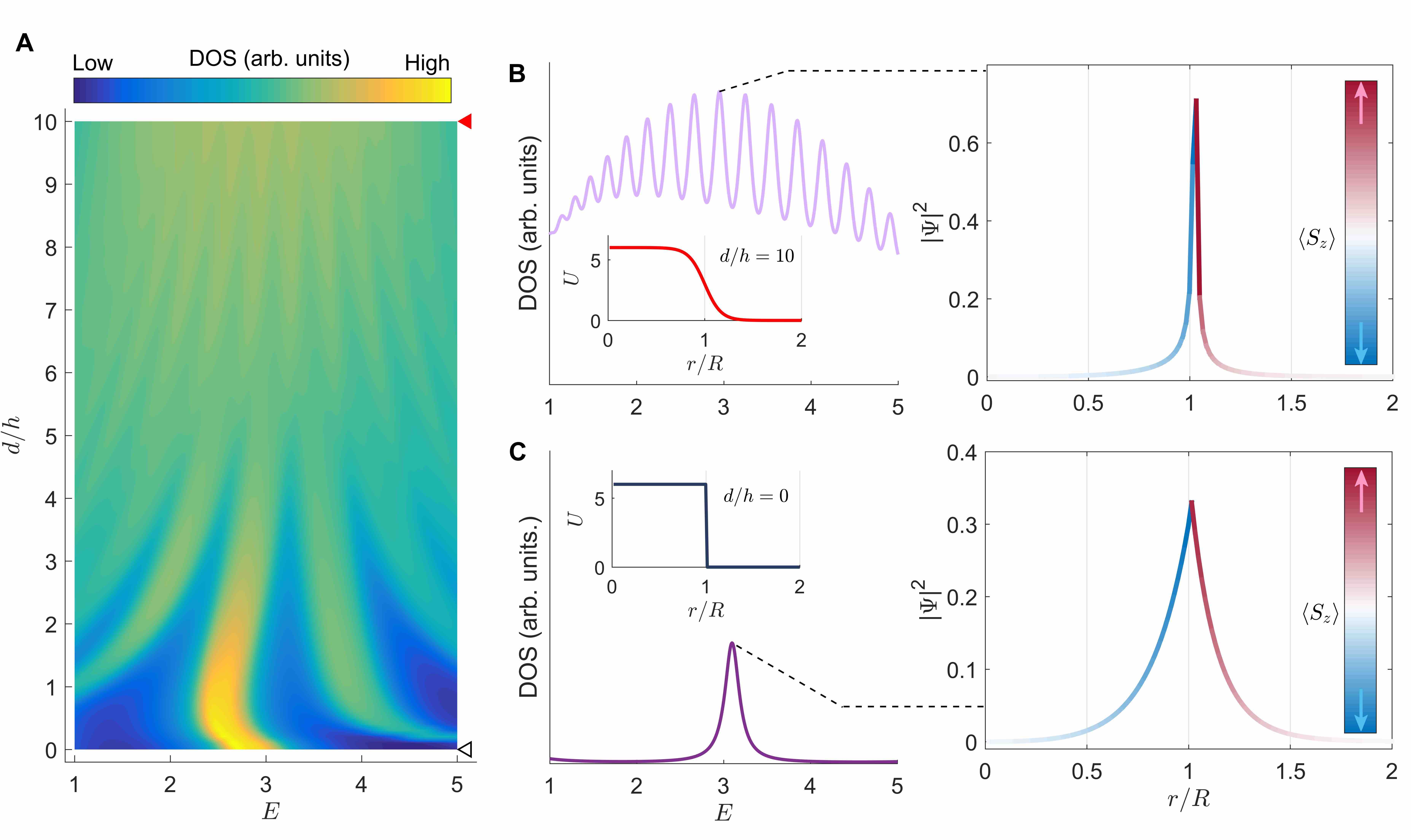}
\caption{ Effect of smooth domain boundaries on the bounded edge 
states. (A) Color coded DOS versus energy $E$ and boundary smoothness $d$. 
(B) Left panel: partial DOS of the $l=-4$ state versus $E$ for a smooth 
potential domain of $d/h=10$ as depicted in the inset. Right panel: wave 
density profile associated with the resonance in the partial DOS. (C) The
corresponding results for the case of infinitely sharp potential domain for
comparison.}
\label{fig:SIfig3}
\end{figure*}

\begin{figure*}[ht!]
\centering
\includegraphics[width=\linewidth]{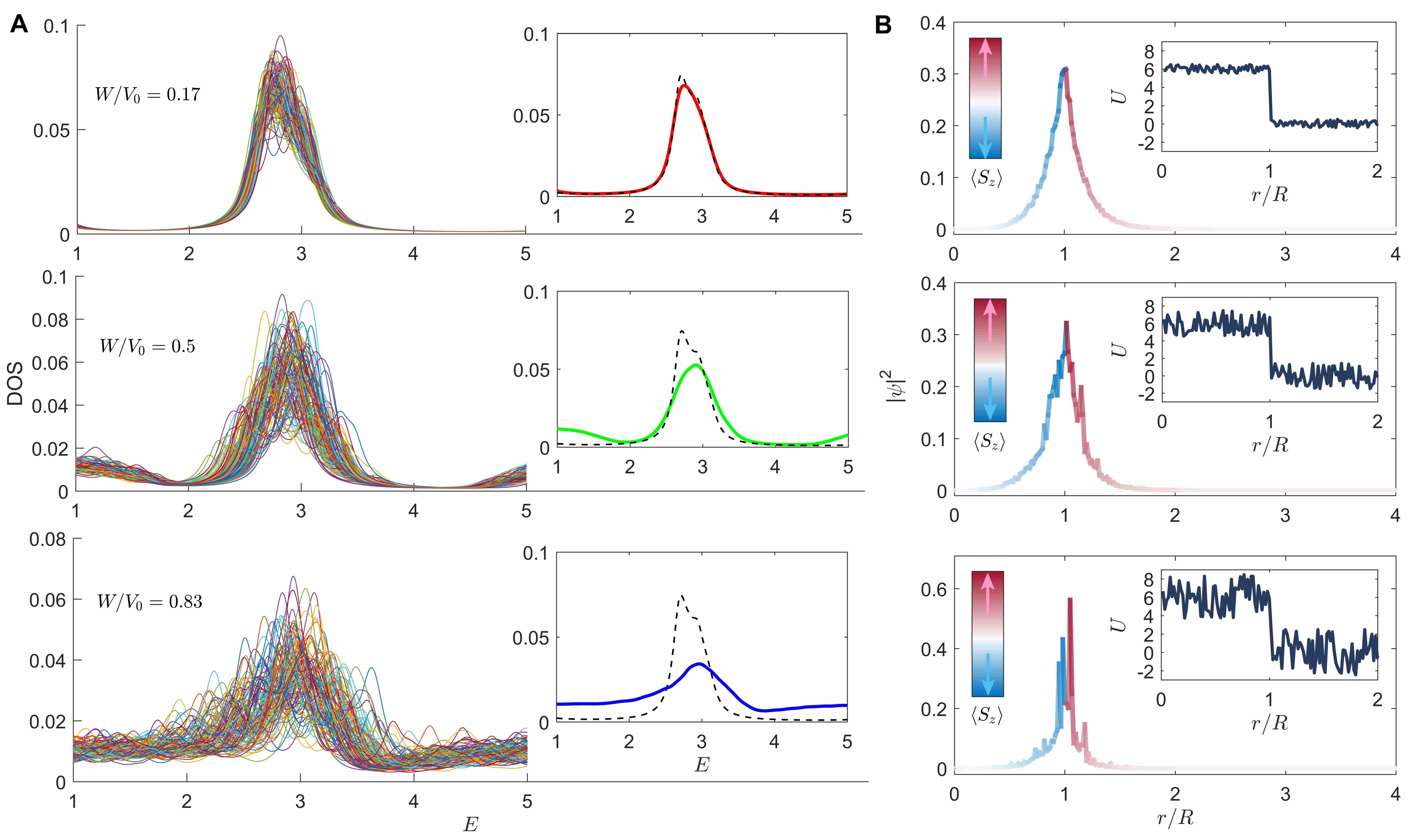}
\caption{ Effect of scalar impurities on the in-gap edge modes. 
(A) DOS as a function of energy for different values of the disorder strength, 
each obtained from $100$ realizations as indicated by multiple colored curves. 
Insets show the corresponding ensemble-averaged DOS versus energy with thick 
solid curves, where the dashed curves are for the case of absence of disorder. 
(B) Typical wave density profiles corresponding to the three cases of 
disorder strength in (A).}
\label{fig:SIfig4}
\end{figure*}

Note that, for a given value of $l$, in the semiclassical limit $p,q\gg1$, 
we have, approximately,
$$I_l(x)\sim \frac{e^x}{\sqrt{2\pi x}}; K_l(x)\sim \sqrt{\frac{\pi}{2x}}e^{-x}.$$ 
From the eigenvalue equation, we have
\begin{equation}
\frac{\epsilon-v_0}{q}+\frac{\epsilon}{p}\approx 0 \Longrightarrow v_0(v_0-2\epsilon)\approx 0 \Longrightarrow \epsilon\sim\frac{v_0}{2},
\end{equation}
which leads to the same in-gap spectral properties as those from the large $l$
regime. The associated semiclassical eigenstates are 
\begin{eqnarray}
\Psi_l(\bm{r})\approx &\frac{i^{-l}C_O}{\sqrt{\pi}}\frac{e^{-\kappa}}{\sqrt{\kappa\rho}} 
e^{-\kappa|\rho-1|}\left\{\left(\begin{array}{c}
\frac{i\kappa}{\epsilon-\delta}e^{-i\theta} \\
\sqrt{2} \\
\frac{i\kappa}{\epsilon+\delta} e^{i\theta}
\end{array}\right)\Theta(\rho-1) \right. \nonumber\\
&+\left.\left(\begin{array}{c}
\frac{i\kappa}{\epsilon+\delta}e^{-i\theta} \\
\sqrt{2} \\
\frac{i\kappa}{\epsilon-\delta}e^{i\theta}
\end{array}\right)\Theta(1-\rho)\right\}e^{il\theta},
\end{eqnarray}
where $\kappa=p\approx q\sim \sqrt{\delta^2-v_0^2/4}\gg1$. We obtain the
resulting spin textures as 
\begin{eqnarray}
\left(\begin{array}{c}
\langle S_x\rangle \\
\langle S_y\rangle \\
\langle S_z\rangle \\
\end{array}\right)\approx &\frac{|C_O|^2e^{-\kappa}}{\pi\kappa\rho}e^{-2\kappa|\rho-1|}\frac{4\delta}{\kappa}
\left(\begin{array}{c}
-\sin\theta \\
\cos\theta \\
\frac{v_0}{\sqrt{4\delta^2-v_0^2}}
\end{array}\right) \nonumber\\
&\times\left[2\Theta(\rho-1)-1\right],
\end{eqnarray}
which exhibit a Bloch-type of domain wall spin ordering about the domain 
boundary as a result of the applied electrostatic potential. Semiclassically,
the in-gap states are thus exponentially localized edge modes with 
spontaneously topological spin textures, which are reminiscent of the 
interfacial Jackiw-Rebbi modes but here the modes have a distinct spectral 
features and an unconventional physical origin.

\section{Effects of smoothly varying electrostatic potential profiles and 
impurities on in-gap modes in massive spin-1 Dirac systems} \label{appendix_C}

Realistically, the applied electrostatic potential will not be infinitely 
sharp at the domain boundary but, rather, the potential file varies smoothly
across the boundary. From an experimental standpoint, it is necessary to 
investigate if the in-gap states can persist when the domain boundary is 
``smeared.'' The test would provide further support for the robustness and 
topological origin of those states. To be concrete, we use the following 
smoothly varying potential profile: 
\begin{equation}
U(\bm{r}) = -\frac{V_0}{2}\tanh\left(\frac{r-R}{d}\right)+\frac{V_0}{2},
\end{equation}
where $d$ ($1/d$) characterizes the boundary smoothness (sharpness) with 
$d=0$ corresponding to the ideal case of an infinitely sharp boundary.
Generally, for a finite value of $d$, it is not feasible to write down 
explicit solutions of the spin-1 Dirac equation. We thus exploit the 
finite difference method (FDM) recently developed for massless spin-1/2 
Dirac fermions~\cite{Zhao672,Nieva2016,Lee2016,Ghahari845} and generalize 
it to massive spin-1 particles. In particular, taking advantage of the 
rotational symmetry of $U(\bm{r})$ and using the polar decomposition ansatz 
\begin{equation}
\psi_l(r,\theta) = \frac{e^{il\theta}}{\sqrt{r}}
\begin{pmatrix}
\mathcal{R}_1(r)e^{-i\theta} \\
\mathcal{R}_2(r) \\
\mathcal{R}_3(r)e^{i\theta}
\end{pmatrix}, 
\end{equation}
we obtain the corresponding radial eigenvalue equation of the three-component 
spinor $\mathcal{R}=[\mathcal{R}_1, \mathcal{R}_2, \mathcal{R}_3]^T$ as 
\begin{equation} \label{eq:rad}
\hat{H}_r\mathcal{R}=E\mathcal{R},
\end{equation}
where 
\begin{eqnarray}
\hat{H}_r&=\left[-iS_x\partial_r + S_y\frac{l}{r} - \right.\nonumber\\
&\left.\frac{1/2}{r}
\begin{pmatrix}
0 & -i/\sqrt{2} & 0 \\
i/\sqrt{2} & 0 & i/\sqrt{2} \\
0 & -i/\sqrt{2} & 0
\end{pmatrix}
+S_z\Delta + U(r)
\right]. \nonumber
\end{eqnarray}
When discretizing this equation on a finite lattice/grid, we need to 
judiciously specify the difference scheme and the boundary conditions at the 
ends of the lattice so as to preserve the Hermiticity of the Hamiltonian. 
A feasible procedure is to use the backward-forward-backward (BFB) difference 
scheme to approximate the derivatives of the three components in 
Eq.~(\ref{eq:rad}): 
\begin{eqnarray}
&\partial_r\mathcal{R}_1\approx\frac{\mathcal{R}(r)-\mathcal{R}(r-h)}{h}, \  
\partial_r\mathcal{R}_2\approx\frac{\mathcal{R}(r+h)-\mathcal{R}(r)}{h}, \nonumber\\  
&\partial_r\mathcal{R}_3\approx\frac{\mathcal{R}(r)-\mathcal{R}(r-h)}{h},
\end{eqnarray} 
where $h=L/(N+1)$ is the discretization step size for the system in the 
range $0<r<L$ with $N+2$ lattice points. The boundary conditions can be 
deduced from the Hermitian constraint of $\hat{H}_r$:
\begin{displaymath}
\int_0^L\left[\mathcal{R}_\alpha^\dag \hat{H}_r\mathcal{R}_\beta - \left(\hat{H}_r\mathcal{R}_\alpha\right)^\dag\mathcal{R}_\beta\right]dr =0, 
\end{displaymath}
which can be explicitly written as 
\begin{equation}
 -\frac{i}{\sqrt{2}}\left[(\mathcal{R}_{1\alpha}+\mathcal{R}_{3\alpha})^*\mathcal{R}_{2\beta}+\mathcal{R}_{2\alpha}^*(\mathcal{R}_{1\beta}+\mathcal{R}_{3\beta})\right]\Big|_0^L=0.
\end{equation}
The specific boundary conditions on $\mathcal{R}(0)$ and $\mathcal{R}(L)$ 
then become $\mathcal{R}_1(0)+\mathcal{R}_3(0)=0$ and $\mathcal{R}_2(L)=0$. 
Implementing this procedure results in an eigenvalue problem for a 
$3N\times3N$ Hermitian matrix $\mathbb{H}_{3N\times3N}=[H_{\mu\nu}]$ with 
entries given by 
\begin{equation}
\left\{
\begin{aligned}
&{H}_{(3n-2)\times(3n-2)}=U_n+\Delta, H_{(3n-1)\times(3n-1)} = U_n, \\
&H_{3n\times3n} = U_n-\Delta, \\
&H_{(3n-2)\times(3n-1)} = \frac{i}{\sqrt{2}h}-i\frac{l-1/2}{\sqrt{2}r_n}, \\
&H_{(3n-1)\times(3n-2)} = \left(H_{(3n-2)\times(3n-1)}\right)^*, \\
&H_{(3n-1)\times3n} = -\frac{i}{\sqrt{2}h}-i\frac{l+1/2}{\sqrt{2}r_n}, \\
&H_{3n\times(3n-1)} = \left(H_{(3n-1)\times3n}\right)^*,
\end{aligned}
\right.
\end{equation} 
for $n=1,\cdots,N$. For $n<N$, the matrix elements are  
\begin{equation}
\left\{
\begin{aligned}
&{H}_{(3n-2)\times(3(n+1)-1)}=-\frac{i}{\sqrt{2}h}, \\
&H_{(3(n+1)-1)\times(3n-2)} = \frac{i}{\sqrt{2}h}, \\
&H_{3n\times(3(n+1)-1)} = -\frac{i}{\sqrt{2}h}, \\
&H_{(3(n+1)-1)\times3n} = \frac{i}{\sqrt{2}h}. 
\end{aligned}
\right.
\end{equation} 
We use the typical experimental values of the local density of states 
(DOS)~\cite{Zhao672,Lee2016,Ghahari845} to measure the spectral features 
and study the effects of the smooth potential profile and impurity on the
in-gap states, where the DOS is defined as 
\begin{equation}
D(E, r_0) = \sum_l\sum_{\nu}\frac{\Gamma}{\pi}\frac{\langle|\mathcal{R}_\nu(r=r_0)|^2\rangle_\lambda}{(E-E_{l\nu})^2+\Gamma^2},
\end{equation} 
with $\nu$ labeling the obtained radial eigenstates for fixed $l$, and 
$$\langle|\mathcal{R}_\nu(r=r_0)|^2\rangle_\lambda=\int_0^Ldr|\mathcal{R}_\nu(r)|^2e^{-(r-r_0)^2/2\lambda}$$ 
represents a spatial average of the wave function centered at $r = r_0$ with 
a Gaussian weight $\lambda$. We approximate the delta function by a 
Lorentzian with the broadening parameter $\Gamma$. In our simulations, we 
use a system of size $L/R = 10$ and discretize it with a uniform lattice 
of $N = 600$ sites. Other parameters are chosen as $\Gamma/E_* = 0.2$ and 
$\lambda=0.01R$. Representative results are shown in Figs.~\ref{fig:SIfig2}, 
\ref{fig:SIfig3} and \ref{fig:SIfig4}, which provide strong support for the
persistence of the in-gap modes in massive spin-1 Dirac systems in realistic
systems with a smooth potential profile and impurities.

\section{Multiple multipoles method: calculation of eigenenergies and 
eigenstates of massive spin-1 Dirac particle in arbitrary domains}
\label{appendix_D}

To test the robustness and to establish the topological origin of the in-gap 
states for massive spin-1 Dirac particles analytically predicted from the 
setting of a circular potential domain, we seek to search for such states in 
systems with a deformed domain. A difficulty that must be overcome 
is to calculate the eigenenergies and eigenstates of massive spin-1 Dirac 
particle in deformed domains of an arbitrarily geometric shape. We have 
succeeded in generalizing the multiple multipole expansion method originally
developed in optics~\cite{LB:1987,Imhof:1996,KA:2002,MEHV:2002,TE:2004} to 
massive spin-1 Dirac particles. The end result of this nontrivial 
generalization is a systematic, reliable, accurate, and efficient computational
paradigm incorporating the evanescent waves to detect and ascertain 
the existence of \emph{in-gap excitations/modes} for arbitrarily shaped 
electrostatic potential domains.

\subsection{Method implementation}

\begin{figure}[ht!]
\centering
\includegraphics[width=\linewidth]{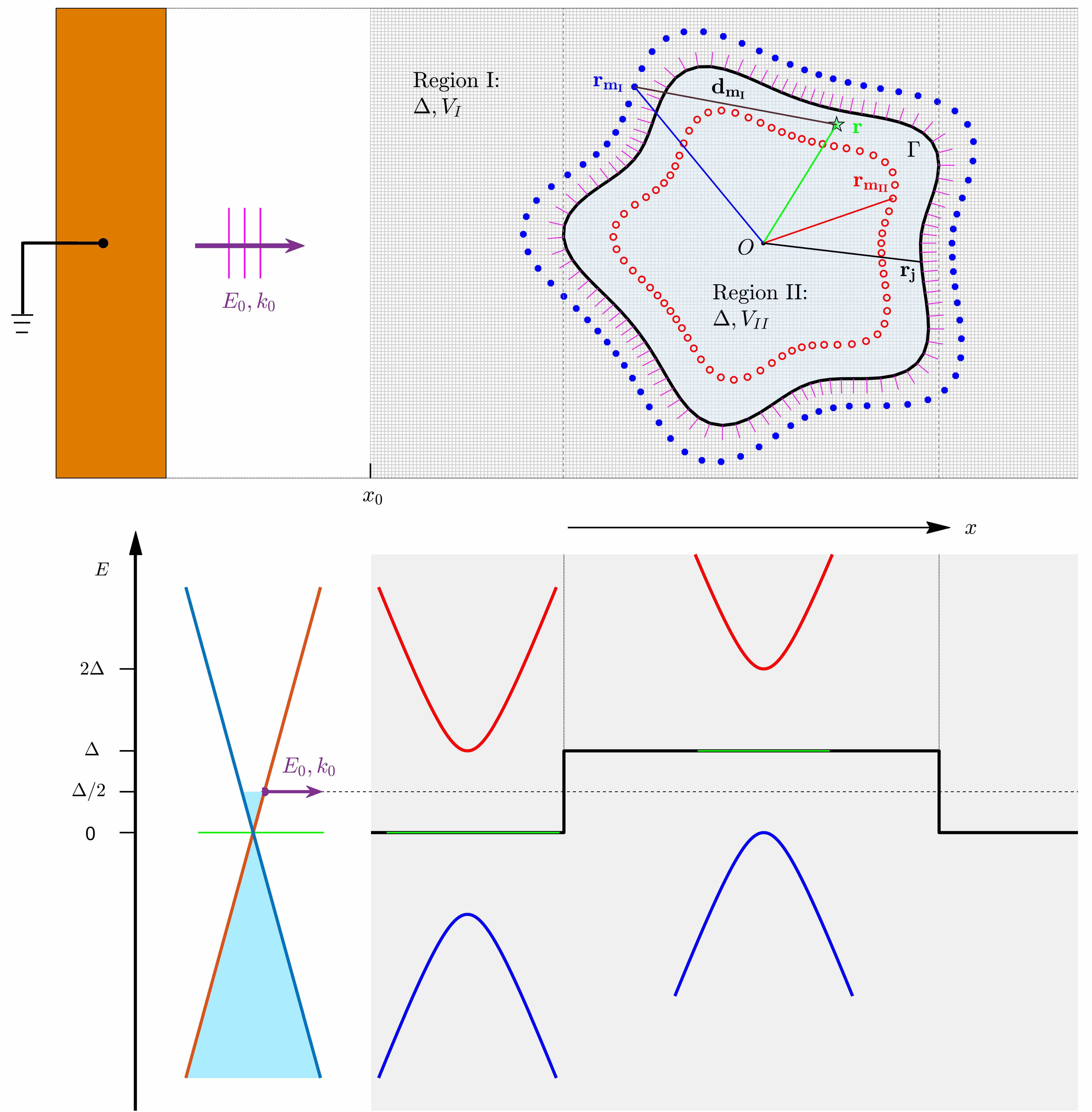}
\caption{ Schematic illustration of the setting of multiple multipole 
expansion method. The domain in which an electrostatic potential is applied
has boundary $\Gamma$ separating regions $I$ and $II$. The basis functions 
originated at $\bm{r}_{m_{I}}$ (blue circular dots) are used to determine 
the wavefunction in region $II$, while those at $\bm{r}_{m_{II}}$ (red circles) 
determine the wavefunction in region $I$. The boundary conditions for the
massive spin-1 Dirac wavefunctions are imposed at the collocation points 
$\bm{r}_j\in\Gamma$.}
\label{fig:SIfig1}
\end{figure}

A concrete setting of a single potential domain of arbitrary shape is 
illustrated in Fig.~\ref{fig:SIfig1}, where the exact shape of the geometric 
boundary is specified according to the superformula in 
botany~\cite{Gielis2003}, a simple but powerful prescription that can generate 
a vast variety of complex geometric shapes. In polar coordinates, the 
superformula is    
\begin{equation}\label{eq:shape}
r(\theta) = \left[\left|\frac{1}{a}\cos\left(\frac{m_1}{4}\theta\right)\right|^{n_2}+\left|\frac{1}{b}\cos\left(\frac{m_2}{4}\theta\right)\right|^{n_3}\right]^{-1/n_1}, 
\end{equation}
where the parameters $(m_1,m_2,n_1,n_2,n_3; a,b)$ control the shape. The
boundary defines two sub-regions, one exterior another interior, denoted 
by $I$ and $II$, respectively, as shown in Fig.~\ref{fig:SIfig1}. The 
three-component spinor wave equation for a massive spin-$1$ Dirac particle 
in each sub-region $\tau\in\{I,II\}$ reads 
\begin{equation}
[\bm{\hat{S}}\cdot\bm{\hat{k}}+\delta S_z]\Psi^{(\tau)}(\bm{r}) 
= \epsilon_\tau\Psi^{(\tau)}(\bm{r}), 
\end{equation}
where $\delta=\Delta/\hbar v_F$ and $\epsilon_{\tau}=(E-V_\tau)/\hbar v_F$. 
In polar coordinates $\bm{r} = (r,\theta)$, the spinor cylindrical wave basis 
of the solutions with angular momentum $l$ is 
\begin{equation}
\Psi_l^{(\tau)}(\bm{r}) = \frac{1}{\sqrt{2}}\begin{pmatrix}
\alpha_\tau B_{l-1}(k_\tau r)e^{-i\theta} \\
i\sqrt{2} B_{l}(k_\tau r) \\
-\beta_\tau B_{l+1}(k_\tau r)e^{i\theta}
\end{pmatrix}e^{il\theta},  
\end{equation} 
where $\alpha_\tau = k_\tau/(\epsilon_\tau - \delta)$, 
$\beta_\tau = k_\tau/(\epsilon_\tau+\delta)$, and 
$k_\tau = \sqrt{\epsilon_\tau^2-\delta^2}$. Choosing 
$B_l(k_\tau r) = H_l^{(1)}(k_\tau r)$ (with $H_l^{(1)}$ being the Hankel 
function of the first kind), we have that the Dirac-type expansion basis 
wavefunctions originated at $\bm{r}_{m_{\overline{\tau}}}$ for the specific 
region $\tau$ are given by 
\begin{equation}
\Psi_l^{(\tau)}(\bm{d}_{m_{\overline\tau}}) = \frac{1}{\sqrt{2}}\begin{pmatrix}
\alpha_\tau H_{l-1}^{(1)}(k_\tau d_{m_{\overline\tau}})e^{-i\theta_{m_{\overline\tau}}} \\
i\sqrt{2} H_{l}^{(1)}(k_\tau d_{m_{\overline\tau}}) \\
-\beta_\tau H_{l+1}^{(1)}(k_\tau d_{m_{\overline\tau}})e^{i\theta_{m_{\overline\tau}}}
\end{pmatrix}e^{il\theta_{m_{\overline\tau}}},
\end{equation}
where $\overline\tau$ denotes the complement of $\tau$,
\begin{displaymath}
d_{m_{\overline\tau}} \equiv |\bm{d}_{m_{\overline\tau}}| 
= |\bm{r} - \bm{r}_{m_{\overline\tau}}|
\end{displaymath}	
and
\begin{displaymath}
\theta_{m_{\overline\tau}}=\textrm{Angle}(\bm{r} - \bm{r}_{m_{\overline\tau}})
\end{displaymath}
with $\bm{r}\in\tau$. Carrying out the expansion in region $II$, we obtain 
the wavefunction as 
\begin{eqnarray}
&\Psi^{(II)}(\bm r) = \sum_{m_I}\sum_{l}C_l^{m_I}\frac{1}{\sqrt{2}}\times \nonumber\\
&\begin{pmatrix}
\alpha_{II} H_{l-1}^{(1)}(k_{II} d_{m_I})e^{-i\theta_{m_I}} \\
i\sqrt{2} H_{l}^{(1)}(k_{II} d_{m_I}) \\
-\beta_{II} H_{l+1}^{(1)}(k_{II}d_{m_I})e^{i\theta_{m_I}}
\end{pmatrix}
e^{il\theta_{m_I}}\equiv 
\begin{pmatrix}
\psi_1^{II} \\
\psi_2^{II} \\
\psi_3^{II}
\end{pmatrix}. 
\end{eqnarray}
The wavefunction in region $I$ has the form 
\begin{eqnarray}\label{eq:swf}
&\Psi^{(I)}(\bm r) = \sum_{m_{II}}\sum_{l}C_l^{m_{II}}\frac{1}{\sqrt{2}}\times \nonumber\\
&\begin{pmatrix}
\alpha_I H_{l-1}^{(1)}(k_{I} d_{m_{II}})e^{-i\theta_{m_{II}}} \\
i\sqrt{2} H_{l}^{(1)}(k_{I} d_{m_{II}}) \\
-\beta_I H_{l+1}^{(1)}(k_{I}d_{m_{II}})e^{i\theta_{m_{II}}}
\end{pmatrix}
e^{il\theta_{m_{II}}}+\Psi^{in}(\bm r) \nonumber\\
&\equiv 
\begin{pmatrix}
\psi_1^{I} \\
\psi_2^{I} \\
\psi_3^{I}
\end{pmatrix},
\end{eqnarray}
where 
\begin{equation}
\Psi^{in}(\bm r) = \frac{1}{2}
\begin{pmatrix}
\alpha_{I} \\
\sqrt{2} \\
\beta_{I}
\end{pmatrix}
e^{i{k_I(x-x_0)}} = 
\begin{pmatrix}
\psi_1^{in} \\
\psi_2^{in} \\
\psi_3^{in}
\end{pmatrix}, 
\end{equation} 
denotes the input source triggered by an applied external excitation outside of
the domain [c.f., top panel of Fig.~\ref{fig:SIfig1}].

\begin{figure}[ht!]
\centering
\includegraphics[width=\linewidth]{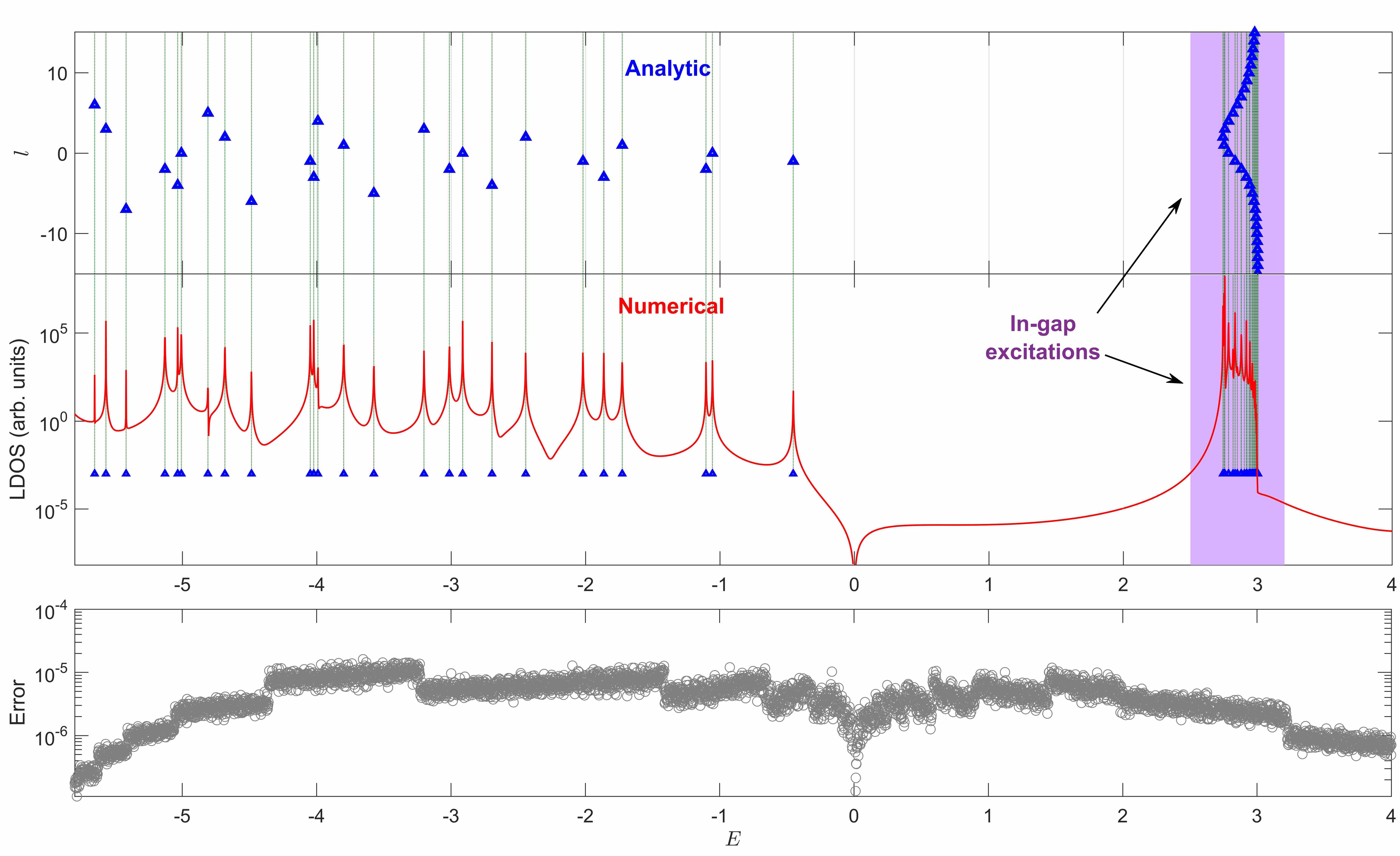}
\caption{ Validation of the multiple multipole method.
For validation purpose, an analytically solvable case of a circular potential
domain is used. Top panel: eigenenergy $E$ versus the angular momentum quantum 
number $l$ calculated analytically from Eq.~(\ref{eq:traeq}). Middle panel: 
the local density of states at a given position inside the domain as a 
function of energy, which are calculated numerically using the multiple 
multipole base expansion method. Bottom panel: the corresponding residual 
error versus energy quantifying the convergence of the numerical method. 
The potential height is $\Delta=V_0=6\hbar v_F/R$.}
\label{fig:SIfig12}
\end{figure}

Imposing the relevant boundary conditions parameterized by the angle $\phi$ 
between the outward normal at any boundary point $\bm{r}_j$ and the $x$-axis:
\begin{subequations}
\begin{equation}
\left.\psi_2^{(I)} \right|_{\bm{r}_j\in\Gamma} 
= \left.\psi_2^{(II)}\right|_{\bm{r}_j\in\Gamma},
\end{equation}
\begin{equation}
\left.\left(\psi_1^{(I)}e^{i\phi} + 
\psi_3^{(I)}e^{-i\phi}\right)
\right|_{\bm{r}_j\in\Gamma}=\left.\left(\psi_1^{(II)}e^{i\phi} 
+ \psi_3^{(II)}e^{-i\phi}\right)\right|_{\bm{r}_j\in\Gamma}, 
\end{equation}
\end{subequations} 
we obtain 
\begin{subequations}
\begin{equation}
\sum_{m_{II}}\sum_l\ ^jA_{lm_{II}}^{(I)}C_l^{m_{II}} - 
\sum_{m_I}\sum_l\ ^jA_{lm_{I}}^{(II)}C_l^{m_I} = -\ ^j\psi_2^{in},
\end{equation}
\begin{equation}
\sum_{m_{II}}\sum_l\ ^jB_{lm_{II}}^{(I)}C_l^{m_{II}} - 
\sum_{m_I}\sum_l\ ^jB_{lm_I}^{(II)}C_l^{m_I} = -\ ^j\chi^{in},
\end{equation}
\end{subequations} 
where the substitutions are given by 
\begin{subequations}
\begin{equation}
\ ^jA_{lm_{II}}^{(I)} = i H_l^{(1)}(k_I|\bm{r}_j - 
\bm{r}_{m_{II}}|)e^{il\theta_{m_{II}}}, 
\end{equation}
\begin{equation}
\ ^jA_{lm_{I}}^{(II)} = i H_l^{(1)}(k_{II}|\bm{r}_j - 
\bm{r}_{m_{I}}|)e^{il\theta_{m_I}},
\end{equation}
\begin{eqnarray}
&\ ^jB_{lm_{II}}^{(I)} = \frac{1}{\sqrt{2}}\left[\alpha_I H_{l-1}^{(1)}(k_I|\bm{r}_j 
- \bm{r}_{m_{II}}|)e^{i(l-1)\theta_{m_{II}}}e^{i\phi} \right.\nonumber\\ 
&\left.- \beta_I H_{l+1}^{(1)}(k_I|\bm{r}_j - 
\bm{r}_{m_{II}}|)e^{i(l+1)\theta_{m_{II}}}e^{-i\phi}\right], 
\end{eqnarray}
\begin{eqnarray}
&\ ^jB_{lm_{I}}^{(II)} = \frac{1}{\sqrt{2}} \left[\alpha_{II}H_{l-1}^{(1)}(k_{II}|\bm{r}_j 
- \bm{r}_{m_{I}}|)e^{i(l-1)\theta_{m_I}}e^{i\phi}  \right.\nonumber\\
&\left.- \beta_{II}H_{l+1}^{(1)}(k_{II}|\bm{r}_j 
- \bm{r}_{m_{I}}|)e^{i(l+1)\theta_{m_I}}e^{-i\phi}\right],
\end{eqnarray}
and 
\begin{equation}
\ ^j\psi_{2}^{in} = \frac{1}{\sqrt{2}} e^{i{k}_I\left(|\bm{r}_j|\cos\theta_j-x_0\right)}, 
\end{equation}
\begin{equation}
\ ^j\chi^{in} = \frac{1}{2}\left[\alpha_Ie^{i\phi} + 
\beta_Ie^{-i\phi}\right] e^{i{k}_I\left(|\bm{r}_j|\cos\theta_j-x_0\right)}.
\end{equation}
\end{subequations}
For the boundary shape defined by Eq.~(\ref{eq:shape}), the associated unit 
normal direction can be written down explicitly: 
\begin{equation}
e^{i\phi} = -ie^{i\theta}\frac{dr(\theta)/d\theta + ir(\theta)}{\left|dr(\theta)/d\theta+ir(\theta)\right|}.
\end{equation}
In principle, the set consists of an infinite number of equations with an 
infinite number of undetermined expansion coefficients $C_l^{m_{II}}$ and 
$C_l^{m_I}$. To solve the system numerically, a finite truncation is necessary,
which turns out to be feasible in practice by discretizing the boundary to a 
finite number of points $J$ and setting the number of basis functions 
$M_\tau$ in the specific region $\tau$ and $l\in[-L, L]$ for all the 
functions. Carrying out the discretization procedure, we arrive at the 
following finite dimensional matrix equation 
\begin{equation}
\mathbb{M}_{2J\times N}\cdot\bm{C}_{N\times1}=-\bm{Y}_{2J\times1},  
\end{equation} 
where $N = (2L+1)\times(M_I + M_{II})=N_I + N_{II}$ and the compact 
substitutions are 
\begin{subequations}
\begin{equation}
\begin{aligned}
&\bm{C}_{N\times1} = [C_{-L}^{1_{II}} \cdots C_{L}^{M_{II}}, C_{-L}^{1_{I}} \cdots C_{L}^{M_{I}}]^T \\
&\bm{Y}_{2J\times1} =[\ ^1\psi_{2}^{in} \cdots \ ^J\psi_{2}^{in}, \ ^1\chi^{in} \cdots \ ^J\chi^{in}]^T,
\end{aligned}
\end{equation}
and
\begin{equation}
\mathbb{M}_{2J\times N} = \left[
\begin{array}{c|c}
\mathbb{A}^{(I)} & -\mathbb{A}^{(II)} \\
\hline
\mathbb{B}^{(I)} & -\mathbb{B}^{(II)}
\end{array}\right]_{2J\times N}, 
\end{equation}
with 
\begin{equation}
\mathbb{A}^{(\tau)}=\begin{pmatrix}
\bm{A}_{-L1_{\overline\tau}}^{(\tau)} & \cdots & \bm{A}_{lM_{\overline\tau}}^{(\tau)} & \cdots & \bm{A}_{LM_{\overline\tau}}^{(\tau)}  \\
\end{pmatrix}_{J\times N_{\overline\tau}},
\end{equation}
\begin{equation}
\mathbb{B}^{(\tau)}=\begin{pmatrix}
\bm{B}_{-L1_{\overline\tau}}^{(\tau)} & \cdots & \bm{B}_{lM_{\overline\tau}}^{(\tau)} & \cdots & \bm{B}_{LM_{\overline\tau}}^{(\tau)}  \\
\end{pmatrix}_{J\times N_{\overline\tau}},
\end{equation}
where 
$$
\begin{aligned}
\bm{B}_{lm_{\overline\tau}}^{(\tau)} &= [\ ^1B_{lm_{\overline\tau}}^{(\tau)}, \ ^2B_{lm_{\overline\tau}}^{(\tau)}, \cdots, \ ^jB_{lm_{\overline\tau}}^{(\tau)}, \cdots, \ ^JB_{lm_{\overline\tau}}^{(\tau)}]^T , \\
\bm{A}_{lm_{\overline\tau}}^{(\tau)} &= [\ ^1A_{lm_{\overline\tau}}^{(\tau)}, \ ^2A_{lm_{\overline\tau}}^{(\tau)}, \cdots, \ ^jA_{lm_{\overline\tau}}^{(\tau)}, \cdots, \ ^JA_{lm_{\overline\tau}}^{(\tau)}]^T.
\end{aligned}
$$
\end{subequations} 
As the expansions are generally nonorthogonal, more equations are required 
than the number of unknowns to enable the deduction of an over-determined matrix 
system with $2J\gg N$, which can be solved by the standard pseudo-inverse 
algorithm: $\bm{C}=-\textrm{pinv}(\mathbb{M})*\bm{Y}$. In particular, we use 
the residual error evaluated at the boundary 
$$\mbox{Error} =\frac{||\mathbb{M}*\bm{C}+\bm{Y}||}{||\bm{Y}||}$$ 
as the criterion to test convergence. We adjust the number, the 
order and/or positions of the multipoles to ensure 
$\mbox{Error} < \mbox{tolerance}$. After the unknown coefficients $\bm{C}$ have 
been obtained, the associated wavefunctions and hence the local density of 
states in the specific region can be calculated accordingly. 

\subsection{Method validation} \label{sec:SME}

To validate the method, we exploit the analytically solvable case of 
circular geometry. Figure~\ref{fig:SIfig12} shows a comparison of the 
eigenenergy spectra obtained analytically and calculated from the multiple
multipole method. The agreement is remarkable. 


\begin{thebibliography}{90}%
\makeatletter
\providecommand \@ifxundefined [1]{%
 \@ifx{#1\undefined}
}%
\providecommand \@ifnum [1]{%
 \ifnum #1\expandafter \@firstoftwo
 \else \expandafter \@secondoftwo
 \fi
}%
\providecommand \@ifx [1]{%
 \ifx #1\expandafter \@firstoftwo
 \else \expandafter \@secondoftwo
 \fi
}%
\providecommand \natexlab [1]{#1}%
\providecommand \enquote  [1]{``#1''}%
\providecommand \bibnamefont  [1]{#1}%
\providecommand \bibfnamefont [1]{#1}%
\providecommand \citenamefont [1]{#1}%
\providecommand \href@noop [0]{\@secondoftwo}%
\providecommand \href [0]{\begingroup \@sanitize@url \@href}%
\providecommand \@href[1]{\@@startlink{#1}\@@href}%
\providecommand \@@href[1]{\endgroup#1\@@endlink}%
\providecommand \@sanitize@url [0]{\catcode `\\12\catcode `\$12\catcode
  `\&12\catcode `\#12\catcode `\^12\catcode `\_12\catcode `\%12\relax}%
\providecommand \@@startlink[1]{}%
\providecommand \@@endlink[0]{}%
\providecommand \url  [0]{\begingroup\@sanitize@url \@url }%
\providecommand \@url [1]{\endgroup\@href {#1}{\urlprefix }}%
\providecommand \urlprefix  [0]{URL }%
\providecommand \Eprint [0]{\href }%
\providecommand \doibase [0]{http://dx.doi.org/}%
\providecommand \selectlanguage [0]{\@gobble}%
\providecommand \bibinfo  [0]{\@secondoftwo}%
\providecommand \bibfield  [0]{\@secondoftwo}%
\providecommand \translation [1]{[#1]}%
\providecommand \BibitemOpen [0]{}%
\providecommand \bibitemStop [0]{}%
\providecommand \bibitemNoStop [0]{.\EOS\space}%
\providecommand \EOS [0]{\spacefactor3000\relax}%
\providecommand \BibitemShut  [1]{\csname bibitem#1\endcsname}%
\let\auto@bib@innerbib\@empty
\bibitem [{\citenamefont {Cao}\ \emph {et~al.}(2018)\citenamefont {Cao},
  \citenamefont {Fatemi}, \citenamefont {Fang}, \citenamefont {Watanabe},
  \citenamefont {Taniguchi}, \citenamefont {Kaxiras},\ and\ \citenamefont
  {Jarillo-Herrero}}]{Caoetal:2018}%
  \BibitemOpen
  \bibfield  {author} {\bibinfo {author} {\bibfnamefont {Y.}~\bibnamefont
  {Cao}}, \bibinfo {author} {\bibfnamefont {V.}~\bibnamefont {Fatemi}},
  \bibinfo {author} {\bibfnamefont {S.}~\bibnamefont {Fang}}, \bibinfo {author}
  {\bibfnamefont {K.}~\bibnamefont {Watanabe}}, \bibinfo {author}
  {\bibfnamefont {T.}~\bibnamefont {Taniguchi}}, \bibinfo {author}
  {\bibfnamefont {E.}~\bibnamefont {Kaxiras}}, \ and\ \bibinfo {author}
  {\bibfnamefont {P.}~\bibnamefont {Jarillo-Herrero}},\ }\bibfield  {title}
  {\enquote {\bibinfo {title} {Unconventional superconductivity in magic-angle
  graphene superlattices},}\ }\href@noop {} {\bibfield  {journal} {\bibinfo
  {journal} {Nature}\ }\textbf {\bibinfo {volume} {556}},\ \bibinfo {pages}
  {43} (\bibinfo {year} {2018})}\BibitemShut {NoStop}%
\bibitem [{\citenamefont {Yankowitz}\ \emph {et~al.}(2019)\citenamefont
  {Yankowitz}, \citenamefont {Chen}, \citenamefont {Polshyn}, \citenamefont
  {Zhang}, \citenamefont {Watanabe}, \citenamefont {Taniguchi}, \citenamefont
  {Graf}, \citenamefont {Young},\ and\ \citenamefont {Dean}}]{YCPZWTGYD:2019}%
  \BibitemOpen
  \bibfield  {author} {\bibinfo {author} {\bibfnamefont {M.}~\bibnamefont
  {Yankowitz}}, \bibinfo {author} {\bibfnamefont {S.}~\bibnamefont {Chen}},
  \bibinfo {author} {\bibfnamefont {H.}~\bibnamefont {Polshyn}}, \bibinfo
  {author} {\bibfnamefont {Y.}~\bibnamefont {Zhang}}, \bibinfo {author}
  {\bibfnamefont {K.}~\bibnamefont {Watanabe}}, \bibinfo {author}
  {\bibfnamefont {T.}~\bibnamefont {Taniguchi}}, \bibinfo {author}
  {\bibfnamefont {D.}~\bibnamefont {Graf}}, \bibinfo {author} {\bibfnamefont
  {A.~F.}\ \bibnamefont {Young}}, \ and\ \bibinfo {author} {\bibfnamefont
  {C.~R.}\ \bibnamefont {Dean}},\ }\bibfield  {title} {\enquote {\bibinfo
  {title} {Tuning superconductivity in twisted bilayer graphene},}\ }\href@noop
  {} {\bibfield  {journal} {\bibinfo  {journal} {Science}\ }\textbf {\bibinfo
  {volume} {363}},\ \bibinfo {pages} {1059} (\bibinfo {year}
  {2019})}\BibitemShut {NoStop}%
\bibitem [{\citenamefont {Sharpe}\ \emph {et~al.}(2019)\citenamefont {Sharpe},
  \citenamefont {Fox}, \citenamefont {Barnard}, \citenamefont {Finney},
  \citenamefont {Watanabe}, \citenamefont {Taniguchi}, \citenamefont
  {Kastner},\ and\ \citenamefont {Goldhaber-Gordon}}]{Sharpeetal:2019}%
  \BibitemOpen
  \bibfield  {author} {\bibinfo {author} {\bibfnamefont {A.~L.}\ \bibnamefont
  {Sharpe}}, \bibinfo {author} {\bibfnamefont {E.~J.}\ \bibnamefont {Fox}},
  \bibinfo {author} {\bibfnamefont {A.~W.}\ \bibnamefont {Barnard}}, \bibinfo
  {author} {\bibfnamefont {J.}~\bibnamefont {Finney}}, \bibinfo {author}
  {\bibfnamefont {K.}~\bibnamefont {Watanabe}}, \bibinfo {author}
  {\bibfnamefont {T.}~\bibnamefont {Taniguchi}}, \bibinfo {author}
  {\bibfnamefont {M.~A.}\ \bibnamefont {Kastner}}, \ and\ \bibinfo {author}
  {\bibfnamefont {D.}~\bibnamefont {Goldhaber-Gordon}},\ }\bibfield  {title}
  {\enquote {\bibinfo {title} {Emergent ferromagnetism near three-quarters
  filling in twisted bilayer graphene},}\ }\href@noop {} {\bibfield  {journal}
  {\bibinfo  {journal} {arXiv}\ }\textbf {\bibinfo {volume} {1901.03520}}
  (\bibinfo {year} {2019})}\BibitemShut {NoStop}%
\bibitem [{\citenamefont {Lu}\ \emph {et~al.}(2019)\citenamefont {Lu},
  \citenamefont {Stepanov}, \citenamefont {Yang}, \citenamefont {Xie},
  \citenamefont {Aamir}, \citenamefont {Das}, \citenamefont {Urgell},
  \citenamefont {Watanabe}, \citenamefont {Taniguchi}, \citenamefont {Zhang},
  \citenamefont {Bachtold}, \citenamefont {MacDonald},\ and\ \citenamefont
  {Efetov}}]{Luetal:2019}%
  \BibitemOpen
  \bibfield  {author} {\bibinfo {author} {\bibfnamefont {X.-B.}\ \bibnamefont
  {Lu}}, \bibinfo {author} {\bibfnamefont {P.}~\bibnamefont {Stepanov}},
  \bibinfo {author} {\bibfnamefont {W.}~\bibnamefont {Yang}}, \bibinfo {author}
  {\bibfnamefont {M.}~\bibnamefont {Xie}}, \bibinfo {author} {\bibfnamefont
  {M.~A.}\ \bibnamefont {Aamir}}, \bibinfo {author} {\bibfnamefont
  {I.}~\bibnamefont {Das}}, \bibinfo {author} {\bibfnamefont {C.}~\bibnamefont
  {Urgell}}, \bibinfo {author} {\bibfnamefont {K.}~\bibnamefont {Watanabe}},
  \bibinfo {author} {\bibfnamefont {T.}~\bibnamefont {Taniguchi}}, \bibinfo
  {author} {\bibfnamefont {G.-Y.}\ \bibnamefont {Zhang}}, \bibinfo {author}
  {\bibfnamefont {A.}~\bibnamefont {Bachtold}}, \bibinfo {author}
  {\bibfnamefont {A.~H.}\ \bibnamefont {MacDonald}}, \ and\ \bibinfo {author}
  {\bibfnamefont {D.~K.}\ \bibnamefont {Efetov}},\ }\bibfield  {title}
  {\enquote {\bibinfo {title} {Superconductors, orbital magnets, and correlated
  states in magic angle bilayer graphene},}\ }\href@noop {} {\bibfield
  {journal} {\bibinfo  {journal} {arXiv}\ }\textbf {\bibinfo {volume}
  {1903.06513}} (\bibinfo {year} {2019})}\BibitemShut {NoStop}%
\bibitem [{\citenamefont {Klitzing}\ \emph {et~al.}(1980)\citenamefont
  {Klitzing}, \citenamefont {Dorda},\ and\ \citenamefont {Pepper}}]{KDP:1980}%
  \BibitemOpen
  \bibfield  {author} {\bibinfo {author} {\bibfnamefont {K.~V.}\ \bibnamefont
  {Klitzing}}, \bibinfo {author} {\bibfnamefont {G.}~\bibnamefont {Dorda}}, \
  and\ \bibinfo {author} {\bibfnamefont {M.}~\bibnamefont {Pepper}},\
  }\bibfield  {title} {\enquote {\bibinfo {title} {New method for high-accuracy
  determination of the fine-structure constant based on quantized {Hall}
  resistance},}\ }\href {\doibase 10.1103/PhysRevLett.45.494} {\bibfield
  {journal} {\bibinfo  {journal} {Phys. Rev. Lett.}\ }\textbf {\bibinfo
  {volume} {45}},\ \bibinfo {pages} {494} (\bibinfo {year} {1980})}\BibitemShut
  {NoStop}%
\bibitem [{\citenamefont {Thouless}\ \emph {et~al.}(1982)\citenamefont
  {Thouless}, \citenamefont {Kohmoto}, \citenamefont {Nightingale},\ and\
  \citenamefont {den Nijs}}]{TKNN:1982}%
  \BibitemOpen
  \bibfield  {author} {\bibinfo {author} {\bibfnamefont {D.~J.}\ \bibnamefont
  {Thouless}}, \bibinfo {author} {\bibfnamefont {M.}~\bibnamefont {Kohmoto}},
  \bibinfo {author} {\bibfnamefont {M.~P.}\ \bibnamefont {Nightingale}}, \ and\
  \bibinfo {author} {\bibfnamefont {M.}~\bibnamefont {den Nijs}},\ }\bibfield
  {title} {\enquote {\bibinfo {title} {Quantized {Hall} conductance in a
  two-dimensional periodic potential},}\ }\href {\doibase
  10.1103/PhysRevLett.49.405} {\bibfield  {journal} {\bibinfo  {journal} {Phys.
  Rev. Lett.}\ }\textbf {\bibinfo {volume} {49}},\ \bibinfo {pages} {405}
  (\bibinfo {year} {1982})}\BibitemShut {NoStop}%
\bibitem [{\citenamefont {Haldane}(1988)}]{Haldane:1988}%
  \BibitemOpen
  \bibfield  {author} {\bibinfo {author} {\bibfnamefont {F.~D.~M.}\
  \bibnamefont {Haldane}},\ }\bibfield  {title} {\enquote {\bibinfo {title}
  {Model for a quantum {Hall} effect without {Landau} levels: Condensed-matter
  realization of the ``parity anomaly''},}\ }\href {\doibase
  10.1103/PhysRevLett.61.2015} {\bibfield  {journal} {\bibinfo  {journal}
  {Phys. Rev. Lett.}\ }\textbf {\bibinfo {volume} {61}},\ \bibinfo {pages}
  {2015} (\bibinfo {year} {1988})}\BibitemShut {NoStop}%
\bibitem [{\citenamefont {Bernevig}\ \emph {et~al.}(2006)\citenamefont
  {Bernevig}, \citenamefont {Hughes},\ and\ \citenamefont {Zhang}}]{BHZ:2006}%
  \BibitemOpen
  \bibfield  {author} {\bibinfo {author} {\bibfnamefont {B.~A.}\ \bibnamefont
  {Bernevig}}, \bibinfo {author} {\bibfnamefont {T.~L.}\ \bibnamefont
  {Hughes}}, \ and\ \bibinfo {author} {\bibfnamefont {S.-C.}\ \bibnamefont
  {Zhang}},\ }\bibfield  {title} {\enquote {\bibinfo {title} {Quantum spin
  {Hall} effect and topological phase transition in {HgTe} quantum wells},}\
  }\href@noop {} {\bibfield  {journal} {\bibinfo  {journal} {Science}\ }\textbf
  {\bibinfo {volume} {314}},\ \bibinfo {pages} {1757} (\bibinfo {year}
  {2006})}\BibitemShut {NoStop}%
\bibitem [{\citenamefont {Fu}\ and\ \citenamefont {Kane}(2007)}]{FK:2007}%
  \BibitemOpen
  \bibfield  {author} {\bibinfo {author} {\bibfnamefont {L.}~\bibnamefont
  {Fu}}\ and\ \bibinfo {author} {\bibfnamefont {C.~L.}\ \bibnamefont {Kane}},\
  }\bibfield  {title} {\enquote {\bibinfo {title} {Topological insulators with
  inversion symmetry},}\ }\href {\doibase 10.1103/PhysRevB.76.045302}
  {\bibfield  {journal} {\bibinfo  {journal} {Phys. Rev. B}\ }\textbf {\bibinfo
  {volume} {76}},\ \bibinfo {pages} {045302} (\bibinfo {year}
  {2007})}\BibitemShut {NoStop}%
\bibitem [{\citenamefont {Zhang}\ \emph {et~al.}(2009)\citenamefont {Zhang},
  \citenamefont {Liu}, \citenamefont {Qi}, \citenamefont {Dai}, \citenamefont
  {Fang},\ and\ \citenamefont {Zhang}}]{ZLQDFZ:2009}%
  \BibitemOpen
  \bibfield  {author} {\bibinfo {author} {\bibfnamefont {H.}~\bibnamefont
  {Zhang}}, \bibinfo {author} {\bibfnamefont {C.-X.}\ \bibnamefont {Liu}},
  \bibinfo {author} {\bibfnamefont {X.-L.}\ \bibnamefont {Qi}}, \bibinfo
  {author} {\bibfnamefont {X.}~\bibnamefont {Dai}}, \bibinfo {author}
  {\bibfnamefont {Z.}~\bibnamefont {Fang}}, \ and\ \bibinfo {author}
  {\bibfnamefont {S.-C.}\ \bibnamefont {Zhang}},\ }\bibfield  {title} {\enquote
  {\bibinfo {title} {Topological insulators in {Bi$_2$Se$_3$, Bi$_2$Te$_3$ and
  Sb$_2$Te$_3$} with a single {Dirac} cone on the surface},}\ }\href@noop {}
  {\bibfield  {journal} {\bibinfo  {journal} {Nat. Phys.}\ }\textbf {\bibinfo
  {volume} {5}},\ \bibinfo {pages} {438} (\bibinfo {year} {2009})}\BibitemShut
  {NoStop}%
\bibitem [{\citenamefont {K{\"o}nig}\ \emph {et~al.}(2007)\citenamefont
  {K{\"o}nig}, \citenamefont {Wiedmann}, \citenamefont {Br{\"u}ne},
  \citenamefont {Roth}, \citenamefont {Buhmann}, \citenamefont {Molenkamp},
  \citenamefont {Qi},\ and\ \citenamefont {Zhang}}]{KWBRBMQZ:2007}%
  \BibitemOpen
  \bibfield  {author} {\bibinfo {author} {\bibfnamefont {M.}~\bibnamefont
  {K{\"o}nig}}, \bibinfo {author} {\bibfnamefont {S.}~\bibnamefont {Wiedmann}},
  \bibinfo {author} {\bibfnamefont {C.}~\bibnamefont {Br{\"u}ne}}, \bibinfo
  {author} {\bibfnamefont {A.}~\bibnamefont {Roth}}, \bibinfo {author}
  {\bibfnamefont {H.}~\bibnamefont {Buhmann}}, \bibinfo {author} {\bibfnamefont
  {L.~W.}\ \bibnamefont {Molenkamp}}, \bibinfo {author} {\bibfnamefont {X.-L.}\
  \bibnamefont {Qi}}, \ and\ \bibinfo {author} {\bibfnamefont {S.-C.}\
  \bibnamefont {Zhang}},\ }\bibfield  {title} {\enquote {\bibinfo {title}
  {Quantum spin {Hall} insulator state in hgte quantum wells},}\ }\href@noop {}
  {\bibfield  {journal} {\bibinfo  {journal} {Science}\ }\textbf {\bibinfo
  {volume} {318}},\ \bibinfo {pages} {766} (\bibinfo {year}
  {2007})}\BibitemShut {NoStop}%
\bibitem [{\citenamefont {Hsieh}\ \emph {et~al.}(2008)\citenamefont {Hsieh},
  \citenamefont {Qian}, \citenamefont {Wray}, \citenamefont {Xia},
  \citenamefont {Hor}, \citenamefont {Cava},\ and\ \citenamefont
  {Hasan}}]{HQWXHCH:2008}%
  \BibitemOpen
  \bibfield  {author} {\bibinfo {author} {\bibfnamefont {D.}~\bibnamefont
  {Hsieh}}, \bibinfo {author} {\bibfnamefont {D.}~\bibnamefont {Qian}},
  \bibinfo {author} {\bibfnamefont {L.}~\bibnamefont {Wray}}, \bibinfo {author}
  {\bibfnamefont {Y.}~\bibnamefont {Xia}}, \bibinfo {author} {\bibfnamefont
  {Y.~S.}\ \bibnamefont {Hor}}, \bibinfo {author} {\bibfnamefont {R.~J.}\
  \bibnamefont {Cava}}, \ and\ \bibinfo {author} {\bibfnamefont {M.~Z.}\
  \bibnamefont {Hasan}},\ }\bibfield  {title} {\enquote {\bibinfo {title} {A
  topological {Dirac} insulator in a quantum spin {Hall} phase},}\ }\href@noop
  {} {\bibfield  {journal} {\bibinfo  {journal} {Nature}\ }\textbf {\bibinfo
  {volume} {452}},\ \bibinfo {pages} {970} (\bibinfo {year}
  {2008})}\BibitemShut {NoStop}%
\bibitem [{\citenamefont {Xia}\ \emph {et~al.}(2009)\citenamefont {Xia},
  \citenamefont {Qian}, \citenamefont {Hsieh}, \citenamefont {Wray},
  \citenamefont {Pal}, \citenamefont {Lin}, \citenamefont {Bansil},
  \citenamefont {Grauer}, \citenamefont {Hor}, \citenamefont {Cava} \emph
  {et~al.}}]{XQHWPLBGHC:2009}%
  \BibitemOpen
  \bibfield  {author} {\bibinfo {author} {\bibfnamefont {Y.}~\bibnamefont
  {Xia}}, \bibinfo {author} {\bibfnamefont {D.}~\bibnamefont {Qian}}, \bibinfo
  {author} {\bibfnamefont {D.}~\bibnamefont {Hsieh}}, \bibinfo {author}
  {\bibfnamefont {L.}~\bibnamefont {Wray}}, \bibinfo {author} {\bibfnamefont
  {A.}~\bibnamefont {Pal}}, \bibinfo {author} {\bibfnamefont {H.}~\bibnamefont
  {Lin}}, \bibinfo {author} {\bibfnamefont {A.}~\bibnamefont {Bansil}},
  \bibinfo {author} {\bibfnamefont {D.}~\bibnamefont {Grauer}}, \bibinfo
  {author} {\bibfnamefont {Y.~S.}\ \bibnamefont {Hor}}, \bibinfo {author}
  {\bibfnamefont {R.~J.}\ \bibnamefont {Cava}},  \emph {et~al.},\ }\bibfield
  {title} {\enquote {\bibinfo {title} {Observation of a large-gap
  topological-insulator class with a single {Dirac} cone on the surface},}\
  }\href@noop {} {\bibfield  {journal} {\bibinfo  {journal} {Nat. Phys.}\
  }\textbf {\bibinfo {volume} {5}},\ \bibinfo {pages} {398} (\bibinfo {year}
  {2009})}\BibitemShut {NoStop}%
\bibitem [{\citenamefont {Moore}(2010)}]{Moore:2010}%
  \BibitemOpen
  \bibfield  {author} {\bibinfo {author} {\bibfnamefont {J.~E.}\ \bibnamefont
  {Moore}},\ }\bibfield  {title} {\enquote {\bibinfo {title} {The birth of
  topological insulators},}\ }\href@noop {} {\bibfield  {journal} {\bibinfo
  {journal} {Nature}\ }\textbf {\bibinfo {volume} {464}},\ \bibinfo {pages}
  {194} (\bibinfo {year} {2010})}\BibitemShut {NoStop}%
\bibitem [{\citenamefont {Hasan}\ and\ \citenamefont {Kane}(2010)}]{HK:2010}%
  \BibitemOpen
  \bibfield  {author} {\bibinfo {author} {\bibfnamefont {M.~Z.}\ \bibnamefont
  {Hasan}}\ and\ \bibinfo {author} {\bibfnamefont {C.~L.}\ \bibnamefont
  {Kane}},\ }\bibfield  {title} {\enquote {\bibinfo {title} {Colloquium:
  Topological insulators},}\ }\href {\doibase 10.1103/RevModPhys.82.3045}
  {\bibfield  {journal} {\bibinfo  {journal} {Rev. Mod. Phys.}\ }\textbf
  {\bibinfo {volume} {82}},\ \bibinfo {pages} {3045} (\bibinfo {year}
  {2010})}\BibitemShut {NoStop}%
\bibitem [{\citenamefont {Qi}\ and\ \citenamefont {Zhang}(2011)}]{QZ:2011}%
  \BibitemOpen
  \bibfield  {author} {\bibinfo {author} {\bibfnamefont {X.-L.}\ \bibnamefont
  {Qi}}\ and\ \bibinfo {author} {\bibfnamefont {S.-C.}\ \bibnamefont {Zhang}},\
  }\bibfield  {title} {\enquote {\bibinfo {title} {Topological insulators and
  superconductors},}\ }\href {\doibase 10.1103/RevModPhys.83.1057} {\bibfield
  {journal} {\bibinfo  {journal} {Rev. Mod. Phys.}\ }\textbf {\bibinfo {volume}
  {83}},\ \bibinfo {pages} {1057} (\bibinfo {year} {2011})}\BibitemShut
  {NoStop}%
\bibitem [{\citenamefont {Benalcazar}\ \emph {et~al.}(2017)\citenamefont
  {Benalcazar}, \citenamefont {Bernevig},\ and\ \citenamefont
  {Hughes}}]{Benalcazar2017}%
  \BibitemOpen
  \bibfield  {author} {\bibinfo {author} {\bibfnamefont {W.~A.}\ \bibnamefont
  {Benalcazar}}, \bibinfo {author} {\bibfnamefont {B.~A.}\ \bibnamefont
  {Bernevig}}, \ and\ \bibinfo {author} {\bibfnamefont {T.~L.}\ \bibnamefont
  {Hughes}},\ }\bibfield  {title} {\enquote {\bibinfo {title} {Quantized
  electric multipole insulators},}\ }\href {\doibase 10.1126/science.aah6442}
  {\bibfield  {journal} {\bibinfo  {journal} {Science}\ }\textbf {\bibinfo
  {volume} {357}},\ \bibinfo {pages} {61} (\bibinfo {year} {2017})}\BibitemShut
  {NoStop}%
\bibitem [{\citenamefont {Song}\ \emph {et~al.}(2017)\citenamefont {Song},
  \citenamefont {Fang},\ and\ \citenamefont {Fang}}]{Song2017}%
  \BibitemOpen
  \bibfield  {author} {\bibinfo {author} {\bibfnamefont {Z.}~\bibnamefont
  {Song}}, \bibinfo {author} {\bibfnamefont {Z.}~\bibnamefont {Fang}}, \ and\
  \bibinfo {author} {\bibfnamefont {C.}~\bibnamefont {Fang}},\ }\bibfield
  {title} {\enquote {\bibinfo {title} {$(d\ensuremath{-}2)$-dimensional edge
  states of rotation symmetry protected topological states},}\ }\href {\doibase
  10.1103/PhysRevLett.119.246402} {\bibfield  {journal} {\bibinfo  {journal}
  {Phys. Rev. Lett.}\ }\textbf {\bibinfo {volume} {119}},\ \bibinfo {pages}
  {246402} (\bibinfo {year} {2017})}\BibitemShut {NoStop}%
\bibitem [{\citenamefont {Schindler}\ \emph {et~al.}(2018)\citenamefont
  {Schindler}, \citenamefont {Cook}, \citenamefont {Vergniory}, \citenamefont
  {Wang}, \citenamefont {Parkin}, \citenamefont {Bernevig},\ and\ \citenamefont
  {Neupert}}]{Schindlereaat2018}%
  \BibitemOpen
  \bibfield  {author} {\bibinfo {author} {\bibfnamefont {F.}~\bibnamefont
  {Schindler}}, \bibinfo {author} {\bibfnamefont {A.~M.}\ \bibnamefont {Cook}},
  \bibinfo {author} {\bibfnamefont {M.~G.}\ \bibnamefont {Vergniory}}, \bibinfo
  {author} {\bibfnamefont {Z.}~\bibnamefont {Wang}}, \bibinfo {author}
  {\bibfnamefont {S.~S.~P.}\ \bibnamefont {Parkin}}, \bibinfo {author}
  {\bibfnamefont {B.~A.}\ \bibnamefont {Bernevig}}, \ and\ \bibinfo {author}
  {\bibfnamefont {T.}~\bibnamefont {Neupert}},\ }\bibfield  {title} {\enquote
  {\bibinfo {title} {Higher-order topological insulators},}\ }\href {\doibase
  10.1126/sciadv.aat0346} {\bibfield  {journal} {\bibinfo  {journal} {Sci.
  Adv.}\ }\textbf {\bibinfo {volume} {4}},\ \bibinfo {pages} {eaat0346}
  (\bibinfo {year} {2018})}\BibitemShut {NoStop}%
\bibitem [{\citenamefont {Pesin}\ and\ \citenamefont
  {MacDonald}(2012)}]{PM:2012}%
  \BibitemOpen
  \bibfield  {author} {\bibinfo {author} {\bibfnamefont {D.}~\bibnamefont
  {Pesin}}\ and\ \bibinfo {author} {\bibfnamefont {A.~H.}\ \bibnamefont
  {MacDonald}},\ }\bibfield  {title} {\enquote {\bibinfo {title} {Spintronics
  and pseudospintronics in graphene and topological insulators},}\ }\href@noop
  {} {\bibfield  {journal} {\bibinfo  {journal} {Nat. Mater.}\ }\textbf
  {\bibinfo {volume} {11}},\ \bibinfo {pages} {409} (\bibinfo {year}
  {2012})}\BibitemShut {NoStop}%
\bibitem [{\citenamefont {Kane}\ and\ \citenamefont
  {Mele}(2005{\natexlab{a}})}]{Kane2005a}%
  \BibitemOpen
  \bibfield  {author} {\bibinfo {author} {\bibfnamefont {C.~L.}\ \bibnamefont
  {Kane}}\ and\ \bibinfo {author} {\bibfnamefont {E.~J.}\ \bibnamefont
  {Mele}},\ }\bibfield  {title} {\enquote {\bibinfo {title} {Quantum spin
  {Hall} effect in graphene},}\ }\href {\doibase 10.1103/PhysRevLett.95.226801}
  {\bibfield  {journal} {\bibinfo  {journal} {Phys. Rev. Lett.}\ }\textbf
  {\bibinfo {volume} {95}},\ \bibinfo {pages} {226801} (\bibinfo {year}
  {2005}{\natexlab{a}})}\BibitemShut {NoStop}%
\bibitem [{\citenamefont {Kane}\ and\ \citenamefont
  {Mele}(2005{\natexlab{b}})}]{Kane2005b}%
  \BibitemOpen
  \bibfield  {author} {\bibinfo {author} {\bibfnamefont {C.~L.}\ \bibnamefont
  {Kane}}\ and\ \bibinfo {author} {\bibfnamefont {E.~J.}\ \bibnamefont
  {Mele}},\ }\bibfield  {title} {\enquote {\bibinfo {title} {${Z}_{2}$
  topological order and the quantum spin {Hall} effect},}\ }\href {\doibase
  10.1103/PhysRevLett.95.146802} {\bibfield  {journal} {\bibinfo  {journal}
  {Phys. Rev. Lett.}\ }\textbf {\bibinfo {volume} {95}},\ \bibinfo {pages}
  {146802} (\bibinfo {year} {2005}{\natexlab{b}})}\BibitemShut {NoStop}%
\bibitem [{\citenamefont {Chang}\ \emph {et~al.}(2013)\citenamefont {Chang},
  \citenamefont {Zhang}, \citenamefont {Feng}, \citenamefont {Shen},
  \citenamefont {Zhang}, \citenamefont {Guo}, \citenamefont {Li}, \citenamefont
  {Ou}, \citenamefont {Wei}, \citenamefont {Wang}, \citenamefont {Ji},
  \citenamefont {Feng}, \citenamefont {Ji}, \citenamefont {Chen}, \citenamefont
  {Jia}, \citenamefont {Dai}, \citenamefont {Fang}, \citenamefont {Zhang},
  \citenamefont {He}, \citenamefont {Wang}, \citenamefont {Lu}, \citenamefont
  {Ma},\ and\ \citenamefont {Xue}}]{Chang2013}%
  \BibitemOpen
  \bibfield  {author} {\bibinfo {author} {\bibfnamefont {C.-Z.}\ \bibnamefont
  {Chang}}, \bibinfo {author} {\bibfnamefont {J.}~\bibnamefont {Zhang}},
  \bibinfo {author} {\bibfnamefont {X.}~\bibnamefont {Feng}}, \bibinfo {author}
  {\bibfnamefont {J.}~\bibnamefont {Shen}}, \bibinfo {author} {\bibfnamefont
  {Z.}~\bibnamefont {Zhang}}, \bibinfo {author} {\bibfnamefont
  {M.}~\bibnamefont {Guo}}, \bibinfo {author} {\bibfnamefont {K.}~\bibnamefont
  {Li}}, \bibinfo {author} {\bibfnamefont {Y.}~\bibnamefont {Ou}}, \bibinfo
  {author} {\bibfnamefont {P.}~\bibnamefont {Wei}}, \bibinfo {author}
  {\bibfnamefont {L.-L.}\ \bibnamefont {Wang}}, \bibinfo {author}
  {\bibfnamefont {Z.-Q.}\ \bibnamefont {Ji}}, \bibinfo {author} {\bibfnamefont
  {Y.}~\bibnamefont {Feng}}, \bibinfo {author} {\bibfnamefont {S.}~\bibnamefont
  {Ji}}, \bibinfo {author} {\bibfnamefont {X.}~\bibnamefont {Chen}}, \bibinfo
  {author} {\bibfnamefont {J.}~\bibnamefont {Jia}}, \bibinfo {author}
  {\bibfnamefont {X.}~\bibnamefont {Dai}}, \bibinfo {author} {\bibfnamefont
  {Z.}~\bibnamefont {Fang}}, \bibinfo {author} {\bibfnamefont {S.-C.}\
  \bibnamefont {Zhang}}, \bibinfo {author} {\bibfnamefont {K.}~\bibnamefont
  {He}}, \bibinfo {author} {\bibfnamefont {Y.}~\bibnamefont {Wang}}, \bibinfo
  {author} {\bibfnamefont {L.}~\bibnamefont {Lu}}, \bibinfo {author}
  {\bibfnamefont {X.-C.}\ \bibnamefont {Ma}}, \ and\ \bibinfo {author}
  {\bibfnamefont {Q.-K.}\ \bibnamefont {Xue}},\ }\bibfield  {title} {\enquote
  {\bibinfo {title} {Experimental observation of the quantum anomalous {Hall}
  effect in a magnetic topological insulator},}\ }\href {\doibase
  10.1126/science.1234414} {\bibfield  {journal} {\bibinfo  {journal}
  {Science}\ }\textbf {\bibinfo {volume} {340}},\ \bibinfo {pages} {167}
  (\bibinfo {year} {2013})}\BibitemShut {NoStop}%
\bibitem [{\citenamefont {Semenoff}\ \emph {et~al.}(2008)\citenamefont
  {Semenoff}, \citenamefont {Semenoff},\ and\ \citenamefont
  {Zhou}}]{Semenoff2008}%
  \BibitemOpen
  \bibfield  {author} {\bibinfo {author} {\bibfnamefont {G.~W.}\ \bibnamefont
  {Semenoff}}, \bibinfo {author} {\bibfnamefont {V.}~\bibnamefont {Semenoff}},
  \ and\ \bibinfo {author} {\bibfnamefont {F.}~\bibnamefont {Zhou}},\
  }\bibfield  {title} {\enquote {\bibinfo {title} {Domain walls in gapped
  graphene},}\ }\href {\doibase 10.1103/PhysRevLett.101.087204} {\bibfield
  {journal} {\bibinfo  {journal} {Phys. Rev. Lett.}\ }\textbf {\bibinfo
  {volume} {101}},\ \bibinfo {pages} {087204} (\bibinfo {year}
  {2008})}\BibitemShut {NoStop}%
\bibitem [{\citenamefont {Martin}\ \emph {et~al.}(2008)\citenamefont {Martin},
  \citenamefont {Blanter},\ and\ \citenamefont {Morpurgo}}]{Martin2008}%
  \BibitemOpen
  \bibfield  {author} {\bibinfo {author} {\bibfnamefont {I.}~\bibnamefont
  {Martin}}, \bibinfo {author} {\bibfnamefont {Y.~M.}\ \bibnamefont {Blanter}},
  \ and\ \bibinfo {author} {\bibfnamefont {A.~F.}\ \bibnamefont {Morpurgo}},\
  }\bibfield  {title} {\enquote {\bibinfo {title} {Topological confinement in
  bilayer graphene},}\ }\href {\doibase 10.1103/PhysRevLett.100.036804}
  {\bibfield  {journal} {\bibinfo  {journal} {Phys. Rev. Lett.}\ }\textbf
  {\bibinfo {volume} {100}},\ \bibinfo {pages} {036804} (\bibinfo {year}
  {2008})}\BibitemShut {NoStop}%
\bibitem [{\citenamefont {Qiao}\ \emph {et~al.}(2011)\citenamefont {Qiao},
  \citenamefont {Jung}, \citenamefont {Niu},\ and\ \citenamefont
  {MacDonald}}]{Qiao2011}%
  \BibitemOpen
  \bibfield  {author} {\bibinfo {author} {\bibfnamefont {Z.}~\bibnamefont
  {Qiao}}, \bibinfo {author} {\bibfnamefont {J.}~\bibnamefont {Jung}}, \bibinfo
  {author} {\bibfnamefont {Q.}~\bibnamefont {Niu}}, \ and\ \bibinfo {author}
  {\bibfnamefont {A.~H.}\ \bibnamefont {MacDonald}},\ }\bibfield  {title}
  {\enquote {\bibinfo {title} {Electronic highways in bilayer graphene},}\
  }\href {\doibase 10.1021/nl201941f} {\bibfield  {journal} {\bibinfo
  {journal} {Nano Lett.}\ }\textbf {\bibinfo {volume} {11}},\ \bibinfo {pages}
  {3453} (\bibinfo {year} {2011})}\BibitemShut {NoStop}%
\bibitem [{\citenamefont {Yasuda}\ \emph {et~al.}(2017)\citenamefont {Yasuda},
  \citenamefont {Mogi}, \citenamefont {Yoshimi}, \citenamefont {Tsukazaki},
  \citenamefont {Takahashi}, \citenamefont {Kawasaki}, \citenamefont {Kagawa},\
  and\ \citenamefont {Tokura}}]{Yasuda2017}%
  \BibitemOpen
  \bibfield  {author} {\bibinfo {author} {\bibfnamefont {K.}~\bibnamefont
  {Yasuda}}, \bibinfo {author} {\bibfnamefont {M.}~\bibnamefont {Mogi}},
  \bibinfo {author} {\bibfnamefont {R.}~\bibnamefont {Yoshimi}}, \bibinfo
  {author} {\bibfnamefont {A.}~\bibnamefont {Tsukazaki}}, \bibinfo {author}
  {\bibfnamefont {K.~S.}\ \bibnamefont {Takahashi}}, \bibinfo {author}
  {\bibfnamefont {M.}~\bibnamefont {Kawasaki}}, \bibinfo {author}
  {\bibfnamefont {F.}~\bibnamefont {Kagawa}}, \ and\ \bibinfo {author}
  {\bibfnamefont {Y.}~\bibnamefont {Tokura}},\ }\bibfield  {title} {\enquote
  {\bibinfo {title} {Quantized chiral edge conduction on domain walls of a
  magnetic topological insulator},}\ }\href {\doibase 10.1126/science.aan5991}
  {\bibfield  {journal} {\bibinfo  {journal} {Science}\ }\textbf {\bibinfo
  {volume} {358}},\ \bibinfo {pages} {1311} (\bibinfo {year}
  {2017})}\BibitemShut {NoStop}%
\bibitem [{\citenamefont {Zhang}\ \emph {et~al.}(2011)\citenamefont {Zhang},
  \citenamefont {Jung}, \citenamefont {Fiete}, \citenamefont {Niu},\ and\
  \citenamefont {MacDonald}}]{Zhang2011}%
  \BibitemOpen
  \bibfield  {author} {\bibinfo {author} {\bibfnamefont {F.}~\bibnamefont
  {Zhang}}, \bibinfo {author} {\bibfnamefont {J.}~\bibnamefont {Jung}},
  \bibinfo {author} {\bibfnamefont {G.~A.}\ \bibnamefont {Fiete}}, \bibinfo
  {author} {\bibfnamefont {Q.}~\bibnamefont {Niu}}, \ and\ \bibinfo {author}
  {\bibfnamefont {A.~H.}\ \bibnamefont {MacDonald}},\ }\bibfield  {title}
  {\enquote {\bibinfo {title} {Spontaneous quantum {Hall} states in chirally
  stacked few-layer graphene systems},}\ }\href {\doibase
  10.1103/PhysRevLett.106.156801} {\bibfield  {journal} {\bibinfo  {journal}
  {Phys. Rev. Lett.}\ }\textbf {\bibinfo {volume} {106}},\ \bibinfo {pages}
  {156801} (\bibinfo {year} {2011})}\BibitemShut {NoStop}%
\bibitem [{\citenamefont {Fu}(2011)}]{Fu2011b}%
  \BibitemOpen
  \bibfield  {author} {\bibinfo {author} {\bibfnamefont {L.}~\bibnamefont
  {Fu}},\ }\bibfield  {title} {\enquote {\bibinfo {title} {Topological
  crystalline insulators},}\ }\href {\doibase 10.1103/PhysRevLett.106.106802}
  {\bibfield  {journal} {\bibinfo  {journal} {Phys. Rev. Lett.}\ }\textbf
  {\bibinfo {volume} {106}},\ \bibinfo {pages} {106802} (\bibinfo {year}
  {2011})}\BibitemShut {NoStop}%
\bibitem [{\citenamefont {Ma\~nes}(2012)}]{MaJL2012}%
  \BibitemOpen
  \bibfield  {author} {\bibinfo {author} {\bibfnamefont {J.~L.}\ \bibnamefont
  {Ma\~nes}},\ }\bibfield  {title} {\enquote {\bibinfo {title} {Existence of
  bulk chiral fermions and crystal symmetry},}\ }\href {\doibase
  10.1103/PhysRevB.85.155118} {\bibfield  {journal} {\bibinfo  {journal} {Phys.
  Rev. B}\ }\textbf {\bibinfo {volume} {85}},\ \bibinfo {pages} {155118}
  (\bibinfo {year} {2012})}\BibitemShut {NoStop}%
\bibitem [{\citenamefont {RomhManyi}\ \emph {et~al.}(2015)\citenamefont
  {RomhManyi}, \citenamefont {Penc},\ and\ \citenamefont
  {Ganesh}}]{RomhMannyi2015}%
  \BibitemOpen
  \bibfield  {author} {\bibinfo {author} {\bibfnamefont {J.}~\bibnamefont
  {RomhManyi}}, \bibinfo {author} {\bibfnamefont {K.}~\bibnamefont {Penc}}, \
  and\ \bibinfo {author} {\bibfnamefont {R.}~\bibnamefont {Ganesh}},\
  }\bibfield  {title} {\enquote {\bibinfo {title} {Hall effect of triplons in a
  dimerized quantum magnet},}\ }\href@noop {} {\bibfield  {journal} {\bibinfo
  {journal} {Nat. Commun.}\ }\textbf {\bibinfo {volume} {6}},\ \bibinfo {pages}
  {6805} (\bibinfo {year} {2015})}\BibitemShut {NoStop}%
\bibitem [{\citenamefont {Zhong}\ \emph {et~al.}(2017)\citenamefont {Zhong},
  \citenamefont {Chen}, \citenamefont {Yu}, \citenamefont {Xie}, \citenamefont
  {Wang}, \citenamefont {Yang},\ and\ \citenamefont {Zhang}}]{Zhong2017}%
  \BibitemOpen
  \bibfield  {author} {\bibinfo {author} {\bibfnamefont {C.}~\bibnamefont
  {Zhong}}, \bibinfo {author} {\bibfnamefont {Y.}~\bibnamefont {Chen}},
  \bibinfo {author} {\bibfnamefont {Z.-M.}\ \bibnamefont {Yu}}, \bibinfo
  {author} {\bibfnamefont {Y.}~\bibnamefont {Xie}}, \bibinfo {author}
  {\bibfnamefont {H.}~\bibnamefont {Wang}}, \bibinfo {author} {\bibfnamefont
  {S.~A.}\ \bibnamefont {Yang}}, \ and\ \bibinfo {author} {\bibfnamefont
  {S.}~\bibnamefont {Zhang}},\ }\bibfield  {title} {\enquote {\bibinfo {title}
  {Three-dimensional pentagon carbon with a genesis of emergent fermions},}\
  }\href@noop {} {\bibfield  {journal} {\bibinfo  {journal} {Nat. Commun.}\
  }\textbf {\bibinfo {volume} {8}},\ \bibinfo {pages} {15641} (\bibinfo {year}
  {2017})}\BibitemShut {NoStop}%
\bibitem [{\citenamefont {Slot}\ \emph {et~al.}(2017)\citenamefont {Slot},
  \citenamefont {Gardenier}, \citenamefont {Jacobse}, \citenamefont {van
  Miert}, \citenamefont {Kempkes}, \citenamefont {Zevenhuizen}, \citenamefont
  {Smith}, \citenamefont {Vanmaekelbergh},\ and\ \citenamefont
  {Swart}}]{Slot2017}%
  \BibitemOpen
  \bibfield  {author} {\bibinfo {author} {\bibfnamefont {M.~R.}\ \bibnamefont
  {Slot}}, \bibinfo {author} {\bibfnamefont {T.~S.}\ \bibnamefont {Gardenier}},
  \bibinfo {author} {\bibfnamefont {P.~H.}\ \bibnamefont {Jacobse}}, \bibinfo
  {author} {\bibfnamefont {G.~C.~P.}\ \bibnamefont {van Miert}}, \bibinfo
  {author} {\bibfnamefont {S.~N.}\ \bibnamefont {Kempkes}}, \bibinfo {author}
  {\bibfnamefont {S.~J.~M.}\ \bibnamefont {Zevenhuizen}}, \bibinfo {author}
  {\bibfnamefont {C.~M.}\ \bibnamefont {Smith}}, \bibinfo {author}
  {\bibfnamefont {D.}~\bibnamefont {Vanmaekelbergh}}, \ and\ \bibinfo {author}
  {\bibfnamefont {I.}~\bibnamefont {Swart}},\ }\bibfield  {title} {\enquote
  {\bibinfo {title} {Experimental realization and characterization of an
  electronic {Lieb} lattice},}\ }\href@noop {} {\bibfield  {journal} {\bibinfo
  {journal} {Nat. Phys.}\ }\textbf {\bibinfo {volume} {13}},\ \bibinfo {pages}
  {672} (\bibinfo {year} {2017})}\BibitemShut {NoStop}%
\bibitem [{\citenamefont {Bradlyn}\ \emph {et~al.}(2016)\citenamefont
  {Bradlyn}, \citenamefont {Cano}, \citenamefont {Wang}, \citenamefont
  {Vergniory}, \citenamefont {Felser}, \citenamefont {Cava},\ and\
  \citenamefont {Bernevig}}]{Bradlynaaf2016}%
  \BibitemOpen
  \bibfield  {author} {\bibinfo {author} {\bibfnamefont {B.}~\bibnamefont
  {Bradlyn}}, \bibinfo {author} {\bibfnamefont {J.}~\bibnamefont {Cano}},
  \bibinfo {author} {\bibfnamefont {Z.}~\bibnamefont {Wang}}, \bibinfo {author}
  {\bibfnamefont {M.~G.}\ \bibnamefont {Vergniory}}, \bibinfo {author}
  {\bibfnamefont {C.}~\bibnamefont {Felser}}, \bibinfo {author} {\bibfnamefont
  {R.~J.}\ \bibnamefont {Cava}}, \ and\ \bibinfo {author} {\bibfnamefont
  {B.~A.}\ \bibnamefont {Bernevig}},\ }\bibfield  {title} {\enquote {\bibinfo
  {title} {Beyond {Dirac and Weyl} fermions: Unconventional quasiparticles in
  conventional crystals},}\ }\href {\doibase 10.1126/science.aaf5037}
  {\bibfield  {journal} {\bibinfo  {journal} {Science}\ }\textbf {\bibinfo
  {volume} {353}},\ \bibinfo {pages} {aaf5037} (\bibinfo {year}
  {2016})}\BibitemShut {NoStop}%
\bibitem [{\citenamefont {Takane}\ \emph {et~al.}(2019)\citenamefont {Takane},
  \citenamefont {Wang}, \citenamefont {Souma}, \citenamefont {Nakayama},
  \citenamefont {Nakamura}, \citenamefont {Oinuma}, \citenamefont {Nakata},
  \citenamefont {Iwasawa}, \citenamefont {Cacho}, \citenamefont {Kim},
  \citenamefont {Horiba}, \citenamefont {Kumigashira}, \citenamefont
  {Takahashi}, \citenamefont {Ando},\ and\ \citenamefont {Sato}}]{Takane2019}%
  \BibitemOpen
  \bibfield  {author} {\bibinfo {author} {\bibfnamefont {D.}~\bibnamefont
  {Takane}}, \bibinfo {author} {\bibfnamefont {Z.}~\bibnamefont {Wang}},
  \bibinfo {author} {\bibfnamefont {S.}~\bibnamefont {Souma}}, \bibinfo
  {author} {\bibfnamefont {K.}~\bibnamefont {Nakayama}}, \bibinfo {author}
  {\bibfnamefont {T.}~\bibnamefont {Nakamura}}, \bibinfo {author}
  {\bibfnamefont {H.}~\bibnamefont {Oinuma}}, \bibinfo {author} {\bibfnamefont
  {Y.}~\bibnamefont {Nakata}}, \bibinfo {author} {\bibfnamefont
  {H.}~\bibnamefont {Iwasawa}}, \bibinfo {author} {\bibfnamefont
  {C.}~\bibnamefont {Cacho}}, \bibinfo {author} {\bibfnamefont
  {T.}~\bibnamefont {Kim}}, \bibinfo {author} {\bibfnamefont {K.}~\bibnamefont
  {Horiba}}, \bibinfo {author} {\bibfnamefont {H.}~\bibnamefont {Kumigashira}},
  \bibinfo {author} {\bibfnamefont {T.}~\bibnamefont {Takahashi}}, \bibinfo
  {author} {\bibfnamefont {Y.}~\bibnamefont {Ando}}, \ and\ \bibinfo {author}
  {\bibfnamefont {T.}~\bibnamefont {Sato}},\ }\bibfield  {title} {\enquote
  {\bibinfo {title} {Observation of chiral fermions with a large topological
  charge and associated {Fermi}-arc surface states in {CoSi}},}\ }\href
  {\doibase 10.1103/PhysRevLett.122.076402} {\bibfield  {journal} {\bibinfo
  {journal} {Phys. Rev. Lett.}\ }\textbf {\bibinfo {volume} {122}},\ \bibinfo
  {pages} {076402} (\bibinfo {year} {2019})}\BibitemShut {NoStop}%
\bibitem [{\citenamefont {Delplace}\ \emph {et~al.}(2017)\citenamefont
  {Delplace}, \citenamefont {Marston},\ and\ \citenamefont
  {Venaille}}]{Delplace2017}%
  \BibitemOpen
  \bibfield  {author} {\bibinfo {author} {\bibfnamefont {P.}~\bibnamefont
  {Delplace}}, \bibinfo {author} {\bibfnamefont {J.~B.}\ \bibnamefont
  {Marston}}, \ and\ \bibinfo {author} {\bibfnamefont {A.}~\bibnamefont
  {Venaille}},\ }\bibfield  {title} {\enquote {\bibinfo {title} {Topological
  origin of equatorial waves},}\ }\href {\doibase 10.1126/science.aan8819}
  {\bibfield  {journal} {\bibinfo  {journal} {Science}\ }\textbf {\bibinfo
  {volume} {358}},\ \bibinfo {pages} {1075} (\bibinfo {year}
  {2017})}\BibitemShut {NoStop}%
\bibitem [{\citenamefont {Jin}\ \emph {et~al.}(2016)\citenamefont {Jin},
  \citenamefont {Lu}, \citenamefont {Wang}, \citenamefont {Fang}, \citenamefont
  {Joannopoulos}, \citenamefont {Soljacic}, \citenamefont {Fu},\ and\
  \citenamefont {Fang}}]{Jin2016}%
  \BibitemOpen
  \bibfield  {author} {\bibinfo {author} {\bibfnamefont {D.}~\bibnamefont
  {Jin}}, \bibinfo {author} {\bibfnamefont {L.}~\bibnamefont {Lu}}, \bibinfo
  {author} {\bibfnamefont {Z.}~\bibnamefont {Wang}}, \bibinfo {author}
  {\bibfnamefont {C.}~\bibnamefont {Fang}}, \bibinfo {author} {\bibfnamefont
  {J.~D.}\ \bibnamefont {Joannopoulos}}, \bibinfo {author} {\bibfnamefont
  {M.}~\bibnamefont {Soljacic}}, \bibinfo {author} {\bibfnamefont
  {L.}~\bibnamefont {Fu}}, \ and\ \bibinfo {author} {\bibfnamefont {N.~X.}\
  \bibnamefont {Fang}},\ }\bibfield  {title} {\enquote {\bibinfo {title}
  {Topological magnetoplasmon},}\ }\href@noop {} {\bibfield  {journal}
  {\bibinfo  {journal} {Nat. Commun.}\ }\textbf {\bibinfo {volume} {7}},\
  \bibinfo {pages} {13486} (\bibinfo {year} {2016})}\BibitemShut {NoStop}%
\bibitem [{\citenamefont {Gomes}\ \emph {et~al.}(2012)\citenamefont {Gomes},
  \citenamefont {Mar}, \citenamefont {Ko}, \citenamefont {Guinea},\ and\
  \citenamefont {Manoharan}}]{Gomes2012}%
  \BibitemOpen
  \bibfield  {author} {\bibinfo {author} {\bibfnamefont {K.~K.}\ \bibnamefont
  {Gomes}}, \bibinfo {author} {\bibfnamefont {W.}~\bibnamefont {Mar}}, \bibinfo
  {author} {\bibfnamefont {W.}~\bibnamefont {Ko}}, \bibinfo {author}
  {\bibfnamefont {F.}~\bibnamefont {Guinea}}, \ and\ \bibinfo {author}
  {\bibfnamefont {H.~C.}\ \bibnamefont {Manoharan}},\ }\bibfield  {title}
  {\enquote {\bibinfo {title} {Designer {Dirac} fermions and topological phases
  in molecular graphene},}\ }\href@noop {} {\bibfield  {journal} {\bibinfo
  {journal} {Nature}\ }\textbf {\bibinfo {volume} {483}},\ \bibinfo {pages}
  {306} (\bibinfo {year} {2012})}\BibitemShut {NoStop}%
\bibitem [{\citenamefont {Tarruell}\ \emph {et~al.}(2012)\citenamefont
  {Tarruell}, \citenamefont {Greif}, \citenamefont {Uehlinger}, \citenamefont
  {Jotzu},\ and\ \citenamefont {Esslinger}}]{Tarruell2012}%
  \BibitemOpen
  \bibfield  {author} {\bibinfo {author} {\bibfnamefont {L.}~\bibnamefont
  {Tarruell}}, \bibinfo {author} {\bibfnamefont {D.}~\bibnamefont {Greif}},
  \bibinfo {author} {\bibfnamefont {T.}~\bibnamefont {Uehlinger}}, \bibinfo
  {author} {\bibfnamefont {G.}~\bibnamefont {Jotzu}}, \ and\ \bibinfo {author}
  {\bibfnamefont {T.}~\bibnamefont {Esslinger}},\ }\bibfield  {title} {\enquote
  {\bibinfo {title} {Creating, moving and merging {Dirac} points with a {Fermi}
  gas in a tunable honeycomb lattice},}\ }\href@noop {} {\bibfield  {journal}
  {\bibinfo  {journal} {Nature}\ }\textbf {\bibinfo {volume} {483}},\ \bibinfo
  {pages} {302} (\bibinfo {year} {2012})}\BibitemShut {NoStop}%
\bibitem [{\citenamefont {Klembt}\ \emph {et~al.}(2018)\citenamefont {Klembt},
  \citenamefont {Harder}, \citenamefont {Egorov}, \citenamefont {Winkler},
  \citenamefont {Ge}, \citenamefont {Bandres}, \citenamefont {Emmerling},
  \citenamefont {Worschech}, \citenamefont {Liew}, \citenamefont {Segev},
  \citenamefont {Schneider},\ and\ \citenamefont {Hafling}}]{Klembt2018}%
  \BibitemOpen
  \bibfield  {author} {\bibinfo {author} {\bibfnamefont {S.}~\bibnamefont
  {Klembt}}, \bibinfo {author} {\bibfnamefont {T.~H.}\ \bibnamefont {Harder}},
  \bibinfo {author} {\bibfnamefont {O.~A.}\ \bibnamefont {Egorov}}, \bibinfo
  {author} {\bibfnamefont {K.}~\bibnamefont {Winkler}}, \bibinfo {author}
  {\bibfnamefont {R.}~\bibnamefont {Ge}}, \bibinfo {author} {\bibfnamefont
  {M.~A.}\ \bibnamefont {Bandres}}, \bibinfo {author} {\bibfnamefont
  {M.}~\bibnamefont {Emmerling}}, \bibinfo {author} {\bibfnamefont
  {L.}~\bibnamefont {Worschech}}, \bibinfo {author} {\bibfnamefont {T.~C.~H.}\
  \bibnamefont {Liew}}, \bibinfo {author} {\bibfnamefont {M.}~\bibnamefont
  {Segev}}, \bibinfo {author} {\bibfnamefont {C.}~\bibnamefont {Schneider}}, \
  and\ \bibinfo {author} {\bibfnamefont {S.}~\bibnamefont {Hafling}},\
  }\bibfield  {title} {\enquote {\bibinfo {title} {Exciton-polariton
  topological insulator},}\ }\href {\doibase 10.1038/s41586-018-0601-5}
  {\bibfield  {journal} {\bibinfo  {journal} {Nature}\ }\textbf {\bibinfo
  {volume} {562}},\ \bibinfo {pages} {552} (\bibinfo {year}
  {2018})}\BibitemShut {NoStop}%
\bibitem [{\citenamefont {Yang}\ \emph {et~al.}(2019)\citenamefont {Yang},
  \citenamefont {Gao}, \citenamefont {Xue}, \citenamefont {Zhang},
  \citenamefont {He}, \citenamefont {Yang}, \citenamefont {Singh},
  \citenamefont {Chong}, \citenamefont {Zhang},\ and\ \citenamefont
  {Chen}}]{Yang2019}%
  \BibitemOpen
  \bibfield  {author} {\bibinfo {author} {\bibfnamefont {Y.}~\bibnamefont
  {Yang}}, \bibinfo {author} {\bibfnamefont {Z.}~\bibnamefont {Gao}}, \bibinfo
  {author} {\bibfnamefont {H.}~\bibnamefont {Xue}}, \bibinfo {author}
  {\bibfnamefont {L.}~\bibnamefont {Zhang}}, \bibinfo {author} {\bibfnamefont
  {M.}~\bibnamefont {He}}, \bibinfo {author} {\bibfnamefont {Z.}~\bibnamefont
  {Yang}}, \bibinfo {author} {\bibfnamefont {R.}~\bibnamefont {Singh}},
  \bibinfo {author} {\bibfnamefont {Y.}~\bibnamefont {Chong}}, \bibinfo
  {author} {\bibfnamefont {B.}~\bibnamefont {Zhang}}, \ and\ \bibinfo {author}
  {\bibfnamefont {H.}~\bibnamefont {Chen}},\ }\bibfield  {title} {\enquote
  {\bibinfo {title} {Realization of a three-dimensional photonic topological
  insulator},}\ }\href {\doibase 10.1038/s41586-018-0829-0} {\bibfield
  {journal} {\bibinfo  {journal} {Nature}\ }\textbf {\bibinfo {volume} {565}},\
  \bibinfo {pages} {622} (\bibinfo {year} {2019})}\BibitemShut {NoStop}%
\bibitem [{\citenamefont {Kane}\ and\ \citenamefont
  {Lubensky}(2013)}]{Kane2013}%
  \BibitemOpen
  \bibfield  {author} {\bibinfo {author} {\bibfnamefont {C.~L.}\ \bibnamefont
  {Kane}}\ and\ \bibinfo {author} {\bibfnamefont {T.~C.}\ \bibnamefont
  {Lubensky}},\ }\bibfield  {title} {\enquote {\bibinfo {title} {Topological
  boundary modes in isostatic lattices},}\ }\href@noop {} {\bibfield  {journal}
  {\bibinfo  {journal} {Nat. Phys.}\ }\textbf {\bibinfo {volume} {10}},\
  \bibinfo {pages} {39} (\bibinfo {year} {2013})}\BibitemShut {NoStop}%
\bibitem [{\citenamefont {Novoselov}\ \emph {et~al.}(2005)\citenamefont
  {Novoselov}, \citenamefont {Geim}, \citenamefont {Morozov}, \citenamefont
  {Jiang}, \citenamefont {Katsnelson}, \citenamefont {Grigorieva},
  \citenamefont {Dubonos},\ and\ \citenamefont {Firsov}}]{Novoselovetal:2005}%
  \BibitemOpen
  \bibfield  {author} {\bibinfo {author} {\bibfnamefont {K.~S.}\ \bibnamefont
  {Novoselov}}, \bibinfo {author} {\bibfnamefont {A.~K.}\ \bibnamefont {Geim}},
  \bibinfo {author} {\bibfnamefont {S.~V.}\ \bibnamefont {Morozov}}, \bibinfo
  {author} {\bibfnamefont {D.}~\bibnamefont {Jiang}}, \bibinfo {author}
  {\bibfnamefont {M.~I.}\ \bibnamefont {Katsnelson}}, \bibinfo {author}
  {\bibfnamefont {I.~V.}\ \bibnamefont {Grigorieva}}, \bibinfo {author}
  {\bibfnamefont {S.~V.}\ \bibnamefont {Dubonos}}, \ and\ \bibinfo {author}
  {\bibfnamefont {A.~A.}\ \bibnamefont {Firsov}},\ }\bibfield  {title}
  {\enquote {\bibinfo {title} {Two-dimensional gas of massless {Dirac} fermions
  in graphene},}\ }\href@noop {} {\bibfield  {journal} {\bibinfo  {journal}
  {Nature}\ }\textbf {\bibinfo {volume} {438}},\ \bibinfo {pages} {197}
  (\bibinfo {year} {2005})}\BibitemShut {NoStop}%
\bibitem [{\citenamefont {Wehling}\ \emph {et~al.}(2014)\citenamefont
  {Wehling}, \citenamefont {Black-Schaffer},\ and\ \citenamefont
  {Balatsky}}]{Wehling:2014}%
  \BibitemOpen
  \bibfield  {author} {\bibinfo {author} {\bibfnamefont {T.}~\bibnamefont
  {Wehling}}, \bibinfo {author} {\bibfnamefont {A.}~\bibnamefont
  {Black-Schaffer}}, \ and\ \bibinfo {author} {\bibfnamefont {A.}~\bibnamefont
  {Balatsky}},\ }\bibfield  {title} {\enquote {\bibinfo {title} {Dirac
  materials},}\ }\href {\doibase 10.1080/00018732.2014.927109} {\bibfield
  {journal} {\bibinfo  {journal} {Adv. Phys.}\ }\textbf {\bibinfo {volume}
  {63}},\ \bibinfo {pages} {1} (\bibinfo {year} {2014})}\BibitemShut {NoStop}%
\bibitem [{\citenamefont {Jackiw}\ and\ \citenamefont
  {Rebbi}(1976)}]{Jackiw1976}%
  \BibitemOpen
  \bibfield  {author} {\bibinfo {author} {\bibfnamefont {R.}~\bibnamefont
  {Jackiw}}\ and\ \bibinfo {author} {\bibfnamefont {C.}~\bibnamefont {Rebbi}},\
  }\bibfield  {title} {\enquote {\bibinfo {title} {Solitons with fermion
  number~$1/2$},}\ }\href {\doibase 10.1103/PhysRevD.13.3398} {\bibfield
  {journal} {\bibinfo  {journal} {Phys. Rev. D}\ }\textbf {\bibinfo {volume}
  {13}},\ \bibinfo {pages} {3398} (\bibinfo {year} {1976})}\BibitemShut
  {NoStop}%
\bibitem [{\citenamefont {Su}\ \emph {et~al.}(1979)\citenamefont {Su},
  \citenamefont {Schrieffer},\ and\ \citenamefont {Heeger}}]{SSH1979}%
  \BibitemOpen
  \bibfield  {author} {\bibinfo {author} {\bibfnamefont {W.~P.}\ \bibnamefont
  {Su}}, \bibinfo {author} {\bibfnamefont {J.~R.}\ \bibnamefont {Schrieffer}},
  \ and\ \bibinfo {author} {\bibfnamefont {A.~J.}\ \bibnamefont {Heeger}},\
  }\bibfield  {title} {\enquote {\bibinfo {title} {Solitons in
  polyacetylene},}\ }\href {\doibase 10.1103/PhysRevLett.42.1698} {\bibfield
  {journal} {\bibinfo  {journal} {Phys. Rev. Lett.}\ }\textbf {\bibinfo
  {volume} {42}},\ \bibinfo {pages} {1698} (\bibinfo {year}
  {1979})}\BibitemShut {NoStop}%
\bibitem [{\citenamefont {Heeger}\ \emph {et~al.}(1988)\citenamefont {Heeger},
  \citenamefont {Kivelson}, \citenamefont {Schrieffer},\ and\ \citenamefont
  {Su}}]{Heeger1988}%
  \BibitemOpen
  \bibfield  {author} {\bibinfo {author} {\bibfnamefont {A.~J.}\ \bibnamefont
  {Heeger}}, \bibinfo {author} {\bibfnamefont {S.}~\bibnamefont {Kivelson}},
  \bibinfo {author} {\bibfnamefont {J.~R.}\ \bibnamefont {Schrieffer}}, \ and\
  \bibinfo {author} {\bibfnamefont {W.~P.}\ \bibnamefont {Su}},\ }\bibfield
  {title} {\enquote {\bibinfo {title} {Solitons in conducting polymers},}\
  }\href {\doibase 10.1103/RevModPhys.60.781} {\bibfield  {journal} {\bibinfo
  {journal} {Rev. Mod. Phys.}\ }\textbf {\bibinfo {volume} {60}},\ \bibinfo
  {pages} {781} (\bibinfo {year} {1988})}\BibitemShut {NoStop}%
\bibitem [{\citenamefont {{Volkov}}\ and\ \citenamefont
  {{Pankratov}}(1985)}]{Volkov1985}%
  \BibitemOpen
  \bibfield  {author} {\bibinfo {author} {\bibfnamefont {B.~A.}\ \bibnamefont
  {{Volkov}}}\ and\ \bibinfo {author} {\bibfnamefont {O.~A.}\ \bibnamefont
  {{Pankratov}}},\ }\bibfield  {title} {\enquote {\bibinfo {title}
  {Two-dimensional massless electrons in an inverted contact},}\ }\href@noop {}
  {\bibfield  {journal} {\bibinfo  {journal} {Sov. J. Exp. Theo. Phys. Lett.}\
  }\textbf {\bibinfo {volume} {42}},\ \bibinfo {pages} {178} (\bibinfo {year}
  {1985})}\BibitemShut {NoStop}%
\bibitem [{\citenamefont {Pankratov}\ \emph {et~al.}(1987)\citenamefont
  {Pankratov}, \citenamefont {Pakhomov},\ and\ \citenamefont
  {Volkov}}]{Pankratov1987}%
  \BibitemOpen
  \bibfield  {author} {\bibinfo {author} {\bibfnamefont {O.~A.}\ \bibnamefont
  {Pankratov}}, \bibinfo {author} {\bibfnamefont {S.~V.}\ \bibnamefont
  {Pakhomov}}, \ and\ \bibinfo {author} {\bibfnamefont {B.~A.}\ \bibnamefont
  {Volkov}},\ }\bibfield  {title} {\enquote {\bibinfo {title} {Supersymmetry in
  heterojunctions: Band-inverting contact on the basis of {Pb1$_{1-x}$Sn$_x$Te
  and Hg$_{1-x}$Cd$_x$Te}},}\ }\href {\doibase
  https://doi.org/10.1016/0038-1098(87)90934-3} {\bibfield  {journal} {\bibinfo
   {journal} {Solid State Commun.}\ }\textbf {\bibinfo {volume} {61}},\
  \bibinfo {pages} {93} (\bibinfo {year} {1987})}\BibitemShut {NoStop}%
\bibitem [{\citenamefont {Wang}\ \emph {et~al.}(2015)\citenamefont {Wang},
  \citenamefont {Lian},\ and\ \citenamefont {Zhang}}]{Wang2015b}%
  \BibitemOpen
  \bibfield  {author} {\bibinfo {author} {\bibfnamefont {J.}~\bibnamefont
  {Wang}}, \bibinfo {author} {\bibfnamefont {B.}~\bibnamefont {Lian}}, \ and\
  \bibinfo {author} {\bibfnamefont {S.-C.}\ \bibnamefont {Zhang}},\ }\bibfield
  {title} {\enquote {\bibinfo {title} {Electrically tunable magnetism in
  magnetic topological insulators},}\ }\href {\doibase
  10.1103/PhysRevLett.115.036805} {\bibfield  {journal} {\bibinfo  {journal}
  {Phys. Rev. Lett.}\ }\textbf {\bibinfo {volume} {115}},\ \bibinfo {pages}
  {036805} (\bibinfo {year} {2015})}\BibitemShut {NoStop}%
\bibitem [{\citenamefont {Shen}(2013)}]{Shen2013}%
  \BibitemOpen
  \bibfield  {author} {\bibinfo {author} {\bibfnamefont {S.}~\bibnamefont
  {Shen}},\ }\href@noop {} {\emph {\bibinfo {title} {Topological Insulators:
  Dirac Equation in Condensed Matters}}},\ Springer Series in Solid-State
  Sciences\ (\bibinfo  {publisher} {Springer Berlin Heidelberg},\ \bibinfo
  {year} {2013})\BibitemShut {NoStop}%
\bibitem [{\citenamefont {Nagaosa}\ and\ \citenamefont
  {Tokura}(2013)}]{Nagaosa2013}%
  \BibitemOpen
  \bibfield  {author} {\bibinfo {author} {\bibfnamefont {N.}~\bibnamefont
  {Nagaosa}}\ and\ \bibinfo {author} {\bibfnamefont {Y.}~\bibnamefont
  {Tokura}},\ }\bibfield  {title} {\enquote {\bibinfo {title} {Topological
  properties and dynamics of magnetic skyrmions},}\ }\href@noop {} {\bibfield
  {journal} {\bibinfo  {journal} {Nat. Nanotech.}\ }\textbf {\bibinfo {volume}
  {8}},\ \bibinfo {pages} {899} (\bibinfo {year} {2013})},\ \bibinfo {note}
  {review Article}\BibitemShut {NoStop}%
\bibitem [{\citenamefont {Salomaa}\ and\ \citenamefont
  {Volovik}(1987)}]{SV:1987}%
  \BibitemOpen
  \bibfield  {author} {\bibinfo {author} {\bibfnamefont {M.~M.}\ \bibnamefont
  {Salomaa}}\ and\ \bibinfo {author} {\bibfnamefont {G.~E.}\ \bibnamefont
  {Volovik}},\ }\bibfield  {title} {\enquote {\bibinfo {title} {Quantized
  vortices in superfluid $^{3}\mathrm{He}$},}\ }\href {\doibase
  10.1103/RevModPhys.59.533} {\bibfield  {journal} {\bibinfo  {journal} {Rev.
  Mod. Phys.}\ }\textbf {\bibinfo {volume} {59}},\ \bibinfo {pages} {533}
  (\bibinfo {year} {1987})}\BibitemShut {NoStop}%
\bibitem [{\citenamefont {Garaud}\ and\ \citenamefont
  {Babaev}(2012)}]{GB:2012}%
  \BibitemOpen
  \bibfield  {author} {\bibinfo {author} {\bibfnamefont {J.}~\bibnamefont
  {Garaud}}\ and\ \bibinfo {author} {\bibfnamefont {E.}~\bibnamefont
  {Babaev}},\ }\bibfield  {title} {\enquote {\bibinfo {title} {Skyrmionic state
  and stable half-quantum vortices in chiral $p$-wave superconductors},}\
  }\href {\doibase 10.1103/PhysRevB.86.060514} {\bibfield  {journal} {\bibinfo
  {journal} {Phys. Rev. B}\ }\textbf {\bibinfo {volume} {86}},\ \bibinfo
  {pages} {060514} (\bibinfo {year} {2012})}\BibitemShut {NoStop}%
\bibitem [{\citenamefont {Gielis}(2003)}]{Gielis2003}%
  \BibitemOpen
  \bibfield  {author} {\bibinfo {author} {\bibfnamefont {J.}~\bibnamefont
  {Gielis}},\ }\bibfield  {title} {\enquote {\bibinfo {title} {A generic
  geometric transformation that unifies a wide range of natural and abstract
  shapes},}\ }\href {\doibase 10.3732/ajb.90.3.333} {\bibfield  {journal}
  {\bibinfo  {journal} {Ame. J. Botany}\ }\textbf {\bibinfo {volume} {90}},\
  \bibinfo {pages} {333} (\bibinfo {year} {2003})}\BibitemShut {NoStop}%
\bibitem [{\citenamefont {Bernevig}\ and\ \citenamefont
  {Zhang}(2006)}]{BZ:2006}%
  \BibitemOpen
  \bibfield  {author} {\bibinfo {author} {\bibfnamefont {B.~A.}\ \bibnamefont
  {Bernevig}}\ and\ \bibinfo {author} {\bibfnamefont {S.-C.}\ \bibnamefont
  {Zhang}},\ }\bibfield  {title} {\enquote {\bibinfo {title} {Quantum spin
  {Hall} effect},}\ }\href {\doibase 10.1103/PhysRevLett.96.106802} {\bibfield
  {journal} {\bibinfo  {journal} {Phys. Rev. Lett.}\ }\textbf {\bibinfo
  {volume} {96}},\ \bibinfo {pages} {106802} (\bibinfo {year}
  {2006})}\BibitemShut {NoStop}%
\bibitem [{\citenamefont {H\"am\"al\"ainen}\ \emph {et~al.}(2011)\citenamefont
  {H\"am\"al\"ainen}, \citenamefont {Sun}, \citenamefont {Boneschanscher},
  \citenamefont {Uppstu}, \citenamefont {Ij\"as}, \citenamefont {Harju},
  \citenamefont {Vanmaekelbergh},\ and\ \citenamefont {Liljeroth}}]{Hamal2011}%
  \BibitemOpen
  \bibfield  {author} {\bibinfo {author} {\bibfnamefont {S.~K.}\ \bibnamefont
  {H\"am\"al\"ainen}}, \bibinfo {author} {\bibfnamefont {Z.}~\bibnamefont
  {Sun}}, \bibinfo {author} {\bibfnamefont {M.~P.}\ \bibnamefont
  {Boneschanscher}}, \bibinfo {author} {\bibfnamefont {A.}~\bibnamefont
  {Uppstu}}, \bibinfo {author} {\bibfnamefont {M.}~\bibnamefont {Ij\"as}},
  \bibinfo {author} {\bibfnamefont {A.}~\bibnamefont {Harju}}, \bibinfo
  {author} {\bibfnamefont {D.}~\bibnamefont {Vanmaekelbergh}}, \ and\ \bibinfo
  {author} {\bibfnamefont {P.}~\bibnamefont {Liljeroth}},\ }\bibfield  {title}
  {\enquote {\bibinfo {title} {Quantum-confined electronic states in atomically
  well-defined graphene nanostructures},}\ }\href {\doibase
  10.1103/PhysRevLett.107.236803} {\bibfield  {journal} {\bibinfo  {journal}
  {Phys. Rev. Lett.}\ }\textbf {\bibinfo {volume} {107}},\ \bibinfo {pages}
  {236803} (\bibinfo {year} {2011})}\BibitemShut {NoStop}%
\bibitem [{\citenamefont {Subramaniam}\ \emph {et~al.}(2012)\citenamefont
  {Subramaniam}, \citenamefont {Libisch}, \citenamefont {Li}, \citenamefont
  {Pauly}, \citenamefont {Geringer}, \citenamefont {Reiter}, \citenamefont
  {Mashoff}, \citenamefont {Liebmann}, \citenamefont {Burgd\"orfer},
  \citenamefont {Busse}, \citenamefont {Michely}, \citenamefont {Mazzarello},
  \citenamefont {Pratzer},\ and\ \citenamefont {Morgenstern}}]{Subra2012}%
  \BibitemOpen
  \bibfield  {author} {\bibinfo {author} {\bibfnamefont {D.}~\bibnamefont
  {Subramaniam}}, \bibinfo {author} {\bibfnamefont {F.}~\bibnamefont
  {Libisch}}, \bibinfo {author} {\bibfnamefont {Y.}~\bibnamefont {Li}},
  \bibinfo {author} {\bibfnamefont {C.}~\bibnamefont {Pauly}}, \bibinfo
  {author} {\bibfnamefont {V.}~\bibnamefont {Geringer}}, \bibinfo {author}
  {\bibfnamefont {R.}~\bibnamefont {Reiter}}, \bibinfo {author} {\bibfnamefont
  {T.}~\bibnamefont {Mashoff}}, \bibinfo {author} {\bibfnamefont
  {M.}~\bibnamefont {Liebmann}}, \bibinfo {author} {\bibfnamefont
  {J.}~\bibnamefont {Burgd\"orfer}}, \bibinfo {author} {\bibfnamefont
  {C.}~\bibnamefont {Busse}}, \bibinfo {author} {\bibfnamefont
  {T.}~\bibnamefont {Michely}}, \bibinfo {author} {\bibfnamefont
  {R.}~\bibnamefont {Mazzarello}}, \bibinfo {author} {\bibfnamefont
  {M.}~\bibnamefont {Pratzer}}, \ and\ \bibinfo {author} {\bibfnamefont
  {M.}~\bibnamefont {Morgenstern}},\ }\bibfield  {title} {\enquote {\bibinfo
  {title} {Wave-function mapping of graphene quantum dots with soft
  confinement},}\ }\href {\doibase 10.1103/PhysRevLett.108.046801} {\bibfield
  {journal} {\bibinfo  {journal} {Phys. Rev. Lett.}\ }\textbf {\bibinfo
  {volume} {108}},\ \bibinfo {pages} {046801} (\bibinfo {year}
  {2012})}\BibitemShut {NoStop}%
\bibitem [{\citenamefont {Raoux}\ \emph {et~al.}(2014)\citenamefont {Raoux},
  \citenamefont {Morigi}, \citenamefont {Fuchs}, \citenamefont {Pi\'echon},\
  and\ \citenamefont {Montambaux}}]{Raoux2014}%
  \BibitemOpen
  \bibfield  {author} {\bibinfo {author} {\bibfnamefont {A.}~\bibnamefont
  {Raoux}}, \bibinfo {author} {\bibfnamefont {M.}~\bibnamefont {Morigi}},
  \bibinfo {author} {\bibfnamefont {J.-N.}\ \bibnamefont {Fuchs}}, \bibinfo
  {author} {\bibfnamefont {F.}~\bibnamefont {Pi\'echon}}, \ and\ \bibinfo
  {author} {\bibfnamefont {G.}~\bibnamefont {Montambaux}},\ }\bibfield  {title}
  {\enquote {\bibinfo {title} {From dia- to paramagnetic orbital susceptibility
  of massless fermions},}\ }\href {\doibase 10.1103/PhysRevLett.112.026402}
  {\bibfield  {journal} {\bibinfo  {journal} {Phys. Rev. Lett.}\ }\textbf
  {\bibinfo {volume} {112}},\ \bibinfo {pages} {026402} (\bibinfo {year}
  {2014})}\BibitemShut {NoStop}%
\bibitem [{\citenamefont {Drost}\ \emph {et~al.}(2017)\citenamefont {Drost},
  \citenamefont {Ojanen}, \citenamefont {Harju},\ and\ \citenamefont
  {Liljeroth}}]{Drost2017}%
  \BibitemOpen
  \bibfield  {author} {\bibinfo {author} {\bibfnamefont {R.}~\bibnamefont
  {Drost}}, \bibinfo {author} {\bibfnamefont {T.}~\bibnamefont {Ojanen}},
  \bibinfo {author} {\bibfnamefont {A.}~\bibnamefont {Harju}}, \ and\ \bibinfo
  {author} {\bibfnamefont {P.}~\bibnamefont {Liljeroth}},\ }\bibfield  {title}
  {\enquote {\bibinfo {title} {Topological states in engineered atomic
  lattices},}\ }\href@noop {} {\bibfield  {journal} {\bibinfo  {journal} {Nat.
  Phys.}\ }\textbf {\bibinfo {volume} {13}},\ \bibinfo {pages} {668} (\bibinfo
  {year} {2017})}\BibitemShut {NoStop}%
\bibitem [{\citenamefont {Vicencio}\ \emph {et~al.}(2015)\citenamefont
  {Vicencio}, \citenamefont {Cantillano}, \citenamefont {Morales-Inostroza},
  \citenamefont {Real}, \citenamefont {Mej\'{\i}a-Cort\'es}, \citenamefont
  {Weimann}, \citenamefont {Szameit},\ and\ \citenamefont
  {Molina}}]{Vicencio:2015}%
  \BibitemOpen
  \bibfield  {author} {\bibinfo {author} {\bibfnamefont {R.~A.}\ \bibnamefont
  {Vicencio}}, \bibinfo {author} {\bibfnamefont {C.}~\bibnamefont
  {Cantillano}}, \bibinfo {author} {\bibfnamefont {L.}~\bibnamefont
  {Morales-Inostroza}}, \bibinfo {author} {\bibfnamefont {B.}~\bibnamefont
  {Real}}, \bibinfo {author} {\bibfnamefont {C.}~\bibnamefont
  {Mej\'{\i}a-Cort\'es}}, \bibinfo {author} {\bibfnamefont {S.}~\bibnamefont
  {Weimann}}, \bibinfo {author} {\bibfnamefont {A.}~\bibnamefont {Szameit}}, \
  and\ \bibinfo {author} {\bibfnamefont {M.~I.}\ \bibnamefont {Molina}},\
  }\bibfield  {title} {\enquote {\bibinfo {title} {Observation of localized
  states in {Lieb} photonic lattices},}\ }\href {\doibase
  10.1103/PhysRevLett.114.245503} {\bibfield  {journal} {\bibinfo  {journal}
  {Phys. Rev. Lett.}\ }\textbf {\bibinfo {volume} {114}},\ \bibinfo {pages}
  {245503} (\bibinfo {year} {2015})}\BibitemShut {NoStop}%
\bibitem [{\citenamefont {Mukherjee}\ \emph {et~al.}(2015)\citenamefont
  {Mukherjee}, \citenamefont {Spracklen}, \citenamefont {Choudhury},
  \citenamefont {Goldman}, \citenamefont {\"Ohberg}, \citenamefont
  {Andersson},\ and\ \citenamefont {Thomson}}]{Mukherjee:2015}%
  \BibitemOpen
  \bibfield  {author} {\bibinfo {author} {\bibfnamefont {S.}~\bibnamefont
  {Mukherjee}}, \bibinfo {author} {\bibfnamefont {A.}~\bibnamefont
  {Spracklen}}, \bibinfo {author} {\bibfnamefont {D.}~\bibnamefont
  {Choudhury}}, \bibinfo {author} {\bibfnamefont {N.}~\bibnamefont {Goldman}},
  \bibinfo {author} {\bibfnamefont {P.}~\bibnamefont {\"Ohberg}}, \bibinfo
  {author} {\bibfnamefont {E.}~\bibnamefont {Andersson}}, \ and\ \bibinfo
  {author} {\bibfnamefont {R.~R.}\ \bibnamefont {Thomson}},\ }\bibfield
  {title} {\enquote {\bibinfo {title} {Observation of a localized flat-band
  state in a photonic {Lieb} lattice},}\ }\href {\doibase
  10.1103/PhysRevLett.114.245504} {\bibfield  {journal} {\bibinfo  {journal}
  {Phys. Rev. Lett.}\ }\textbf {\bibinfo {volume} {114}},\ \bibinfo {pages}
  {245504} (\bibinfo {year} {2015})}\BibitemShut {NoStop}%
\bibitem [{\citenamefont {Li}\ \emph {et~al.}(2014)\citenamefont {Li},
  \citenamefont {Guo}, \citenamefont {Zhang},\ and\ \citenamefont
  {Zhang}}]{Li2014}%
  \BibitemOpen
  \bibfield  {author} {\bibinfo {author} {\bibfnamefont {W.}~\bibnamefont
  {Li}}, \bibinfo {author} {\bibfnamefont {M.}~\bibnamefont {Guo}}, \bibinfo
  {author} {\bibfnamefont {G.}~\bibnamefont {Zhang}}, \ and\ \bibinfo {author}
  {\bibfnamefont {Y.-W.}\ \bibnamefont {Zhang}},\ }\bibfield  {title} {\enquote
  {\bibinfo {title} {Gapless {${\text{MoS}}_{2}$} allotrope possessing both
  massless dirac and heavy fermions},}\ }\href {\doibase
  10.1103/PhysRevB.89.205402} {\bibfield  {journal} {\bibinfo  {journal} {Phys.
  Rev. B}\ }\textbf {\bibinfo {volume} {89}},\ \bibinfo {pages} {205402}
  (\bibinfo {year} {2014})}\BibitemShut {NoStop}%
\bibitem [{\citenamefont {Wang}\ \emph {et~al.}(2018)\citenamefont {Wang},
  \citenamefont {Liu}, \citenamefont {Yu}, \citenamefont {Sheng}, \citenamefont
  {Zhu}, \citenamefont {Guan},\ and\ \citenamefont {Yang}}]{Wangss2018}%
  \BibitemOpen
  \bibfield  {author} {\bibinfo {author} {\bibfnamefont {S.-S.}\ \bibnamefont
  {Wang}}, \bibinfo {author} {\bibfnamefont {Y.}~\bibnamefont {Liu}}, \bibinfo
  {author} {\bibfnamefont {Z.-M.}\ \bibnamefont {Yu}}, \bibinfo {author}
  {\bibfnamefont {X.-L.}\ \bibnamefont {Sheng}}, \bibinfo {author}
  {\bibfnamefont {L.}~\bibnamefont {Zhu}}, \bibinfo {author} {\bibfnamefont
  {S.}~\bibnamefont {Guan}}, \ and\ \bibinfo {author} {\bibfnamefont {S.~A.}\
  \bibnamefont {Yang}},\ }\bibfield  {title} {\enquote {\bibinfo {title}
  {Monolayer {${\mathrm{Mg}}_{2}\mathrm{C}$}: Negative {Poisson}'s ratio and
  unconventional two-dimensional emergent fermions},}\ }\href {\doibase
  10.1103/PhysRevMaterials.2.104003} {\bibfield  {journal} {\bibinfo  {journal}
  {Phys. Rev. Materials}\ }\textbf {\bibinfo {volume} {2}},\ \bibinfo {pages}
  {104003} (\bibinfo {year} {2018})}\BibitemShut {NoStop}%
\bibitem [{\citenamefont {Giovannetti}\ \emph {et~al.}(2015)\citenamefont
  {Giovannetti}, \citenamefont {Capone}, \citenamefont {van~den Brink},\ and\
  \citenamefont {Ortix}}]{Gio2015}%
  \BibitemOpen
  \bibfield  {author} {\bibinfo {author} {\bibfnamefont {G.}~\bibnamefont
  {Giovannetti}}, \bibinfo {author} {\bibfnamefont {M.}~\bibnamefont {Capone}},
  \bibinfo {author} {\bibfnamefont {J.}~\bibnamefont {van~den Brink}}, \ and\
  \bibinfo {author} {\bibfnamefont {C.}~\bibnamefont {Ortix}},\ }\bibfield
  {title} {\enquote {\bibinfo {title} {Kekul\'e textures, pseudospin-one
  {Dirac} cones, and quadratic band crossings in a graphene-hexagonal indium
  chalcogenide bilayer},}\ }\href {\doibase 10.1103/PhysRevB.91.121417}
  {\bibfield  {journal} {\bibinfo  {journal} {Phys. Rev. B}\ }\textbf {\bibinfo
  {volume} {91}},\ \bibinfo {pages} {121417} (\bibinfo {year}
  {2015})}\BibitemShut {NoStop}%
\bibitem [{\citenamefont {Green}\ \emph {et~al.}(2010)\citenamefont {Green},
  \citenamefont {Santos},\ and\ \citenamefont {Chamon}}]{Green2010}%
  \BibitemOpen
  \bibfield  {author} {\bibinfo {author} {\bibfnamefont {D.}~\bibnamefont
  {Green}}, \bibinfo {author} {\bibfnamefont {L.}~\bibnamefont {Santos}}, \
  and\ \bibinfo {author} {\bibfnamefont {C.}~\bibnamefont {Chamon}},\
  }\bibfield  {title} {\enquote {\bibinfo {title} {Isolated flat bands and
  spin-1 conical bands in two-dimensional lattices},}\ }\href {\doibase
  10.1103/PhysRevB.82.075104} {\bibfield  {journal} {\bibinfo  {journal} {Phys.
  Rev. B}\ }\textbf {\bibinfo {volume} {82}},\ \bibinfo {pages} {075104}
  (\bibinfo {year} {2010})}\BibitemShut {NoStop}%
\bibitem [{\citenamefont {D\'ora}\ \emph {et~al.}(2011)\citenamefont {D\'ora},
  \citenamefont {Kailasvuori},\ and\ \citenamefont {Moessner}}]{Dora2011}%
  \BibitemOpen
  \bibfield  {author} {\bibinfo {author} {\bibfnamefont {B.}~\bibnamefont
  {D\'ora}}, \bibinfo {author} {\bibfnamefont {J.}~\bibnamefont {Kailasvuori}},
  \ and\ \bibinfo {author} {\bibfnamefont {R.}~\bibnamefont {Moessner}},\
  }\bibfield  {title} {\enquote {\bibinfo {title} {Lattice generalization of
  the {Dirac} equation to general spin and the role of the flat band},}\ }\href
  {\doibase 10.1103/PhysRevB.84.195422} {\bibfield  {journal} {\bibinfo
  {journal} {Phys. Rev. B}\ }\textbf {\bibinfo {volume} {84}},\ \bibinfo
  {pages} {195422} (\bibinfo {year} {2011})}\BibitemShut {NoStop}%
\bibitem [{\citenamefont {Goda}\ \emph {et~al.}(2006)\citenamefont {Goda},
  \citenamefont {Nishino},\ and\ \citenamefont {Matsuda}}]{Goda2006}%
  \BibitemOpen
  \bibfield  {author} {\bibinfo {author} {\bibfnamefont {M.}~\bibnamefont
  {Goda}}, \bibinfo {author} {\bibfnamefont {S.}~\bibnamefont {Nishino}}, \
  and\ \bibinfo {author} {\bibfnamefont {H.}~\bibnamefont {Matsuda}},\
  }\bibfield  {title} {\enquote {\bibinfo {title} {Inverse {Anderson}
  transition caused by flatbands},}\ }\href {\doibase
  10.1103/PhysRevLett.96.126401} {\bibfield  {journal} {\bibinfo  {journal}
  {Phys. Rev. Lett.}\ }\textbf {\bibinfo {volume} {96}},\ \bibinfo {pages}
  {126401} (\bibinfo {year} {2006})}\BibitemShut {NoStop}%
\bibitem [{\citenamefont {Bodyfelt}\ \emph {et~al.}(2014)\citenamefont
  {Bodyfelt}, \citenamefont {Leykam}, \citenamefont {Danieli}, \citenamefont
  {Yu},\ and\ \citenamefont {Flach}}]{Bodyfelt2014}%
  \BibitemOpen
  \bibfield  {author} {\bibinfo {author} {\bibfnamefont {J.~D.}\ \bibnamefont
  {Bodyfelt}}, \bibinfo {author} {\bibfnamefont {D.}~\bibnamefont {Leykam}},
  \bibinfo {author} {\bibfnamefont {C.}~\bibnamefont {Danieli}}, \bibinfo
  {author} {\bibfnamefont {X.}~\bibnamefont {Yu}}, \ and\ \bibinfo {author}
  {\bibfnamefont {S.}~\bibnamefont {Flach}},\ }\bibfield  {title} {\enquote
  {\bibinfo {title} {Flatbands under correlated perturbations},}\ }\href
  {\doibase 10.1103/PhysRevLett.113.236403} {\bibfield  {journal} {\bibinfo
  {journal} {Phys. Rev. Lett.}\ }\textbf {\bibinfo {volume} {113}},\ \bibinfo
  {pages} {236403} (\bibinfo {year} {2014})}\BibitemShut {NoStop}%
\bibitem [{\citenamefont {Taie}\ \emph {et~al.}(2015)\citenamefont {Taie},
  \citenamefont {Ozawa}, \citenamefont {Ichinose}, \citenamefont {Nishio},
  \citenamefont {Nakajima},\ and\ \citenamefont {Takahashi}}]{Taiee2015}%
  \BibitemOpen
  \bibfield  {author} {\bibinfo {author} {\bibfnamefont {S.}~\bibnamefont
  {Taie}}, \bibinfo {author} {\bibfnamefont {H.}~\bibnamefont {Ozawa}},
  \bibinfo {author} {\bibfnamefont {T.}~\bibnamefont {Ichinose}}, \bibinfo
  {author} {\bibfnamefont {T.}~\bibnamefont {Nishio}}, \bibinfo {author}
  {\bibfnamefont {S.}~\bibnamefont {Nakajima}}, \ and\ \bibinfo {author}
  {\bibfnamefont {Y.}~\bibnamefont {Takahashi}},\ }\bibfield  {title} {\enquote
  {\bibinfo {title} {Coherent driving and freezing of bosonic matter wave in an
  optical {Lieb} lattice},}\ }\href {\doibase 10.1126/sciadv.1500854}
  {\bibfield  {journal} {\bibinfo  {journal} {Sci. Adv.}\ }\textbf {\bibinfo
  {volume} {1}} (\bibinfo {year} {2015}),\ 10.1126/sciadv.1500854}\BibitemShut
  {NoStop}%
\bibitem [{\citenamefont {Julku}\ \emph {et~al.}(2016)\citenamefont {Julku},
  \citenamefont {Peotta}, \citenamefont {Vanhala}, \citenamefont {Kim},\ and\
  \citenamefont {T\"orm\"a}}]{Julku2016}%
  \BibitemOpen
  \bibfield  {author} {\bibinfo {author} {\bibfnamefont {A.}~\bibnamefont
  {Julku}}, \bibinfo {author} {\bibfnamefont {S.}~\bibnamefont {Peotta}},
  \bibinfo {author} {\bibfnamefont {T.~I.}\ \bibnamefont {Vanhala}}, \bibinfo
  {author} {\bibfnamefont {D.-H.}\ \bibnamefont {Kim}}, \ and\ \bibinfo
  {author} {\bibfnamefont {P.}~\bibnamefont {T\"orm\"a}},\ }\bibfield  {title}
  {\enquote {\bibinfo {title} {Geometric origin of superfluidity in the
  {Lieb}-lattice flat band},}\ }\href {\doibase 10.1103/PhysRevLett.117.045303}
  {\bibfield  {journal} {\bibinfo  {journal} {Phys. Rev. Lett.}\ }\textbf
  {\bibinfo {volume} {117}},\ \bibinfo {pages} {045303} (\bibinfo {year}
  {2016})}\BibitemShut {NoStop}%
\bibitem [{\citenamefont {Roy}\ and\ \citenamefont {Juri\ifmmode \check{c}\else
  \v{c}\fi{}i\ifmmode~\acute{c}\else \'{c}\fi{}}(2019)}]{Roy2019}%
  \BibitemOpen
  \bibfield  {author} {\bibinfo {author} {\bibfnamefont {B.}~\bibnamefont
  {Roy}}\ and\ \bibinfo {author} {\bibfnamefont {V.}~\bibnamefont {Juri\ifmmode
  \check{c}\else \v{c}\fi{}i\ifmmode~\acute{c}\else \'{c}\fi{}}},\ }\bibfield
  {title} {\enquote {\bibinfo {title} {Unconventional superconductivity in
  nearly flat bands in twisted bilayer graphene},}\ }\href {\doibase
  10.1103/PhysRevB.99.121407} {\bibfield  {journal} {\bibinfo  {journal} {Phys.
  Rev. B}\ }\textbf {\bibinfo {volume} {99}},\ \bibinfo {pages} {121407}
  (\bibinfo {year} {2019})}\BibitemShut {NoStop}%
\bibitem [{\citenamefont {Guo}\ \emph {et~al.}(2018)\citenamefont {Guo},
  \citenamefont {Zhu}, \citenamefont {Feng},\ and\ \citenamefont
  {Scalettar}}]{Guo2018}%
  \BibitemOpen
  \bibfield  {author} {\bibinfo {author} {\bibfnamefont {H.}~\bibnamefont
  {Guo}}, \bibinfo {author} {\bibfnamefont {X.}~\bibnamefont {Zhu}}, \bibinfo
  {author} {\bibfnamefont {S.}~\bibnamefont {Feng}}, \ and\ \bibinfo {author}
  {\bibfnamefont {R.~T.}\ \bibnamefont {Scalettar}},\ }\bibfield  {title}
  {\enquote {\bibinfo {title} {Pairing symmetry of interacting fermions on a
  twisted bilayer graphene superlattice},}\ }\href {\doibase
  10.1103/PhysRevB.97.235453} {\bibfield  {journal} {\bibinfo  {journal} {Phys.
  Rev. B}\ }\textbf {\bibinfo {volume} {97}},\ \bibinfo {pages} {235453}
  (\bibinfo {year} {2018})}\BibitemShut {NoStop}%
\bibitem [{\citenamefont {Yu}(1965)}]{Yu:1965}%
  \BibitemOpen
  \bibfield  {author} {\bibinfo {author} {\bibfnamefont {L.}~\bibnamefont
  {Yu}},\ }\bibfield  {title} {\enquote {\bibinfo {title} {Bound state in
  superconductors with paramagnetic impurities},}\ }\href {\doibase
  10.7498/aps.21.75} {\bibfield  {journal} {\bibinfo  {journal} {Acta Phys.
  Sinica}\ }\textbf {\bibinfo {volume} {21}},\ \bibinfo {pages} {75} (\bibinfo
  {year} {1965})}\BibitemShut {NoStop}%
\bibitem [{\citenamefont {Shiba}(1968)}]{Shiba:1968}%
  \BibitemOpen
  \bibfield  {author} {\bibinfo {author} {\bibfnamefont {H.}~\bibnamefont
  {Shiba}},\ }\bibfield  {title} {\enquote {\bibinfo {title} {Classical spins
  in superconductors},}\ }\href {\doibase 10.1143/PTP.40.435} {\bibfield
  {journal} {\bibinfo  {journal} {Prog. Theo. Phys.}\ }\textbf {\bibinfo
  {volume} {40}},\ \bibinfo {pages} {435} (\bibinfo {year} {1968})}\BibitemShut
  {NoStop}%
\bibitem [{\citenamefont {Rusinov}(1969)}]{Rusinov:1969}%
  \BibitemOpen
  \bibfield  {author} {\bibinfo {author} {\bibfnamefont {A.~I.}\ \bibnamefont
  {Rusinov}},\ }\bibfield  {title} {\enquote {\bibinfo {title}
  {Superconductivity near a paramagnetic impurity},}\ }\href@noop {} {\bibfield
   {journal} {\bibinfo  {journal} {JETP Lett.}\ }\textbf {\bibinfo {volume}
  {9}},\ \bibinfo {pages} {85} (\bibinfo {year} {1969})}\BibitemShut {NoStop}%
\bibitem [{\citenamefont {Castro}\ \emph {et~al.}(2008)\citenamefont {Castro},
  \citenamefont {Peres}, \citenamefont {Lopes~dos Santos}, \citenamefont
  {Neto},\ and\ \citenamefont {Guinea}}]{CPLNG:2008}%
  \BibitemOpen
  \bibfield  {author} {\bibinfo {author} {\bibfnamefont {E.~V.}\ \bibnamefont
  {Castro}}, \bibinfo {author} {\bibfnamefont {N.~M.~R.}\ \bibnamefont
  {Peres}}, \bibinfo {author} {\bibfnamefont {J.~M.~B.}\ \bibnamefont
  {Lopes~dos Santos}}, \bibinfo {author} {\bibfnamefont {A.~H.~C.}\
  \bibnamefont {Neto}}, \ and\ \bibinfo {author} {\bibfnamefont
  {F.}~\bibnamefont {Guinea}},\ }\bibfield  {title} {\enquote {\bibinfo {title}
  {Localized states at zigzag edges of bilayer graphene},}\ }\href {\doibase
  10.1103/PhysRevLett.100.026802} {\bibfield  {journal} {\bibinfo  {journal}
  {Phys. Rev. Lett.}\ }\textbf {\bibinfo {volume} {100}},\ \bibinfo {pages}
  {026802} (\bibinfo {year} {2008})}\BibitemShut {NoStop}%
\bibitem [{\citenamefont {Castro}\ \emph {et~al.}(2010)\citenamefont {Castro},
  \citenamefont {L\'opez-Sancho},\ and\ \citenamefont {Vozmediano}}]{CLV:2010}%
  \BibitemOpen
  \bibfield  {author} {\bibinfo {author} {\bibfnamefont {E.~V.}\ \bibnamefont
  {Castro}}, \bibinfo {author} {\bibfnamefont {M.~P.}\ \bibnamefont
  {L\'opez-Sancho}}, \ and\ \bibinfo {author} {\bibfnamefont {M.~A.~H.}\
  \bibnamefont {Vozmediano}},\ }\bibfield  {title} {\enquote {\bibinfo {title}
  {New type of vacancy-induced localized states in multilayer graphene},}\
  }\href {\doibase 10.1103/PhysRevLett.104.036802} {\bibfield  {journal}
  {\bibinfo  {journal} {Phys. Rev. Lett.}\ }\textbf {\bibinfo {volume} {104}},\
  \bibinfo {pages} {036802} (\bibinfo {year} {2010})}\BibitemShut {NoStop}%
\bibitem [{\citenamefont {Lu}\ \emph {et~al.}(2011)\citenamefont {Lu},
  \citenamefont {Shan}, \citenamefont {Lu},\ and\ \citenamefont
  {Shen}}]{LSLS:2011}%
  \BibitemOpen
  \bibfield  {author} {\bibinfo {author} {\bibfnamefont {J.}~\bibnamefont
  {Lu}}, \bibinfo {author} {\bibfnamefont {W.-Y.}\ \bibnamefont {Shan}},
  \bibinfo {author} {\bibfnamefont {H.-Z.}\ \bibnamefont {Lu}}, \ and\ \bibinfo
  {author} {\bibfnamefont {S.-Q.}\ \bibnamefont {Shen}},\ }\bibfield  {title}
  {\enquote {\bibinfo {title} {Non-magnetic impurities and in-gap bound states
  in topological insulators},}\ }\href {\doibase
  10.1088/1367-2630/13/10/103016} {\bibfield  {journal} {\bibinfo  {journal}
  {New J. Phys.}\ }\textbf {\bibinfo {volume} {13}},\ \bibinfo {pages} {103016}
  (\bibinfo {year} {2011})}\BibitemShut {NoStop}%
\bibitem [{\citenamefont {Shtanko}\ and\ \citenamefont
  {Levitov}(2018)}]{SL:2018}%
  \BibitemOpen
  \bibfield  {author} {\bibinfo {author} {\bibfnamefont {O.}~\bibnamefont
  {Shtanko}}\ and\ \bibinfo {author} {\bibfnamefont {L.}~\bibnamefont
  {Levitov}},\ }\bibfield  {title} {\enquote {\bibinfo {title} {Robustness and
  universality of surface states in {Dirac} materials},}\ }\href {\doibase
  10.1073/pnas.1722663115} {\bibfield  {journal} {\bibinfo  {journal} {Proc.
  Natl. Acad. Sci. (USA)}\ }\textbf {\bibinfo {volume} {115}},\ \bibinfo
  {pages} {5908} (\bibinfo {year} {2018})}\BibitemShut {NoStop}%
\bibitem [{\citenamefont {Abramowitz}\ and\ \citenamefont
  {Stegun}(2012)}]{AbrSte2012}%
  \BibitemOpen
  \bibfield  {author} {\bibinfo {author} {\bibfnamefont {M.}~\bibnamefont
  {Abramowitz}}\ and\ \bibinfo {author} {\bibfnamefont {I.}~\bibnamefont
  {Stegun}},\ }\href@noop {} {\emph {\bibinfo {title} {Handbook of Mathematical
  Functions: with Formulas, Graphs, and Mathematical Tables}}},\ Dover Books on
  Mathematics\ (\bibinfo  {publisher} {Dover Publications},\ \bibinfo {year}
  {2012})\BibitemShut {NoStop}%
\bibitem [{\citenamefont {Zhao}\ \emph {et~al.}(2015)\citenamefont {Zhao},
  \citenamefont {Wyrick}, \citenamefont {Natterer}, \citenamefont
  {Rodriguez-Nieva}, \citenamefont {Lewandowski}, \citenamefont {Watanabe},
  \citenamefont {Taniguchi}, \citenamefont {Levitov}, \citenamefont
  {Zhitenev},\ and\ \citenamefont {Stroscio}}]{Zhao672}%
  \BibitemOpen
  \bibfield  {author} {\bibinfo {author} {\bibfnamefont {Y.}~\bibnamefont
  {Zhao}}, \bibinfo {author} {\bibfnamefont {J.}~\bibnamefont {Wyrick}},
  \bibinfo {author} {\bibfnamefont {F.~D.}\ \bibnamefont {Natterer}}, \bibinfo
  {author} {\bibfnamefont {J.~F.}\ \bibnamefont {Rodriguez-Nieva}}, \bibinfo
  {author} {\bibfnamefont {C.}~\bibnamefont {Lewandowski}}, \bibinfo {author}
  {\bibfnamefont {K.}~\bibnamefont {Watanabe}}, \bibinfo {author}
  {\bibfnamefont {T.}~\bibnamefont {Taniguchi}}, \bibinfo {author}
  {\bibfnamefont {L.~S.}\ \bibnamefont {Levitov}}, \bibinfo {author}
  {\bibfnamefont {N.~B.}\ \bibnamefont {Zhitenev}}, \ and\ \bibinfo {author}
  {\bibfnamefont {J.~A.}\ \bibnamefont {Stroscio}},\ }\bibfield  {title}
  {\enquote {\bibinfo {title} {Creating and probing electron whispering-gallery
  modes in graphene},}\ }\href {\doibase 10.1126/science.aaa7469} {\bibfield
  {journal} {\bibinfo  {journal} {Science}\ }\textbf {\bibinfo {volume}
  {348}},\ \bibinfo {pages} {672} (\bibinfo {year} {2015})}\BibitemShut
  {NoStop}%
\bibitem [{\citenamefont {Rodriguez-Nieva}\ and\ \citenamefont
  {Levitov}(2016)}]{Nieva2016}%
  \BibitemOpen
  \bibfield  {author} {\bibinfo {author} {\bibfnamefont {J.~F.}\ \bibnamefont
  {Rodriguez-Nieva}}\ and\ \bibinfo {author} {\bibfnamefont {L.~S.}\
  \bibnamefont {Levitov}},\ }\bibfield  {title} {\enquote {\bibinfo {title}
  {{Berry} phase jumps and giant nonreciprocity in dirac quantum dots},}\
  }\href {\doibase 10.1103/PhysRevB.94.235406} {\bibfield  {journal} {\bibinfo
  {journal} {Phys. Rev. B}\ }\textbf {\bibinfo {volume} {94}},\ \bibinfo
  {pages} {235406} (\bibinfo {year} {2016})}\BibitemShut {NoStop}%
\bibitem [{\citenamefont {Lee}\ \emph {et~al.}(2016)\citenamefont {Lee},
  \citenamefont {Wong}, \citenamefont {Velasco~Jr}, \citenamefont
  {Rodriguez-Nieva}, \citenamefont {Kahn}, \citenamefont {Tsai}, \citenamefont
  {Taniguchi}, \citenamefont {Watanabe}, \citenamefont {Zettl}, \citenamefont
  {Wang}, \citenamefont {Levitov},\ and\ \citenamefont {Crommie}}]{Lee2016}%
  \BibitemOpen
  \bibfield  {author} {\bibinfo {author} {\bibfnamefont {J.}~\bibnamefont
  {Lee}}, \bibinfo {author} {\bibfnamefont {D.}~\bibnamefont {Wong}}, \bibinfo
  {author} {\bibfnamefont {J.}~\bibnamefont {Velasco~Jr}}, \bibinfo {author}
  {\bibfnamefont {J.~F.}\ \bibnamefont {Rodriguez-Nieva}}, \bibinfo {author}
  {\bibfnamefont {S.}~\bibnamefont {Kahn}}, \bibinfo {author} {\bibfnamefont
  {H.-Z.}\ \bibnamefont {Tsai}}, \bibinfo {author} {\bibfnamefont
  {T.}~\bibnamefont {Taniguchi}}, \bibinfo {author} {\bibfnamefont
  {K.}~\bibnamefont {Watanabe}}, \bibinfo {author} {\bibfnamefont
  {A.}~\bibnamefont {Zettl}}, \bibinfo {author} {\bibfnamefont
  {F.}~\bibnamefont {Wang}}, \bibinfo {author} {\bibfnamefont {L.~S.}\
  \bibnamefont {Levitov}}, \ and\ \bibinfo {author} {\bibfnamefont {M.~F.}\
  \bibnamefont {Crommie}},\ }\bibfield  {title} {\enquote {\bibinfo {title}
  {Imaging electrostatically confined {Dirac} fermions in graphene quantum
  dots},}\ }\href@noop {} {\bibfield  {journal} {\bibinfo  {journal} {Nat.
  Phys.}\ }\textbf {\bibinfo {volume} {12}},\ \bibinfo {pages} {1032} (\bibinfo
  {year} {2016})}\BibitemShut {NoStop}%
\bibitem [{\citenamefont {Ghahari}\ \emph {et~al.}(2017)\citenamefont
  {Ghahari}, \citenamefont {Walkup}, \citenamefont {Guti{\'e}rrez},
  \citenamefont {Rodriguez-Nieva}, \citenamefont {Zhao}, \citenamefont
  {Wyrick}, \citenamefont {Natterer}, \citenamefont {Cullen}, \citenamefont
  {Watanabe}, \citenamefont {Taniguchi}, \citenamefont {Levitov}, \citenamefont
  {Zhitenev},\ and\ \citenamefont {Stroscio}}]{Ghahari845}%
  \BibitemOpen
  \bibfield  {author} {\bibinfo {author} {\bibfnamefont {F.}~\bibnamefont
  {Ghahari}}, \bibinfo {author} {\bibfnamefont {D.}~\bibnamefont {Walkup}},
  \bibinfo {author} {\bibfnamefont {C.}~\bibnamefont {Guti{\'e}rrez}}, \bibinfo
  {author} {\bibfnamefont {J.~F.}\ \bibnamefont {Rodriguez-Nieva}}, \bibinfo
  {author} {\bibfnamefont {Y.}~\bibnamefont {Zhao}}, \bibinfo {author}
  {\bibfnamefont {J.}~\bibnamefont {Wyrick}}, \bibinfo {author} {\bibfnamefont
  {F.~D.}\ \bibnamefont {Natterer}}, \bibinfo {author} {\bibfnamefont {W.~G.}\
  \bibnamefont {Cullen}}, \bibinfo {author} {\bibfnamefont {K.}~\bibnamefont
  {Watanabe}}, \bibinfo {author} {\bibfnamefont {T.}~\bibnamefont {Taniguchi}},
  \bibinfo {author} {\bibfnamefont {L.~S.}\ \bibnamefont {Levitov}}, \bibinfo
  {author} {\bibfnamefont {N.~B.}\ \bibnamefont {Zhitenev}}, \ and\ \bibinfo
  {author} {\bibfnamefont {J.~A.}\ \bibnamefont {Stroscio}},\ }\bibfield
  {title} {\enquote {\bibinfo {title} {An on/off {Berry} phase switch in
  circular graphene resonators},}\ }\href {\doibase 10.1126/science.aal0212}
  {\bibfield  {journal} {\bibinfo  {journal} {Science}\ }\textbf {\bibinfo
  {volume} {356}},\ \bibinfo {pages} {845} (\bibinfo {year}
  {2017})}\BibitemShut {NoStop}%
\bibitem [{\citenamefont {Leviatan}\ and\ \citenamefont
  {Boag}(1987)}]{LB:1987}%
  \BibitemOpen
  \bibfield  {author} {\bibinfo {author} {\bibfnamefont {Y.}~\bibnamefont
  {Leviatan}}\ and\ \bibinfo {author} {\bibfnamefont {A.}~\bibnamefont
  {Boag}},\ }\bibfield  {title} {\enquote {\bibinfo {title} {Analysis of
  electromagnetic scattering from dielectric cylinders using a multifilament
  current model},}\ }\href {\doibase 10.1109/TAP.1987.1143994} {\bibfield
  {journal} {\bibinfo  {journal} {IEEE Trans. Anten. Propa.}\ }\textbf
  {\bibinfo {volume} {35}},\ \bibinfo {pages} {1119} (\bibinfo {year}
  {1987})}\BibitemShut {NoStop}%
\bibitem [{\citenamefont {Imhof}(1996)}]{Imhof:1996}%
  \BibitemOpen
  \bibfield  {author} {\bibinfo {author} {\bibfnamefont {M.~G.}\ \bibnamefont
  {Imhof}},\ }\bibfield  {title} {\enquote {\bibinfo {title} {Multiple
  multipole expansions for elastic scattering},}\ }\href {\doibase
  10.1121/1.417109} {\bibfield  {journal} {\bibinfo  {journal} {J. Acous. Soc.
  Am.}\ }\textbf {\bibinfo {volume} {100}},\ \bibinfo {pages} {2969} (\bibinfo
  {year} {1996})}\BibitemShut {NoStop}%
\bibitem [{\citenamefont {Kaklamani}\ and\ \citenamefont
  {Anastassiu}(2002)}]{KA:2002}%
  \BibitemOpen
  \bibfield  {author} {\bibinfo {author} {\bibfnamefont {D.~I.}\ \bibnamefont
  {Kaklamani}}\ and\ \bibinfo {author} {\bibfnamefont {H.~T.}\ \bibnamefont
  {Anastassiu}},\ }\bibfield  {title} {\enquote {\bibinfo {title} {Aspects of
  the method of auxiliary sources ({MAS}) in computational electromagnetics},}\
  }\href {\doibase 10.1109/MAP.2002.1028734} {\bibfield  {journal} {\bibinfo
  {journal} {IEEE Anten. Propag. Maga.}\ }\textbf {\bibinfo {volume} {44}},\
  \bibinfo {pages} {48} (\bibinfo {year} {2002})}\BibitemShut {NoStop}%
\bibitem [{\citenamefont {Moreno}\ \emph {et~al.}(2002)\citenamefont {Moreno},
  \citenamefont {Erni}, \citenamefont {Hafner},\ and\ \citenamefont
  {Vahldieck}}]{MEHV:2002}%
  \BibitemOpen
  \bibfield  {author} {\bibinfo {author} {\bibfnamefont {E.}~\bibnamefont
  {Moreno}}, \bibinfo {author} {\bibfnamefont {D.}~\bibnamefont {Erni}},
  \bibinfo {author} {\bibfnamefont {C.}~\bibnamefont {Hafner}}, \ and\ \bibinfo
  {author} {\bibfnamefont {R.}~\bibnamefont {Vahldieck}},\ }\bibfield  {title}
  {\enquote {\bibinfo {title} {Multiple multipole method with automatic
  multipole setting applied to the simulation of surface plasmons in metallic
  nanostructures},}\ }\href {\doibase 10.1364/JOSAA.19.000101} {\bibfield
  {journal} {\bibinfo  {journal} {J. Opt. Soc. Am. A}\ }\textbf {\bibinfo
  {volume} {19}},\ \bibinfo {pages} {101} (\bibinfo {year} {2002})}\BibitemShut
  {NoStop}%
\bibitem [{\citenamefont {Tayeb}\ and\ \citenamefont {Enoch}(2004)}]{TE:2004}%
  \BibitemOpen
  \bibfield  {author} {\bibinfo {author} {\bibfnamefont {G.}~\bibnamefont
  {Tayeb}}\ and\ \bibinfo {author} {\bibfnamefont {S.}~\bibnamefont {Enoch}},\
  }\bibfield  {title} {\enquote {\bibinfo {title} {Combined
  fictitious-sources--scattering-matrix method},}\ }\href {\doibase
  10.1364/JOSAA.21.001417} {\bibfield  {journal} {\bibinfo  {journal} {J. Opt.
  Soc. Am. A}\ }\textbf {\bibinfo {volume} {21}},\ \bibinfo {pages} {1417}
  (\bibinfo {year} {2004})}\BibitemShut {NoStop}%
\end{thebibliography}

%
\end{document}